\documentclass[11pt]{article}
\usepackage[a4paper,bottom=4.2cm,top=1cm,head=3cm,width=18cm,dvipdfm]{geometry}
\evensidemargin=\oddsidemargin

\usepackage[dotinlabels]{titletoc}
\usepackage{titlesec}

\usepackage[linktocpage=true,bookmarks=true,bookmarksnumbered=true,breaklinks=true,pdfpagemode=Fullscreen,pdfstartview=FitBH]{hyperref}

\usepackage{graphicx}
\usepackage{dcolumn}
\usepackage{bm}
\usepackage{amssymb}
\usepackage{mathrsfs}
\usepackage{amsmath}
\usepackage{epsfig}
\usepackage{here}
\usepackage[dvips]{color}
\usepackage{hhline}
\usepackage{srcltx}
\usepackage{multirow} 

\allowdisplaybreaks[4]

\renewcommand{\thefootnote}{\fnsymbol{footnote}}
\long\def\symbolfootnote[#1]#2{\begingroup
\def\thefootnote{\fnsymbol{footnote}}\footnote[#1]{#2}\endgroup}

\newcommand{\beq}{\begin{equation}}
\newcommand{\eeq}{\end{equation}}
\newcommand{\bq}{\begin{equation}}
\newcommand{\eq}{\end{equation}}
\newcommand{\ba}{\begin{array}}
\newcommand{\ea}{\end{array}}
\newcommand{\beqa}{\begin{eqnarray}}
\newcommand{\eeqa}{\end{eqnarray}}
\newcommand{\beqs}{\begin{subequations}}
\newcommand{\eeqs}{\end{subequations}}

\def\nn{\nonumber}

\def\dis{\displaystyle}
\def\f{\frac}
\def\({\left(}
\def\){\right)}
\def\LB{\left[}
\def\RB{\right]}
\def\hf{\frac{1}{2}}
\def\deg{\circ}
\def\End{\end{document}}

\def\DM{\textrm{DM}}
\def\B{\textrm{B}}
\def\SM{\textrm{SM}}
\def\DVs{\Delta V_{\textrm{soft}}}
\def\Vh{\widehat{V}}
\def\N{\mathcal{N}}
\def\NN{\mathbb{N}}
\def\ga{\gamma}
\def\Ga{\Gamma}

\def\De{\Delta}
\def\tm{\widetilde{m}}
\def\bm{\overline{m}}
\def\MP{M_{\rm P}^{}}

\def\over{\overline}

\def\leqq{\leqslant}
\def\geqq{\geqslant}
\def\tolr{\leftrightarrow}

\def\hf{\frac{1}{2}}

\def\be{\beta}

\def\la{\lambda}
\def\X{\chi}
\def\vX{v_{\chi}^{}}

\def\Xc{\chi_c^{}}
\def\vp{v_{\phi}^{}}

\def\vpb{v_{\phi'}^{}}

\def\O{{\cal O}}

\def\nuR{\nu_R^{}}

\def\lambdaT{\widetilde{\lambda}}
\def\laT{\widetilde{\lambda}}
\def\mut{\widetilde{\mu}}

\def\mutau{\mu\!-\!\tau}

\def\ep{\epsilon}
\def\phih{\hat{\phi}}
\def\chih{\hat{\chi}}
\def\hh{\hat{h}}
\def\hph{\hat{h}'}
\def\Xh{\hat{\chi}}
\def\rN{r_N^{}}

\numberwithin{equation}{section}

\begin{document}
 \thispagestyle{empty}
 \setcounter{footnote}{0}
 \titlelabel{\thetitle.\quad \hspace{-0.8em}}
\titlecontents{section}
              [1.5em]
              {\vspace{4mm} \large \bf}
              {\contentslabel{1em}}
              {\hspace*{-1em}}
              {\titlerule*[.5pc]{.}\contentspage}
\titlecontents{subsection}
              [3.5em]
              {\vspace{2mm}}
              {\contentslabel{1.8em}}
              {\hspace*{.3em}}
              {\titlerule*[.5pc]{.}\contentspage}
\titlecontents{subsubsection}
              [5.5em]
              {\vspace{2mm}}
              {\contentslabel{2.5em}}
              {\hspace*{.3em}}
              {\titlerule*[.5pc]{.}\contentspage}

 \begin{center}
 {\bf {\Large
 Spontaneous Mirror Parity Violation, Common Origin of \\[1mm]
Matter and Dark Matter, and the LHC Signatures}}

 \vspace*{8mm}

 {\sc Jian-Wei Cui}\,$^a$,~
 {\sc Hong-Jian He}\,$^{a,b,c}$,~ 
 {\sc Lan-Chun L\"{u}}\,$^a$,~
 {\sc Fu-Rong Yin}\,$^a$\,

\vspace*{3mm}

$^a$\,Institute of Modern Physics and Center for High Energy Physics,
\\
Tsinghua University, Beijing 100084, China
\\[1mm]
$^b$\,Center for High Energy Physics, Peking University, Beijing 100871, China
\\[1mm]
$^c$\,Kavli Institute for Theoretical Physics China, CAS, Beijing 100190, China\\

\vspace{25mm}
\end{center}

\begin{abstract}
\baselineskip 17pt
\noindent
Existence of a mirror world in the universe is a fundamental way to restore 
the observed parity violation in weak interactions and provides the lightest 
mirror nucleon as a unique GeV-scale dark matter particle candidate.
The visible and mirror worlds share the same spacetime of the universe and are connected
by a unique space-inversion symmetry --- the mirror parity ($P$).
We conjecture that the mirror parity is respected by the fundamental interaction
Lagrangian, and study its spontaneous breaking from minimizing the Higgs vacuum potential.
The domain wall problem is resolved by a unique soft breaking linear-term from the $P$-odd
weak-singlet Higgs field. We also derive constraint from the Big-Bang nucleosynthesis.
We then analyze the neutrino seesaw for both visible and mirror worlds, and demonstrate
that the desired amounts of visible matter and mirror dark matter in the universe arise
from a common origin of CP violation in the neutrino sector via leptogenesis.
We derive the Higgs mass-spectrum and Higgs couplings with gauge bosons and fermions. We show their consistency with the direct Higgs searches and the indirect precision constraints.
We further study the distinctive signatures of the predicted non-standard Higgs bosons at the LHC.
Finally, we analyze the direct detections of GeV-scale mirror dark matter
by TEXONO and CDEX experiments.
\\[4mm]
PACS numbers: 95.35.+d, 11.30.Er, 14.60.Pq, 98.80.Ft, 12.60.-i
\\[2mm]
Phys.\ Rev.\ D (2012),\, in Press. [\,{\tt arXiv:1110.6893}\,]

\end{abstract}


\newpage
 \setcounter{page}{2}

 \tableofcontents

 \newpage
 \setcounter{footnote}{0}
 \renewcommand{\thefootnote}{\arabic{footnote}}
 \baselineskip 18.5pt

\section{Introduction}
\label{sec:1}
\vspace*{1.5mm}

The experimental fact that weak force in our visible world only invokes
left-handed fermions does not necessarily imply the parity violation
in the whole universe. The possible existence of a hidden mirror world
in the universe as a fundamental way of restoring parity was first conceived 
by Lee and Yang in their seminal work in 1956\,\cite{LY1956}.
This truly simple and beautiful idea was further developed
by several groups in the following decades\,\cite{EMirror1,EMirror2,MRev},
where a mirror parity was introduced to connect the visible and mirror worlds.

On the other hand, astronomy and cosmology observations have pointed to
the existence of mystery dark matter which constitutes about 23\% of the
total energy density of the present universe. This is five times larger than
all the visible matter, $\,\Omega_{\rm DM}:\Omega_{\rm B}\simeq 5:1\,$,\,
but they are still {\it comparable} within one order of magnitude.
In parallel to the visible world, the mirror world conserves mirror baryon number
and thus protects the stability of the lightest mirror nucleon, providing
a natural GeV-scale dark matter candidate \cite{MRev,Kolb:1985bf,MDM2,An:2009vq}.
This raises an intriguing possibility that the right amount of dark matter is generated
via the mirror leptogenesis under mirror neutrino seesaw,
just like that the visible matter is generated
via ordinary leptogenesis\,\cite{LepG0,LepG-Rev}.

In this work, we will demonstrate that the mirror parity ($P$) can play a key role to
quantitatively connect the visible and mirror neutrino seesaws, including the
associated CP violations.  We conjecture that {\it the mirror parity is respected
by the fundamental interaction Lagrangian,} so its violation only arises from
spontaneous breaking of the Higgs vacuum, and the possible soft breaking can
only be linear or bilinear terms; we further conjecture that {\it all possible
soft breakings of mirror parity simply arise from the gauge-singlet sector.}
With this conceptually simple and attractive conjecture,
we will present a minimal model with spontaneous mirror parity violation,
and the domain wall problem is evaded by a unique soft breaking term in
the singlet Higgs sector.
This is unlike most of previous studies where the mirror parity
is assumed to be unbroken\,\cite{MRev}.
With this we can realize both the visible and dark matter geneses from a common
origin of CP violation in neutrino seesaws via leptogenesis,
as ensured by mirror parity between the two neutrino sectors.
Our minimal Higgs potential can generate
spontaneous mirror parity violation in the weak interaction, where
the visible and mirror Higgs bosons develop different vacuum expectation values (VEVs).
Our neutrino seesaw sector has another unique soft breaking
term with unequal masses of visible and mirror singlet heavy Majorana neutrinos.
These will make the masses of mirror particles differ from the corresponding visible particles
in the standard model (SM), and also cause a different efficiency factor of out-of-equilibrium
decays for the heavy singlet mirror neutrinos. We then demonstrate how the right amount of
visible and dark matter can be generated from a common origin of CP violation.

Our model has two self-contained seesaw sectors, the visible seesaw and mirror seesaw,
with the corresponding visible and mirror singlet Majorana neutrinos.
It is the mirror parity that plays the key role to quantitatively connect the two seesaws
(including the exactly equal CP-phases) and thus ensures the common origin of generating
the right amount of visible and mirror dark matter.
We will present systematical analysis of the minimal Higgs potential, and quantitatively
realize the spontaneous breaking of both the mirror parity and the electroweak gauge symmetry.
The mirror Higgs VEV is found to be about a factor-2 smaller than the visible Higgs VEV.
We then derive the distinctive mass-spectrum of Higgs bosons and their couplings, leading to
new collider phenomenology, different from all previous mirror model signals.
We also analyze the existing low energy constraints via electroweak precision measurements
and direct production.
We further study the new signatures of predicted non-standard Higgs bosons
at the LHC. Finally, we analyze the direct detections of mirror dark matter.
Our construction also fully differs from a recent interesting study\,\cite{An:2009vq}
with resonant leptogensis for the matter and dark matter genesis,
where the visible and mirror sectors share the same right-handed neutrinos with
inverse seesaw, and the ratio $\,\Omega_{\rm DM}/\Omega_{\rm B}\simeq 5\,$ arises from
an assumed large ratio (about 1000) of the two VEVs for the mirror and visible Higgs bosons
which causes the mirror nucleon about a factor-5 heavier than the visible nucleon.
Two Higgs doubelts and one Higgs triplet with soft mirror parity breaking
are introduced for both sectors, which will generate a mass $\sim 50$\,MeV for mirror photons
and masses $\gtrsim\! 100$\,MeV for light mirror neutrinos.
The existence of proper Higgs potential and its minimum are assumed in \cite{An:2009vq}.

This paper is organized as follows. In Sec.\,\ref{sec:2}, we show that a unique mirror parity can be
introduced to connect the visible and mirror worlds, as a fundamental way to restore the parity
in weak interactions. We then construct a minimal Higgs potential and derive conditions for
its physical vacuum to realize both the spontaneous mirror parity violation and
spontaneous electroweak symmetry breaking. In Sec.\,\ref{sec:3}, we analyze the visible and mirror
leptogeneses via neutrino seesaws, with a common origin of CP violations. We then derive the
conditions for generating the right amount of visible and mirror dark matter.
In Sec.\,\ref{sec:4}, we study the analytical structure of the vacuum Higgs potential, and then
present three numerical samples for the Higgs vacuum and the corresponding
Higgs mass-spectrum that obey the conditions for desired matter and dark matter geneses.
We further demonstrate their consistency with the current low energy precision constraints.
In Sec.\,\ref{sec:5}, we study the distinctive collider signatures of
the non-standard Higgs bosons at the LHC.
We further analyze the direct detections of GeV-scale mirror dark matter by
TEXONO \cite{TEX} and CDEX \cite{JP} experiments in Sec.\,\ref{sec:6}.
Finally, we conclude in Sec.\,\ref{sec:7}.

\vspace*{3mm}
\section{Spontaneous Mirror Parity Violation}
\label{sec:2}
\vspace*{1.5mm}

In Sec.\,\ref{sec:2-1}, we will first analyze the structure of the mirror model with unbroken
mirror parity, and then we discuss the connections between the visible and mirror
worlds in Sec.\,\ref{sec:2-2}.
We conjecture that {\it the mirror parity is respected by the fundamental interaction
Lagrangian,} so its violation only arises from spontaneous breaking of the Higgs vacuum,
and the possible soft breaking can only be linear or bilinear terms;
we further conjecture that {\it all possible soft breakings simply arise from
the gauge-singlet sector alone.}
With this conceptually simple and attractive conjecture,
we will present a minimal model with spontaneous mirror parity violation
(Sec.\,\ref{sec:3}), where the Higgs sector includes the SM Higgs doublet, the mirror Higgs doublet
and a $P$-odd weak singlet scalar. We find that the possible soft breakings can be
uniquely realized via the $P$-odd weak singlet scalar in the Higgs potential
and via the Majorana mass-terms of heavy singlet neutrinos in the seesaw sector.
As we will show, the unique soft breaking in Higgs potential nicely evades
the domain wall problem\,\cite{DW}
associated with spontaneous mirror parity violation (Sec.\,\ref{sec:2-3}),
and the unique soft breaking in the heavy Majorana mass term will play a key
role to realize the desired dark matter density (Sec.\,\ref{sec:3-1}-\ref{sec:3-2}).

\vspace*{3mm}
\subsection{Structure of the Model}
\label{sec:2-1}
\vspace*{1.5mm}

The visible and mirror worlds share the same spacetime of the universe,
this leads to a unique space-inversion symmetry -- the mirror parity.
We know that the representations of Lorentz group can be characterized
by $SU(2)\otimes SU(2)$ with generators
$\,A_{i}=\frac{1}{2}(J_{i}+iK_{i})\,$ and
$\,B_{i}=\frac{1}{2}(J_{i}-iK_{i})\,$, ($i,j=1, 2, 3$), where
$J_{i}$ is angular momentum and $K_{i}$ the Lorentz boost.
So each representation is labeled by two angular momenta $(j,\ j^{\prime})$,
corresponding to $A$ and $B$, respectively.
Under the parity transformation ${\cal P}$, we have
$\,\mathcal{P}J_{i}\mathcal{P}^{-1}=J_{i}\,$ and
$\,\mathcal{P}K_{i}\mathcal{P}^{-1}=-K_{i}\,$.\,
This means the exchange $\,A_i\leftrightarrow B_i\,$,\, i.e.,
the parity operator transforms a representation $(j,\, j')$
into $(j',\, j)$. In the SM, the left-handed fermions
belong to $(\frac{1}{2},\, 0)$ and group into $SU(2)_L$
doublets, while the right-handed fermions belong to $(0,\, \frac{1}{2})$
and are $SU(2)_L$ singlets. Hence the parity symmetry is explicitly broken
in the SM by the weak interaction.

There are two fundamental ways to restore parity symmetry.
One is to enlarge the weak gauge group $SU(2)_L$ into a left-right symmetric form,
$SU(2)_L\otimes SU(2)_R$. Assigning the left-handed
fermions to $SU(2)_L$ doublets (but $SU(2)_R$ singlets), and right-handed
fermions to $SU(2)_R$ doublets (but $SU(2)_L$ singlets). Then, assigning
$SU(2)_L\leftrightarrow SU(2)_R$ under
under the Parity transformation, one sees that the parity symmetry is restored.
Adding the $B\!-\!L$ gauge group, one has the gauge group
$SU(2)_L\otimes SU(2)_R\otimes U(1)_{B-L}$, which
is just the conventional left-right model\,\cite{LRM}.
Another fundamental way for parity restoration is to enlarge the matter contents of
the SM.  To be explicit, we can assign that under
the parity transformation left-handed fermions $f_{L}$ transform
into corresponding new right-handed fermions $f_{R}^{\prime}$ which also group into
doublets of a new gauge group $SU(2)_{R}^{\prime}$, and the right-handed
fermions $f_{R}$ transform into corresponding new left-handed fermions $f_{L}^{\prime}$,
which are singlets of group $SU(2)_R^{\prime}$. This means that we should
enlarge the SM gauge group $G_{SM}^{}=SU(3)_c\otimes SU(2)_L\otimes U(1)_Y$
to  $\,G_{SM}^{}\otimes G_{SM}^{\prime}$,\,
where the new gauge group
$\,G_{SM}' = SU(3)_c'\otimes SU(2)_R'\otimes U(1)_Y'$ \cite{MRev}
is a mirror of $G_{SM}^{}$ with identical gauge couplings,
under which the matter contents switch their chiralities.
Hence, the parity is restored in the universe
where the visible and mirror worlds coexist in the same spacetime.

In fact, the mirror world is a hidden sector of particles and interactions,
as a mirror-duplicate of our visible world. The fermionic matter contents of
the mirror model can be summarized below,
\beq
\label{eq:QNdef}
\ba{ll}
Q_{L}^{i}\sim (3,\ 2,\ \frac{1}{6})(1,\ 1,\ 0)',
&~~~ (Q_{R}^{\prime})^{i}\sim(1,\ 1,\ 0)(3,\ 2,\ \frac{1}{6})',
\\[2mm]
u_{R}^{i}\sim (3,\ 1,\ \frac{2}{3})(1,\ 1,\ 0)',
&~~~ (u_{L}^{\prime})^{i}\sim(1,\ 1,\ 0)(3,\ 1,\ \frac{2}{3})',
\\[2mm]
d_{R}^{i}\sim (3,\ 1,\ -\frac{1}{3})(1,\ 1,\ 0)',
&~~~ (d_{L}^{\prime})^{i}\sim(1,\ 1,\ 0)(3,\ 1,\ -\frac{1}{3})',
\\[2mm]
L_{L}^{i}\sim (1,\ 2,\ -\frac{1}{2})(1,\ 1,\ 0)',
&~~~ (L_{R}^{\prime})^{i}\sim(1,\ 1,\ 0)(1,\ 2,\ -\frac{1}{2})',
\\[2mm]
e_{R}^{i}\sim (1,\ 1,\ -1)(1,\ 1,\ 0)',
&~~~ (e_{L}^{\prime})^{i}\sim(1,\ 1,\ 0)(1,\ 1,\ -1)',
\\[2mm]
\nu_{R}^{i}\sim (1,\ 1,\ 0)(1,\ 1,\ 0)',
&~~~ (\nu_{L}^{\prime})^{i} \sim (1,\ 1,\ 0)(1,\ 1,\ 0)',
\\[2mm]
\phi\sim (1,\ 2,\ \frac{1}{2})(1,\ 1,\ 0)',
&~~~ \,\phi^{\prime}\sim(1,\ 1,\ 0)(1,\ 2,\ \frac{1}{2})',
\ea
\eeq
where $i$ stands for the family index, and the assigned gauge quantum numbers under
$G_{SM}^{}\otimes G_{SM}'$ are also given in the parentheses
(with hypercharge $Y$ defined via $Q=I_3+Y$).
Since light neutrinos are massive,
we have also included the right-handed (left-handed) neutrinos in the visible
(mirror) sector, which are gauge singlets of $G_{SM}^{}\otimes G_{SM}'$.\,
So, under the parity operation  $\,(\vec{x},\,t)\to (-\vec{x},\,t)\,$,\, we have
the following transformations for fermions, gauge bosons and Higgs doublets and their
mirror partners,
\begin{eqnarray}
& &
Q_{L}^i \tolr (Q_{R}')^i,\,~~~
u_{R}^i \tolr (u_L')^i,\,~~~
d_{R}^i \tolr (d_L')^i \,,~~~
L_{L}^i \tolr (L_{R}')^i,\,~~~
e_{R}^i \tolr (e_{L}')^i,\,~~~
\nu_R^i \tolr (\nu_L')^i,\,~~~~~~~~~
\nn\\[2mm]
& &
G_{\mu}^\alpha \tolr (G_{\mu}^\alpha )',\,~~~
W_{\mu}^a \tolr (W_{\mu}^a)',\,~~~
B_{\mu} \tolr B_{\mu}',\,~~~
\phi \tolr \phi' \,.
\label{eq:mirrorTF-fields}
\end{eqnarray}
Furthermore, the parity invariance of the interaction Lagrangian requires the same strengths
of the corresponding gauge (Yukawa) couplings between the visible and mirror sectors;
besides, the heavy Majorana mass-matrices for gauge-singlet neutrinos should be equal
between the two sectors as well.  For our construction,
we will further include a $P$-odd gauge-singlet scalar (Sec.\,\ref{sec:2-3}) to realize
spontaneous mirror parity violation, and allow a unique soft-breaking term in the
singlet-sector of the Higgs potential to evade the domain wall problem. We will also
allow the visible and mirror heavy Majorana mass-matrices to be unequal, as
another unique soft breaking in the gauge-singlet sector of neutrino seesaw,
which will play a key role for realizing the desired dark matter density (Sec.\,\ref{sec:3-1}-\ref{sec:3-2}).

\vspace*{3mm}
\subsection{Communications between Visible and Mirror Worlds}
\label{sec:2-2}
\vspace*{1.5mm}

As we see, the mirror parity symmetry also doubles the particle contents of the SM,
but in a much simpler way than what supersymmetry does.
All the mirror particles have not been seen so far, because the ``communication"
between visible and mirror worlds is hard. If the mirror parity exactly holds, all
mirror particles have the same masses as their SM partners as well as an independent
set of gauge interactions (except sharing the gravity force, which is extremely weak at
ordinary laboratory scales). So the mirror sector consists of a ``hidden world"
and thus provides a generic dark matter candidate
in the universe\,\cite{MRev,Kolb:1985bf,MDM2}.

Nevertheless, besides gravitational interaction,
there are three fundamental ways by which the mirror world can communicate
with our visible world:
(i) interaction between visible and mirror Higgs doubelts;
(ii) mass mixings between singlet visible and mirror neutrinos;
(iii) kinetic gauge mixing of $\,B_\mu \!-\! B_\mu'$\,.

\vspace*{2mm}
\noindent
{{\Large $\bullet$} Interaction between Visible and Mirror Higgs Doublets:\,}
\vspace*{1.5mm}

Gauge invariance also allows the following quartic interaction term between
the visible and mirror Higgs doublets ($\phi$ and $\phi^{\prime}$) \cite{EMirror2,HH'-mix},
\begin{eqnarray}
\label{eq:phi-phi'}
\mathcal{L}_{\phi\phi'}~ =~
\laT \,(\phi^{\dagger}\phi)(\phi^{\prime\dagger}\phi') \,.
\end{eqnarray}
After the electroweak symmetry breaking in the visible and mirror sectors, (\ref{eq:phi-phi'})
can induce a mixing between the Higgs boson $h$ and its mirror partner $h'$.
This will then modify the gauge and Yukawa couplings of both $h$ and $h'$,
giving rise to distinct signatures at the LHC. As will be shown in the next subsection,
our model construction also generates spontaneous mirror parity violation via
$\,\left<\phi\right>\neq\left<\phi'\right>\,$, and thus gives different masses for
mirror gauge bosons $(W',\,Z')$ and all mirror fermions. All these will have important
phenomenological consequences, as to be analyzed in Sec.\,\ref{sec:3}-\ref{sec:6}.

\vspace*{2mm}
\noindent
{{\Large $\bullet$} Mixing between Visible and Mirror Singlet Neutrinos:\,}
\vspace*{1.5mm}

Since $\nuR$ and $\nu_{L}^{\prime}$ are pure gauge singlets, we can
write down the following dimension-3 Dirac mass term\,\cite{nu-mix},
\begin{eqnarray}
\mathcal{L}_{\nu\nu^{\prime}}~ =~
\delta m\, \bar{\nu}_{R}^{}\nu_{L}^{\prime} + \textrm{h.c.}
\end{eqnarray}

\vspace*{2mm}
\noindent
{{\Large $\bullet$} Kinetic Mixing between Visible and Mirror Photons:\,}
\vspace*{1.5mm}

Since the Abelian field strength tensors
$B_{\mu\nu}$ and $B_{\mu\nu}^{\prime}$ are gauge-invariant,
the Lagrangian will generally include the following
dimension-4 mixing operator\,\cite{EMirror2,kmix-more},
\begin{eqnarray}
\label{eq:B-B'}
\mathcal{L}_{BB^{\prime}} ~=\, -\f{\epsilon_0^{}}{2} B^{\mu\nu}B_{\mu\nu}' \,.
\end{eqnarray}
After electroweak symmetry breaking, this term gives rise to a kinetic
mixing between the electromagnetic field strength tensors for visible and mirror photons,
\begin{eqnarray}
\label{eq:Kmix-ga-ga'}
\mathcal{L}_{\textrm{mix}}^{\gamma\gamma'}
~=\, -\f{\epsilon}{2} F^{\mu\nu}F_{\mu\nu}' \,,
\end{eqnarray}
where $\,\epsilon=\epsilon_0^{} \cos^{2}\theta_{W}\,$
and $\theta_W$ is the weak mixing angle. (Since the mirror parity requires
gauge groups $G_{SM}$ and $G_{SM}'$ to have identical gauge couplings, the weak mixing
angle $\theta_W$ remains the same for both visible and mirror sectors.)
Although this kinetic mixing is not suppressed
by known symmetry, an experimental limit can be inferred from
the orthopositronium annihilation into mirror orthopositronium, which imposed
a tight upper bound \cite{AA'mix-Exp2003,Foot:2003eq},
$\,\epsilon < 5\times 10^{-7}\,$.\,
A more recent measurement of the invisible decay
of orthopositronium reduced the upper limit to $\,\epsilon < 1.55\times 10^{-7}\,$
\cite{AA'mix-Exp2007}.
Last year a new experimental proposal plans to reach a sensitivity down to
$\,\epsilon < 4\times 10^{-9}\,$ \cite{AA'mix-plan2010}.
But, we note that this limit only applies to the case where the mirror parity is unbroken,
which can generate oscillation between visible and mirror orthopositroniums.
Our model predicts spontaneous mirror parity violation, so we will reanalyze the
orthopositronium bound in Sec.\,\ref{sec:4-3}.
The operator (\ref{eq:B-B'}) can also induce $Z\!-\!Z'$, $\ga\!-\!Z'$ and
$Z\!-\!\ga'$ mixings, which are all proportional to the tiny $\epsilon$ parameter.
Since these mixings invoke the massive gauge bosons in weak interaction, they
cause no stronger bounds than the $\ga\!-\!\ga'$ mixing.

\vspace*{2mm}
\subsection{Higgs Potential and Spontaneous Symmetry Breaking}
\label{sec:2-3}
\vspace*{1.5mm}

We have conjectured that {\it the mirror parity is respected by the fundamental interaction
Lagrangian,} so its violation only arises from spontaneous breaking of the Higgs
vacuum, and the possible soft breaking can only be linear or bilinear terms;
we further conjecture that {\it all possible soft breakings simply arise from
the gauge-singlet sector alone.}   In this subsection,
we present a minimal model with spontaneous mirror parity violation,
where the Higgs sector includes the SM Higgs doublet, the mirror Higgs doublet
and a $P$-odd weak singlet scalar. We show that the possible soft breaking can be
uniquely realized via the $P$-odd weak singlet scalar in the Higgs potential,
which evades the domain wall problem.

For the minimal construction, we introduce a weak-singlet real scalar field $\chi$
which is $P$-odd. So, the Higgs sector consists of two Higgs doublets ($\phi$ and $\phi'$)
and a real singlet ($\chi$). Under the mirror parity, they transform as follows,
\begin{eqnarray}
\phi ~\leftrightarrow~ \phi^{\prime}\,,
&~~~~~&
\chi ~\leftrightarrow\, -\chi \,.
\end{eqnarray}
Then, we can write down the most general renormalizable form of the minimal Higgs potential
$\,V\,$ for $(\phi,\,\phi',\,\chi)$,
which preserves the gauge group $\,G_{SM}^{}\otimes G_{SM}'\,$
and the mirror parity \,$P$\,,
\beqs
\label{eq:V-tot}
\begin{eqnarray}
\label{eq:V}
V(\phi,\phi^{\prime},\chi)  & \!=\! &
-\mu_{\phi}^{2}\( |\phi|^{2}\!+\!|\phi^{\prime}|^{2} \)
+\lambda_{\phi}^{+}\( |\phi|^{2}\!+\!|\phi^{\prime}|^{2}\)^{2}
+\lambda_{\phi}^{-}\( |\phi|^{2}\!-\!|\phi^{\prime}|^{2}\)^{2}
\nonumber \\
 &  &
-\hf\mu_{\chi}^{2}\chi^{2} +\f{1}{4}\lambda_{\chi}^{}\chi^{4}
 +\beta_{\chi\phi}^{} \,\chi\(|\phi|^{2}\!-\!|\phi^{\prime}|^{2}\)
 +\hf\la_{\chi\phi}^{} \,\chi^{2}\(|\phi|^{2}\!+\!|\phi^{\prime}|^{2}\) ,
\\[1mm]
\label{eq:V-soft}
\Delta V_{\textrm{soft}}(\X) & \!=\! & \beta_\X^{}\X\,,
\end{eqnarray}
\eeqs
where we also included the allowed soft $P$-breaking term
$\,\Delta V_{\textrm{soft}}\,$ from the singlet sector,
which is unique and must be linear in the gauge-singlet field $\,\X$\, because
we have conjectured that all interactions are naturally $P$-invariant.
The Higgs vacuum expectation values (VEVs) are defined as,
\begin{eqnarray}
\label{eq:VEV}
\qquad\langle\phi\rangle ~\equiv~\left(\!\!
\begin{array}{c}
0\\
v_{\phi}^{}
\end{array}
\!\!\right),
\qquad
\langle\phi^{\prime}\rangle ~\equiv~\left(\!\!
\begin{array}{c}
0\\
v_{\phi^{\prime}}^{}
\end{array} \!\!\right),
\qquad
\langle\chi\rangle ~\equiv ~ v_{\chi}^{}\,.
\end{eqnarray}
As we will see, the $\beta_{\chi\phi}^{}$ term in (\ref{eq:V}) is the key to
realize $\,v_{\phi}^{}\neq v_{\phi'}^{}\,$,\, and thus generate the spontaneous
mirror parity violation. [Some early studies
considered spontaneous mirror parity violation
via different approaches, such as setting $\,v_{\phi}^{}\ll v_{\phi'}^{}\,$ and
assuming the coupling of symmetry-allowed mixing interaction
$|\phi|^2|\phi'|^2$ to be highly suppressed down to the level of
$10^{-7}$ \cite{SMPVold}.  This is not the case for our model.]
Then, comparing the operator (\ref{eq:phi-phi'}) with the $\lambda_{\phi+}^{}$
and $\lambda_{\phi-}^{}$ terms in Eq.\,(\ref{eq:V}), we have the relation,
\beqa
\lambdaT ~=~ 2(\lambda_{\phi}^{+} - \lambda_{\phi}^{-}) \,.
\eeqa

Since we are considering the spontaneous $P$ violation,
we may concern the domain wall problem\,\cite{DW} which occurs
for spontaneous breaking of a discrete symmetry (such as
parity) with scalar fields. There are different ways to avoid this problem in
the literature\,\cite{DWsol}.
We have derived the unique soft $P$-breaking term (\ref{eq:V-soft})
as the simplest resolution here to remove the domain wall problem.
This is because Eq.\,(\ref{eq:V-soft}) lifts the degenerate vacua of
the Higgs potential (\ref{eq:V}).
It is natural to consider the soft breaking to be relatively small, i.e.,
the dimension-3 coefficient $\,\beta_\X^{}\,$ is in the range,
$\,\beta_\X^{}\ll \mu_\X^3\,$.\,

With (\ref{eq:V-tot}) and (\ref{eq:VEV}), we infer the full vacuum Higgs potential,
\beqs
\label{eq:V0-tot}
\begin{eqnarray}
\label{eq:V0-sum}
\langle\widehat{V}(\phi,\,\phi',\,\chi)\rangle & \!\equiv\! &
\langle V(\phi,\,\phi',\,\chi)\rangle + \langle{\DVs} (\chi)\rangle \,,
\\[1mm]
\label{eq:V0}
 \langle V\rangle
& \!=\! &
-\mu_{\phi}^{2}\(v_{\phi}^{2}+v_{\phi^{\prime}}^{2}\)
+\lambda_{\phi}^+\(v_{\phi}^{2}+v_{\phi^{\prime}}^{2}\)^{2}
+\lambda_{\phi}^-\(v_{\phi}^{2}-v_{\phi^{\prime}}^{2}\)^{2}
\nonumber \\[1mm]
 &  & -\frac{1}{2}\mu_{\chi}^{2}v_{\chi}^{2}+\frac{1}{4}\lambda_{\chi}^{}v_{\chi}^{4}
 +\beta_{\chi\phi}^{} v_{\chi}\(v_{\phi}^{2} - v_{\phi^{\prime}}^{2}\)
 +\frac{1}{2}\lambda_{\chi\phi}^{} v_{\chi}^{2}\(v_{\phi}^{2}+v_{\phi^{\prime}}^{2}\) ,
 ~~~~~~
\\[2mm]
\label{eq:Vsoft-vac}
\langle\DVs\rangle  & \!=\! &  \beta_\X^{}\vX \,.
\end{eqnarray}
\eeqs
So we see that the full potential $\,\widehat{V}(\phi,\,\phi',\,\chi)$\,
no longer has degenerate vacua,
\beqa
\widehat{V}(\vp ,\,\vpb,\,\vX ) -
\widehat{V}(\vpb ,\,\vp,\, -\vX )
~=~ 2 \beta_\X^{}\vX ~\neq~ 0 \,,
\eeqa
and thus removes the domain wall problem.
The minimal conditions of the vacuum potential give,
\begin{eqnarray}
\frac{\partial\langle \Vh\rangle}{\partial v_{\phi}} \,=\, 0\,,
\qquad
\frac{\partial\langle \Vh\rangle}{\partial v_{\phi^{\prime}}} \,=\, 0\,,
\qquad
\frac{\partial\langle \Vh\rangle}{\partial v_{\chi}} \,=\, 0 \,,
\end{eqnarray}
which lead to the following equations for the nontrivial vacuum,
\beqs
\label{eq:MCs}
\beqa
\label{eq:MC-vphi}
\hspace*{-8mm} &&
-\mu_{\phi}^{2} + 2\lambda_{\phi}^+(v_{\phi}^{2} \!+\! v_{\phi^{\prime}}^{2})
+2\lambda_{\phi}^-(v_{\phi}^{2} \!-\! v_{\phi^{\prime}}^{2})
+\be_{\chi\phi}^{}v_{\chi} + \frac{1}{2}\lambda_{\chi\phi}v_{\chi}^{2} \,=\,  0 \,,
\\[1mm]
\label{eq:MC-vphi'}
\hspace*{-8mm} &&
-\mu_{\phi}^{2} + 2\lambda_{\phi}^+(v_{\phi}^{2} \!+\! v_{\phi^{\prime}}^{2})
-2\lambda_{\phi}^-(v_{\phi}^{2} \!-\! v_{\phi^{\prime}}^{2})
-\be_{\chi\phi}^{}v_{\chi} + \frac{1}{2}\lambda_{\chi\phi}^{} v_{\chi}^{2} \,=\, 0\,,
\\[1mm]
\label{eq:MC-vx}
\hspace*{-8mm} &&
-\mu_{\chi}^{2}v_{\chi}+\lambda_{\chi}v_{\chi}^{3}
+\be_{\chi\phi}^{}(v_{\phi}^{2} \!-\! v_{\phi^{\prime}}^{2})
+\lambda_{\chi\phi}^{}v_{\chi}(v_{\phi}^{2}\!+\!v_{\phi^{\prime}}^{2})
+\beta_\X^{}
\,=\, 0 \,.
\eeqa
\eeqs
From the conditions (\ref{eq:MC-vphi}) and (\ref{eq:MC-vphi'}) we immediately deduce,
\beqs
\label{eq:vphivphi'}
\beqa
\label{eq:vphi-vphi'}
v_\phi^2 - v_{\phi'}^2  & \!=\! & -\f{\be_{\X\phi}^{}}{\,2\la_\phi^-}v_\X^{} \,,
\\
\label{eq:vphi+vphi'}
v_\phi^2 + v_{\phi'}^2  & \!=\! &
\f{\,\mu_\phi^2 -\frac{1}{2}\la_{\X\phi}^{}v_\X^2\,}{2\la_\phi^+} \,.
\eeqa
\eeqs
Inspecting Eq.\,(\ref{eq:vphi-vphi'}),
we see that the VEV $\,v_\X^{}\,$ of the $P$-odd scalar $\X$
together with its trilinear coupling $\,\be_{\X\phi}^{}\,$ is the key to generate
$\,v_\phi^{}\neq v_{\phi'}^{}\,$ and thus the spontaneous mirror parity violation.
We further rewrite (\ref{eq:vphi-vphi'}) as
\beqa
\label{eq:v'/v-ratio}
\(\f{\vpb}{\vp}\)^{\!\!2} ~=~
1 + \f{\,\beta_{\X\phi}^{}\vX\,}{\,2\la_{\phi}^- v_\phi^2\,}\,,
\eeqa
which shows that choosing the ratio $\,\vpb /\vp = 0.1~(0.01)\,$ requires
a fine-tuned cancellation down to the level of
$\,10^{-2}~(10^{-4})\,$ on the right-hand-side (RHS).
Hence, the naturalness of our parameter space puts a lower limit on this ratio,
\beqa
\label{eq:v'/v-lowerB}
\f{\,\vpb\,}{\vp} ~>~ 0.1 \,,
\eeqa
by allowing the fine-tuned cancellation on the RHS of (\ref{eq:v'/v-ratio})
to be better than $\,1\%\,$.\,

Using (\ref{eq:MCs}) we can analytically solve the three VEVs in terms of two mass-parameters
and five couplings in the Higgs potential (\ref{eq:V}),
\beqs
\label{eq:VEVs}
\beqa
\label{eq:v-phi}
v_\phi^2 & \!=\! &
\f{1}{4}\(
\frac{\mu_{\phi}^{2}-\frac{1}{2}\lambda_{\chi\phi}^{}v_{\chi}^{2}}{\lambda_{\phi}^{+}}
-\f{\be_{\X\phi}}{\la_\phi^-}v_\X^{}
\) \!,
\\[1.5mm]
\label{eq:v-phi'}
v_{\phi'}^2 & \!=\! &
\f{1}{4}\(
\frac{\mu_{\phi}^{2}-\frac{1}{2}\lambda_{\chi\phi}^{}v_{\chi}^{2}}{\lambda_{\phi}^{+}}
+\f{\be_{\X\phi}^{}}{\la_\phi^-}v_\X^{}
\) \!,
\\[1.5mm]
\label{eq:v-x-0}
v_{\X0}^2 & \!=\! &
2\f{\,\la_{\X\phi}^{}\mu_\phi^2-2\la_\phi^+\mut_\X^2\,}
   {\la_{\X\phi}^2-4\la_\X\la_\phi^+} \,,
\eeqa
\eeqs
where $\,\mut_\X^2 \equiv \mu_\X^2 + \f{\be_{\X\phi}^2}{2\la_\phi^-}\,$,\, and
in the last equation, $\,v_{\X0}^2\,$ is derived under
the vanishing soft breaking parameter $\,\beta_\X^{}=0\,$.\,
To include a nonzero $\,\beta_\X^{}\,$,\,  we can recast (\ref{eq:MC-vx})
into the form,
\beqs
\label{eq:vX-cubic}
\beqa
\label{eq:vX-cubic1}
\hspace*{-15mm}&&
c_3^{}v_\X^3 - c_1^{}v_\X^{} - \beta_\X^{} ~=~ 0
\\[1mm]
\label{eq:vX-cubic2}
\hspace*{-15mm}&&
c_3^{} ~\equiv~ \f{\,\la_{\X\phi}^2\,}{\,4\la_\phi^+\,} - \la_\X^{}\,,~~~~~
c_1^{} ~\equiv~ \f{\,\la_{\X\phi}^{}\mu_\phi^2\,}{2\la_\phi^+}
- \f{\,\beta_{\X\phi}^2\,}{\,2\la_\phi^-\,} - \mu_\X^2 \,.
\eeqa
\eeqs
where we have made use of (\ref{eq:vphi-vphi'})-(\ref{eq:vphi+vphi'}).
Since $\,\beta_\X^{}\ll \mu_\X^3,\mu_\phi^3\,$,\, we can solve (\ref{eq:vX-cubic})
perturbatively to the first nontrivial order,
\beqa
\label{eq:vX-soft-sol}
\vX ~=~ v_{\X0}^{} +\f{\beta_\X^{}}{\,2c_1^{}\,}
+ O\!\(\f{\beta_\X^2}{\mu_{\X,\phi}^6}\) ,
\eeqa
which reduces back to our leading order solution (\ref{eq:v-x-0}) in the
$\,\beta_\X^{}\to 0\,$ limit.

Choosing unitary gauge and physical vacuum, we see that the doublet $\phi$
($\phi'$) contains a (mirror) neutral Higgs boson, while the $P$-odd $\chi$ gives
a singlet scalar particle.
So denoting $\,\Phi = (\phi,\,\phi',\,\X)^T\,$, we can write down the Higgs mass-term
$\,\Phi^T{\cal M}^2\,\Phi\,$,\, and derive the $3\times 3$ symmetric mass-matrix as follows,
\begin{eqnarray}
\label{eq:Mhiggs-3x3}
{\cal M}^2 ~=
\left(
\begin{array}{ccc}
m_{\phi\phi}^{2} & m_{\phi\phi^{\prime}}^{2} & m_{\phi\chi}^{2}
\\[1.5mm]
m_{\phi\phi^{\prime}}^{2} & m_{\phi^{\prime}\phi^{\prime}}^{2} & m_{\phi^{\prime}\chi}^{2}
\\[1.5mm]
m_{\phi\chi}^{2} & m_{\phi^{\prime}\chi}^{2} & m_{\chi\chi}^{2}
\end{array}\right) \!,
\end{eqnarray}
with the six elements,
%
\begin{eqnarray}
\label{eq:Mhiggs-ij}
m_{\phi\phi}^{2} & \!\!=\!\! &
4( \lambda_{\phi}^{+}+\lambda_{\phi}^{-}) v_{\phi}^{2}\,,
~~~~~~
\nn\\[1mm]
m_{\phi^{\prime}\phi^{\prime}}^{2} & \!\!=\!\! &
4( \lambda_{\phi}^{+}+\lambda_{\phi}^{-}) v_{\phi^{\prime}}^{2} \,,
\nn\\[1mm]
m_{\chi\chi}^{2} & \!\!=\!\! & -\frac{1}{2}\mu_{\chi}^{2}+ \frac{3}{2}\lambda_{\chi}v_{\chi}^{2}+\frac{1}{2}\lambda_{\chi\phi}
(v_{\phi}^{2}+v_{\phi^{\prime}}^{2})\,,
\\[2mm]
m_{\phi\phi^{\prime}}^{2} & \!\!=\!\! &
4(\lambda_{\phi}^+ - \lambda_{\phi}^-) v_{\phi}^{}v_{\phi^{\prime}}^{} \,,
\nn\\[1mm]
m_{\phi\chi}^{2} & \!\!=\!\! &
(\lambda_{\chi\phi}v_{\chi}^{} + \be_{\chi\phi})v_{\phi}^{} \,,
\nn\\[1mm]
m_{\phi^{\prime}\chi}^{2} & \!\!=\!\! &
(\lambda_{\chi\phi}v_{\chi}^{} -\be_{\chi\phi})v_{\phi^{\prime}}^{} \,,
\nn
\end{eqnarray}
%
where we have made use of the vacuum solution (\ref{eq:VEVs}) for simplification.
In Sec.\,\ref{sec:4}, we will present numerical
samples for the vacuum solution (\ref{eq:VEVs}) and derive the physical
Higgs mass-spectrum from diagonalizing the mass-matrix (\ref{eq:Mhiggs-3x3}).

Finally, we comment on the self-interactions of mirror dark matter
due to the unbroken mirror electromagnetism which may be
a concern for all mirror models with unbroken mirror Abelian gauge group.
As clarified in \cite{Foot-Mstar}, the MACHO collaboration\footnote
{MACHO stands for Massive Astrophysical Compact Halo Objects \cite{MACHO}.}
found statistically strong evidence for dark matter in the form of invisible
star-sized objects \cite{MACHO},  which is just one would expect if a significant
amount of mirror dark matter exists in our galaxy.
Another survey analyzed stars across the face of M31 and found significant
evidence for a population of halo microlensing dark matter objects, showing
a halo mass fraction of $\,f=0.29_{-0.13}^{+0.30}\,$
\cite{Mstar2}. This is consistent with the results of MACHO collaboration \cite{MACHO}.
For dark matter in the form of MACHO, it will not show self-interactions in
the bullet cluster.
In addition, it is worth to note that although astronomical observations
have put nontrivial constraints on the possible long range self-interactions
of the dark matter, they are valid only for
the assumed homogeneous dark matter distributions
and thus need not to be directly applicable to the mirror dark matter as it
can form non-homogeneous type of structures\,\cite{Ciarcelluti:2010zz}.
For the bullet cluster, the observations showed that after a collision of two galaxies the
dark matter can pass through each as if no much collisions between them.
But other observations exist with opposite implication. For instance,
studies on the Abell-520 cluster (also known as MS~0451$+$02) 
\cite{Mahdavi:2007yp}\cite{Jee:2012sr}, performed the weak-lensing analysis and
subsequent comparison with the optical and X-ray properties of the cluster.
It was found \cite{Mahdavi:2007yp}\cite{Jee:2012sr} 
that the massive dark core coincides with the central X-ray
emission peak, but is largely devoid of galaxies, indicating certain self-interactions
of the dark matter, which may be explained by the mirror dark matter \cite{Silagadze:2008fa}.  
Astronomers are making further efforts to explore the true natures
of dark matter, including unusual clusters such as the Abell-520 and alike.

\vspace*{3mm}
\section{Common Origin of Matter and Dark Matter via Leptogenesis}
\label{sec:3}
\vspace*{1.5mm}

In this section we study the generation of visible matter and mirror dark matter
from a common origin in neutrino seesaw via leptogenesis.
In Sec.\,\ref{sec:3-1}, with the spontaneous mirror parity violation, we derive the ratio
of visible and mirror nucleon masses as a function of the VEVs of visible and mirror
Higgs bosons. Then, in Sec.\,\ref{sec:3-2}, we connect the visible and mirror leptogeneses,
and compute the ratio of visible and mirror baryon asymmetries in terms of the
unique soft $P$-breaking parameter in the neutrino seesaw.
We will present two seesaw constructions for both
the visible and mirror neutrino sectors, which explicitly realize the common origin
of the matter and dark matter geneses. Then, we demonstrate the realization of
the astrophysical observation,  $\,{\Omega_{\rm DM}}:{\Omega_{\rm M}} \simeq 5:1\,$.\,
Finally, we analyze the consistency of our mirror model with the constraint from
Big-Bang nucleosynthesis in Sec.\,\ref{sec:3-3}, which puts nontrivial limit on the ratio
of the visible and mirror Higgs VEVs.

\vspace*{3mm}
\subsection{Common Origin of Visible and Dark Matter from Leptogenesis}
\label{sec:3-1}
\vspace*{1.5mm}

Observations reveal our visible world to be exclusively populated with baryonic matter
instead of antimatter.  The genesis of net baryon asymmetry requires baryon number violating
interactions, $C$ and $CP$ violations and departure from thermal equilibrium\,\cite{BG}.
This can be naturally realized via leptogenesis \cite{LepG0,LepG-Rev},
where the leptonic $CP$ violations arise from neutrino seesaw and the out-of-equilibrium decays
of heavy Majorana neutrino into lepton-Higgs pair and its conjugate produce the lepton-number
asymmetry. Because of the electroweak sphaleron process\,\cite{SP}, the lepton asymmetry is
partially converted into the desired baryon asymmetry and can explain the observed baryon density
today\,\cite{Omega-B-DM},
\beqa
\label{eq:Omega-B-exp}
\Omega_{\rm B}^{} ~=~ 0.0458\pm 0.0016 \,.
\eeqa
As shown in (\ref{eq:mirrorTF-fields}),
the mirror parity connects the particle contents of the visible and mirror sectors with
one-to-one correspondence. Furthermore, it requires identical gauge (Yukawa)
couplings between the two sectors, as well as the same Majorana mass-matrix for
the gauge-singlet heavy Majorana neutrinos. Thus, it is very natural to generate the
the baryonic mirror matter-antimatter asymmetry from mirror leptogenesis.
With the spontaneous mirror parity violation (Sec.\,\ref{sec:2-3}) and a unique soft breaking
in the singlet sector of neutrino seesaw, we will generate desired mass-splittings
between the visible and mirror nucleus, as well as a different efficiency factor
of out-of-equilibrium decays for the heavy singlet mirror neutrinos,
such that the baryonic mirror matter can naturally provide the observed dark matter
density in the universe\,\cite{Omega-B-DM},
\beqa
\label{eq:Omega-DM-exp}
\Omega_{\textrm{DM}} ~=~ 0.229 \pm 0.015 \,,
\eeqa
which is only about a factor five larger than $\Omega_{\rm B}^{}$.\,
With (\ref{eq:Omega-B-exp})-(\ref{eq:Omega-DM-exp}), we derive the ratio,
$\,\Omega_{\textrm{DM}}/\Omega_{\textrm{B}}= 5.00\pm 0.37\,$,\,
which gives the $2\sigma$ limit:
$\,4.26< \Omega_{\textrm{DM}}/\Omega_{\textrm{B}} < 5.74\,$.\,
For the mirror model, we have the visible matter density
$\,\Omega_{\textrm{M}}\simeq \Omega_{\textrm{B}}\,$
and the mirror dark matter density
$\,\Omega_{\textrm{DM}}\simeq\Omega_{\textrm{B}'}\,$.

With the mirror baryons serving as natural dark matter, we can thus derive the
ratio of dark matter density relative to that of visible matter,
\begin{eqnarray}
\label{eq:DM-B-ratio}
\frac{\Omega_{\DM}}{\Omega_{\rm M}}
~\simeq~ \f{\Omega_{\B'}}{\Omega_{\B}}
~=~ \frac{\,\N_{\B'}\,}{\N_{\B}}\frac{\,m_{N'}^{}}{m_{N}^{}} \,,
\end{eqnarray}
where $\,m_{N}^{}$ denotes the visible nucleon mass and $\,m_{N'}^{}$
the mirror nucleon mass.
In (\ref{eq:DM-B-ratio}), $\,\N_{\B}\,$ ($\,\N_{\B'}\,$) is the baryon number
(mirror baryon number) computed in a portion of comoving volume [which contains
one photon (mirror photon) before the onset of (mirror) leptogenesis].

As shown in Sec.\,\ref{sec:2-3}, the spontaneous mirror parity violation makes visible Higgs VEV
differ from that of the mirror Higgs, $\,v_\phi^{}\neq v_{\phi'}^{}\,$.\,
So, the masses of visible and mirror nucleus also differ from each other,
$\,m_{N}^{}\neq m_{N'}^{}\,$.
For further analysis,
let us derive the relation between nucleon mass and the Higgs boson VEV.
The running of QCD gauge coupling $\,\alpha_{s}^{}(\mu)$\, is given by
\begin{eqnarray}
\label{eq:alpha-s}
\alpha_{s}^{}(\mu)
~=~ \frac{2\pi}{\,11-\frac{2}{3}n_f^{}\,}\frac{1}{\,\ln(\mu/\Lambda)\,} \,,
\end{eqnarray}
while the mirror QCD$'$ has its running gauge coupling $\,\alpha_{s}'(\mu)$\, behave as,
\begin{eqnarray}
\label{eq:alpha-s'}
\alpha_{s}'(\mu)
~=~ \frac{2\pi}{\,11-\frac{2}{3}n_f'\,}\frac{1}{\,\ln(\mu/\Lambda')\,} \,,
\end{eqnarray}
where we denote
$\,\Lambda =\Lambda_{\textrm{QCD}}^{}\,$ for visible QCD
and $\,\Lambda' =\Lambda_{\textrm{QCD}}'\,$ for mirror QCD$'$,
which are renormalization group invariants.
The $n_f^{}$ ($n_f'$) counts the number of (mirror) quark flavors involved at a
given scale $\mu$.\,
Then, we can match $\alpha_{s}^{}(\mu)$ at the scale $\,\mu = m_{t}^{}\,$
with $\,n_f^{}=5\,$ and $\,n_f^{}=6\,$,\,
\begin{eqnarray}
\alpha_{s}^{(5)}(m_{t}^{}) ~=~ \alpha_{s}^{(6)}(m_{t}^{})\,,
\end{eqnarray}
which leads to
\begin{eqnarray}
\Lambda_{(5)}^{} ~=~ (m_t)_{}^{{2}/{23}}(\Lambda_{(6)})_{}^{{21}/{23}} \,.
\end{eqnarray}
Similarly, matching $\alpha_{s}^{}(\mu)$ at $\,\mu = m_b^{}\,$ and $\,\mu =m_c^{}\,$,
respectively, we deduce,
\begin{eqnarray}
\Lambda_{(4)}^{} &\!=\!& (m_b^{})^{2/25}(\Lambda_{(5)}^{})^{23/25} ,
\\[1mm]
\Lambda_{(3)}^{} &\!=\!& (m_c^{})^{2/27}(\Lambda_{(4)}^{})^{25/27} .
\end{eqnarray}
From the above relations, we further arrive at
\begin{eqnarray}
\label{eq:Lambda-3}
\Lambda_{(3)}^{} ~=~ (m_{c}^{}m_{b}^{}m_{t}^{})^{2/27}(\Lambda_{(6)}^{})^{21/27}
~\propto~ v_{\phi}^{2/9}(\Lambda_{(6)}^{})^{21/27} ,
\end{eqnarray}
where we note that the current quark masses for $(c,\,b,\,t)$ are generated from
the Yukawa interactions with Higgs boson $\phi$, and thus proportional to the
Higgs VEV $\,v_{\phi}^{}\,$.\,  The nucleon consists of up and down quarks and its mass
is dominated by the dynamical mass instead of the current masses of $u$ and $d$ quarks
(which are only a few MeV and thus negligible here).
So, the nucleon mass should be proportional to the dynamical QCD scale
$\,\Lambda_{(3)}^{}$.\, Thus, using (\ref{eq:Lambda-3}) we finally derive,
\begin{eqnarray}
\label{eq:mN-Lambda3}
m_{N}^{} ~\propto~ v_{\phi}^{2/9}(\Lambda_{(6)}^{})^{21/27} .
\end{eqnarray}
In parallel, we can infer the relation for mirror nucleon mass,
\begin{eqnarray}
\label{eq:mN'-Lambda3}
m_{N'}^{} ~\propto~ v_{\phi'}^{2/9}(\Lambda_{(6)}')^{21/27} .
\end{eqnarray}
Note that the visible (mirror) sector contains only six (mirror) quark flavors,
so the renormalization group invariant $\Lambda_{(6)}^{}$ (or $\Lambda_{(6)}'$)
holds for all scales above $\,m_t^{}\,$ (or $\,m_t'\,$).
At sufficiently high scales $\mu\gg \,m_t^{},m_t' \!\sim\! v_{\phi}^{},v_{\phi}'$,\,
the renormalization group invariants $\Lambda_{(6)}^{}$ and $\Lambda_{(6)}'$
are determined by the corresponding strong gauge couplings alone.
As the mirror symmetry requires $\,\alpha_s^{}(\mu)=\alpha_s'(\mu)\,$,\,
it leads to $\,\Lambda_{(6)}^{} = \Lambda_{(6)}'$.\,
With this we can deduce from (\ref{eq:mN-Lambda3})-(\ref{eq:mN'-Lambda3}),
\begin{eqnarray}
\label{eq:ratio-mN'/mN}
\frac{\,m_{N'}^{}}{m_{N}^{}} ~=~ \(\frac{v_{\phi'}^{}}{v_{\phi}}^{}\)^{2/9} \,.
\end{eqnarray}

Next, we analyze the ratio of visible and mirror baryon numbers,
$\,\N_{\B'}/\N_{\B}^{}\,$,\,
as appeared in Eq.\,(\ref{eq:DM-B-ratio}).
It is natural and attractive to produce $\N_{\B}$ and $\N_{\B'}$ from
the visible and mirror leptogeneses via neutrino seesaws, respectively.
As we will show, due to the mirror parity, the visible and mirror neutrino
seesaws are quantitatively connected; especially, they share the same $CP$
phases in addition to the same Yukawa couplings and singlet heavy Majorana
mass-matrix. This naturally provides a common origin for the visible matter
and mirror dark matter via leptogeneses. As we mentioned earlier, we will
allow soft breaking of mirror parity in the gauge-singlet sectors which
include the heavy singlet Majorana mass-terms in the neutrino
seesaw\footnote{Note that the ordinary lepton number and mirror lepton number
are also softly broken by the singlet Majorana mass-terms of heavy
right-handed neutrinos and left-handed mirror neutrinos.}

This means that we will allow unequal singlet heavy Majorana mass-matrices,
$\,M_N^{}\neq M_N'\,$,\,  between the visible and mirror seesaws; but we will
maintain this soft breaking to be minimal, i.e., both $\,M_N^{}$ and $\,M_N'\,$
still have identical structure (as required by mirror parity) except
differing by an overall scaling factor,
$\,M_N^{} \varpropto M_N'\,$.\,
Hence, we can write down the seesaw Lagrangian,
\beqs
\label{eq:L-seesaw}
\begin{eqnarray}
\label{eq:L-seesaw-a}
\mathcal{L}_{\textrm{ss}} & \!=\! &
-\overline{L}\,Y_{\ell}^{}\phi\,\ell_{R}^{}
-\overline{L}\,Y_{\nu}\widetilde{\phi}\,\mathbb{N}
+\f{1}{2}{\NN}^{T}M_N^{}\widehat{C}\,\mathbb{N}
-\overline{R'}\,Y_{\ell}'\phi'\,\ell_{L}'
-\overline{R'}\,Y_{\nu}'\widetilde{\phi'}\,\mathbb{N}'
+\f{1}{2}{\NN}^{\prime T}M_N'\widehat{C}\,\mathbb{N}'~~~~~~~
\nonumber\\
 & \!\! &
 +\frac{1}{2}\mathbb{N}^T\delta m\,\widehat{C}\,\mathbb{N}' + \textrm{h.c.}
 \\
 & \!=\! &
 -\overline{\ell_{L}^{}}\,M_{\ell}\,\ell_{R}^{}
 -\overline{\nu_{L}^{}}\,m_{D}^{}\,\mathbb{N}
 +\frac{1}{2}\mathbb{N}^{T}M_N^{}\widehat{C}\,\mathbb{N}
 -\overline{\ell_{R}'}\,M_{\ell}'\,\ell_{L}'  -\overline{\nu_{R}'}\,m_{D}'\,\mathbb{N}'
 +\frac{1}{2}{\NN'}^{T}M_N'\widehat{C}\,{\NN'}~~~~~
 \nonumber \\
 &  &
 +\frac{1}{2}\mathbb{N}^T\delta m\,\widehat{C}\,\mathbb{N}'
 + \textrm{h.c.} +(\textrm{interactions})\,,
\label{eq:L-seesaw-b}
\end{eqnarray}
\eeqs
where $\, L\,$ denotes three left-handed neutrino-lepton weak doublets,
$\,\ell=(e,\,\mu,\,\tau)^{T}\,$ contains charged leptons, $\,\nu=(\nu_{e}^{},\,\nu_{\mu}^{},\,\nu_{\tau}^{})^{T}\,$
is for the light flavor neutrinos, and ${\NN}=(N_1,\, N_2,\,N_3)^{T}$
represents two heavy right-handed singlet neutrinos for the visible sector,
while
$R'$,\, $\ell'=(e',\,\mu',\,\tau')^{T}$,\,
$\nu' = (\nu_{e}',\,\nu_{\mu}',\,\nu_{\tau}')^{T}$,\, and
$\,{\NN}' = (N_1',\, N_2',\,N_3')^T$\,
are the corresponding fields in the mirror sector.
In (\ref{eq:L-seesaw}), we have used
$\widetilde{\phi}$ (\,$\widetilde{\phi'}$\,) to denote the charge-conjugation of
Higgs doublet $\phi$ (mirror Higgs doublet $\phi'$), and
$\,\widehat{C}=i\gamma^2\gamma^0\,$ is the charge-conjugation operator for spinors.
Also, the notations for singlet heavy Majorana neutrinos
${\NN}$ and ${\NN}'$ are connected to that in (\ref{eq:QNdef}) via
$\,{\NN}\sim \nu_R^{}\,$ and $\,{\NN}'\sim \nu_L'\,$,\,
where ${\NN}$ and ${\NN}'$ are Majorana spinors, so we have
$\,{\NN}={\NN}^c\,$ and $\,{\NN}'={\NN}^{\prime\,c}\,$.

Imposing the mirror symmetry $P$ on the interaction Lagrangian in (\ref{eq:L-seesaw-a})
of neutrino seesaw and allowing the minimal soft $P$-breaking for the heavy singlet
Majorana mass-terms in (\ref{eq:L-seesaw-a}),
we deduce the following relations between the visible and mirror sectors,
\begin{eqnarray}
\label{eq:Yl-Ynu-MR-mirror}
Y_{\ell}^{}\,=\, Y_{\ell}'\,,~~~~~
Y_{\nu}^{} \,=\, Y_{\nu}'\,, ~~~~~
M_N^{} \,=\, \rN M_N' \,,
\label{M-O-1}
\end{eqnarray}
where the ratio  $\,r_N^{}\equiv M_N/M_N' \neq 1\,$
characterizes the minimal soft $P$-breaking and its value will be determined
later by generating the desired dark matter density,
$\,\Omega_{\textrm{DM}}\simeq 5\Omega_{\textrm{M}}\,$.\,
The mass-eigenvalues of $\,M_N^{}\,$ and $\,M_N'\,$ will be denoted as
$\,M_j^{}\,$ and  $\,M_j'\,$,\, respectively.
Thus we also have $\,M_j^{}/M_j' = r_N^{}\,$.\,
Since the Dirac mass-matrices $\,m_D^{}=Y_\nu^{} v_{\phi}^{}\,$ and
$\,m_D'=Y_\nu' v_{\phi'}^{}\,$,\, we have
\begin{eqnarray}
\label{eq:mD-mD'}
\frac{\,m_{D}'\,}{m_{D}^{}} ~=~ \frac{\,v_{\phi'}\,}{v_{\phi}} \,.
\label{M-O-2}
\end{eqnarray}
Since the mixing mass-term $\delta m$ between ${\NN}-{\NN}'$ in (\ref{eq:L-seesaw})
will lead to mixings between light visible and mirror neutrinos
after the heavy singlet neutrinos ${\NN}$ and ${\NN}'$ are integrated out,
this term has to be very small due to the tight constraints for sterile neutrinos.
So we have, $\,\delta m\ll\mid M_{R}^{}-M_{R}'\mid\,$,\,
and thus for the present analysis it is safe to neglect $\,\delta m\,$.\,
This means that we have separate seesaw mass formulas
for the light visible and mirror neutrinos,
\beqs
\label{eq:seesaw-M-M'}
\beqa
\label{eq:seesaw-Mnu}
M_\nu &\!\simeq\!& m_D^{}M_N^{-1}m_D^T \,,
\\
\label{eq:seesaw-Mnu'}
M_\nu' &\!\simeq\!& m_D' M_N^{\prime\,-1}m_D^{\prime\,T}
~=~ \f{\,v_{\phi'}^{2}\,}{v_\phi^{2}} \rN M_\nu \,,
\eeqa
\eeqs
where in the second equation we have used the mirror symmetric relations
(\ref{eq:Yl-Ynu-MR-mirror})-(\ref{eq:mD-mD'}).
These show that the visible and mirror neutrino sectors must share the same
$CP$ violation phase(s) as well as the same flavor mixing structure.

In the visible sector,  the baryon number density ${\N}_B$ and the amount
of $B-L$ asymmetry ${\N}_{B-L}$, as defined in a portion of comoving volume
containing one photon at the onset of leptogenesis, are given by\,\cite{LepG-Rev},
\begin{eqnarray}
\label{eq:N-B}
{\N}_B^{} ~=~ \xi\, {\N}_{B-L}^{} ~=~ \frac{3}{4}\xi \kappa_f^{} \epsilon_1^{}\,,
\end{eqnarray}
where the parameter $\,\xi={28}/{79}\,$ is the fraction of $B-L$ asymmetry
converted from $\,{\N}_{B-L}^{}\,$ into a net baryon number $\,{\N}_{B}^{}\,$
by sphaleron processes, and is determined by the number of fermion generations
and Higgs doublets in the SM\,\cite{HT}.
The factor $\kappa_f^{}$ in (\ref{eq:N-B}) measures the efficiency of
out-of-equilibrium $N_1$-decays, and $\epsilon_1^{}$ characterizes the $CP$ asymmetry
produced by the decays of the lighter singlet neutrino $N_1$ at the scale of $M_1$.
In parallel, for the mirror sector we have,
\begin{eqnarray}
\label{eq:N-B'}
{\N}_B' ~=~ \xi' {\N}_{B-L}' ~=~ \frac{3}{4}\xi' \kappa_f' \epsilon_1' \,,
\end{eqnarray}
where $\,\xi' =\xi ={28}/{79},$ since the mirror sector has the same number of
fermion generations and Higgs doublets as the visible sector.
The $P$-odd singlet scalar $\X$ is real and thus has zero chemical potential.
Also, in the Higgs potential (\ref{eq:V-tot})
all the mixing terms among $\phi$, $\phi'$ and $\X$ have vanishing chemical potential.
So they do not affect the conversion efficiencies $\,\xi\,$ and $\,\xi'\,$ in both
visible and mirror sectors, and we have $\,\xi =\xi'\,$.\,

The efficiency factor $\kappa_f^{}$ in (\ref{eq:N-B}) can be solved from the
Boltzmann equations\,\cite{kappa-BDP},
\beqs
\label{eq:BE}
\beqa
\f{\,d\N_{N_1}^{}\,}{dz}
&\!=\!\!& -(D+S)\(\N_{N_1}^{} - \N_{N_1}^{\textrm{eq}}\) \,,
\\[1.5mm]
\f{\,d\N_{B-L}^{}\,}{dz}
&\!=\!\!& -\ep_1^{}D\(\N_{N_1}^{} - \N_{N_1}^{\textrm{eq}}\)
- W\N_{B-L}^{} \,,
\eeqa
\eeqs
where $\,z=M_1/T\,$,\, and
$\,(D,\,S,\,W)=(\Gamma_D^{},\,\Gamma_S^{},\,\Gamma_W^{})/(Hz)\,$
are dimensionless functions of $\,z$\,.\,
The Hubble expansion rate $\,H\,$ is given by,
$\,H\simeq (8\pi^3g_*^{}/90)^{\hf}({M_1^2}/{\MP})z^{-2}_{}\,$,\,
where $\,\MP \simeq 1.22\times 10^{19}\,$GeV equals the Planck mass,
and $\,g_*^{}\,$  represents the relativistic degrees of freedom
at the temperature \,$T$.\,
The rate $\,\Gamma_D^{}\,$ denotes the decays and inverse decays of $N_1$,
$\,\Gamma_S^{}\,$ accounts for $N_1$ scattering rate, and
$\,\Gamma_W^{}\,$ is the washout rate including contributions from
the inverse decays and the $\,\De L=1,2\,$ processes, where the
contribution of $\,\De L=2\,$ processes is denoted by
$\,\De\Gamma_W^{}\equiv (Hz)\De W \,$.\,
It is found\,\cite{kappa-BDP}
that the dimensionless functions $\,(D,\,S,\,W,\,\De W)$
have the following simple scalings,
\beqa
\label{eq:DSW}
D,\, S,\, W \!-\! \De W ~\varpropto~ \f{\,\MP\tm_1^{}\,}{v_\phi^2} \,,
~~~~~~~
\De W ~\varpropto~ \f{\,\MP M_1\bm^2\,}{v_\phi^4} \,,
\eeqa
where $\,\tm_{1}^{}\,$ is the effective light neutrino mass,
\begin{equation}
\label{eq:m1t}
\tm_{1}^{} ~\equiv~
\f{(\widetilde{m}_{D}^{\dagger}\widetilde{m}_D^{})_{11}^{}}{M_{1}}\,,
\end{equation}
and $\,\widetilde{m}_{D}^{}\equiv m_{D}^{}V_{R}\,$, with $V_{R}$ being the
unitary rotation matrix which diagonalizes $M_{R}^{}$\,.\,
In the last relation of (\ref{eq:DSW}), the light neutrino mass-parameter $\bm$
is given by the trace
$\,\bm = [\textrm{tr}(M_\nu^\dag M_\nu^{})]^{\hf}
       = \sqrt{m_1^2+m_2^2+m_3^2\,}\,$.\,
Inspecting (\ref{eq:DSW})-(\ref{eq:m1t}) and (\ref{eq:seesaw-Mnu}), we note that
the functions $\,(D,\,S,\,W,\,\De W)$\, do not actually depend on the Higgs VEV
$\,\vp\,$,\, but depend on relevant products of Yukawa couplings and the heavy
singlet neutrino mass $\,M_1\,$ as well as the Planck mass $\,\MP\,$.\,
We can see this explicitly by examining the analytical solution\,\cite{kappa-BDP}
to the Boltzmann equations (\ref{eq:BE}),
\beqs
\label{eq:kappaf-analytic}
\beqa
\kappa_f^{} & \!\simeq\! & \f{2}{\,z_B^{}(K)K\,}
\left[ 1 -\exp \!\(-\hf z_B^{}(K)K\)\right] \,,
\\[1mm]
z_B^{}(K) & \!\simeq\! &
1 +\hf \ln\left[ 1+\f{\,\pi K^2\,}{1024}
             \(\ln\f{\,3125\pi K^2\,}{1024}\)^{\!\!5}\right],
\eeqa
\eeqs
which agrees with the exact numerical solution very well. This shows that the
efficiency factor $\,\kappa_f^{}\,$ depends only the parameter $K$,
\beqs
\label{eq:K-m*}
\beqa
\label{eq:K}
K & \!=\! & \f{\,\Ga_D^{}(z=\infty )\,}{H(z=1)}
~=~ \f{\tm_1^{}}{m_*^{}} ~\varpropto~ \f{1}{M_1}\,,
\\[1mm]
\label{eq:m*}
m_*^{} & \!=\! & \f{\,16\pi^{\f{5}{2}}\sqrt{g_*^{}}\,}{3\sqrt{5}}\f{v_\phi^2}{\MP}
~\simeq~ 1.5\times 10^{-3}\,\textrm{eV}\,,
\eeqa
\eeqs
where the VEV $\,\vp \simeq 174\,$GeV is responsible for the electroweak symmetry
breaking (cf.\ Sec.\,\ref{sec:4}). At the temperature $\,T\sim M_1\,$,\,
we note that the effective degrees of freedom $\,g_*^{}=O(200)\,$
contains $\,106.75\,$ from SM particles and $\,\f{7}{4}\,$
from Majorana neutrino $N_1$, and in addition, the mirror partners contribute
another $\,106.75\,$ to $\,g_*^{}\,$ and the real scalar $\X$ contributes $1$.\,
So we have $\,g_*^{}=216.25\,$ in total. (Here we do not count on the lightest mirror
singlet neutrino $N_1'$ since its mass is much larger than $N_1$
[cf.\ (\ref{eq:rN-bound}) below] and already decays at a higher temperature.)
Inspecting (\ref{eq:m1t}) and (\ref{eq:K-m*}), we note that
$\,K\,\varpropto\, M_1^{-1}\,$,\, but has no dependence on the Higgs VEV
$\,\vp\,$ since both $\,\tm_1^{}\,$ and $\,m_*^{}\,$ are proportional
to $\,v_\phi^2\,$.\,

For practical applications, it is more convenient to use the fitting formula
for the efficiency factor $\kappa_f^{}$ in the power-law form
under $\,\tm_1^{} > m_*^{}\,$ \cite{kappa-BDP},
\beqa
\label{eq:kapa-f-BDP}
\kappa_f^{} & \!=\! &
{(2\pm 1)\!\times\! 10^{-2}}
\left(\f{\,0.01\,\textrm{eV}\,}{\tm_{1}^{}}\right)^{\!1.1\pm 0.1}
\,\varpropto~ M_1^{(1.1\pm 0.1)} \,,
\eeqa
where the effective light neutrino mass $\,\tm_{1}^{}\,$ is defined in
(\ref{eq:m1t}) and we have extracted the scaling behavior
$\,\kappa_f^{} \varpropto M_{1}^{(1.1\pm 0.1)}$.\,
The formula (\ref{eq:kapa-f-BDP}) is found in good agreement
with the exact numerical solution \cite{kappa-BDP}.
We expect that, without accidental cancellation, effective mass
$\,\tm_1^{}\,$ should be the typical mass scale of light neutrinos, i.e.,
$\,\tm_1^{} = O(10^{-1}\!-\!10^{-2})\,$eV.\,
For natural Yukawa couplings $\,Y_\nu \leqq O(1)\,$,\, one can infer\,\cite{kappa-BDP},
$\,m_1^{}\leqq \overline{m}_1^{}\leqq m_3^{}\,$ or
$\,m_3^{}\leqq \overline{m}_1^{}\leqq m_1^{}\,$.
As will be shown in Sec.\,\ref{sec:3-2} [cf.\ Eq.\,(\ref{eq:m1bar})],
for our explicit seesaw realizations, we can deduce,
$\,\tm_1^{}\sim \sqrt{\Delta m_{13}^2} = O(10^{-1}-10^{-2})$eV,
where $\,\Delta m_{13}^2\,$ is the atmospheric mass-squared difference as measured by
the oscillation experiments\,\cite{nu-data}.
For computing the ratio of the two efficiency factors
$\,\kappa_f^{}\,$ and $\,\kappa_f'\,$ in the visible and mirror sectors,
we see that the overall coefficient on the RHS of
(\ref{eq:kapa-f-BDP}) is irrelevant, only
the scaling behaviors, $\,\kappa_f^{} \varpropto M_{1}^{(1.1\pm 0.1)}\,$
and $\,\kappa_f' \varpropto {M_1'}^{(1.1\pm 0.1)}\,$,\, will matter.
So, we can deduce the ratio,
\beqa
\label{eq:kappaf'/kappaf}
\f{\kappa_f'}{\kappa_f^{}} ~=\,
\(\f{\,M_1'\,}{\,M_1^{}\,}\)^{\!1.1\pm 0.1}
=\, \(\f{1}{\,\rN\,}\)^{\!1.1\pm 0.1}\,,
\eeqa
which depends on the mass-ratio $\,r_N^{}\,$
of the visible/mirror heavy singlet neutrinos,
and does not equal one due to the soft breaking of mirror parity in the
singlet sector, $\,M_1^{}\neq M_1'\,$,\, as in (\ref{M-O-1}).

The $CP$ asymmetry parameter $\,\epsilon_1^{}\,$ can be expressed as,
\begin{eqnarray}
\label{eq:epsilon-1}
 \epsilon_{1}^{} ~=~
 \f{\Gamma[N_{1}\rightarrow \ell H]-\Gamma[N_{1}\rightarrow\bar{\ell}H^{*}]}
   {\Gamma[N_{1}\rightarrow \ell H]+\Gamma[N_{1}\rightarrow\bar{\ell}H^{*}]}
 ~=~ \f{1}{\,4\pi v_{\phi}^{2}\,}F\!\(\!\frac{M_{2}}{M_{1}}\!\)\f{\,\Im\mathfrak{m}\!
 \left\{ [(\widetilde{m}_{D}^{\dagger}\widetilde{m}_{D})_{12}^{}]^{2}\right\}\,}
 {(\widetilde{m}_{D}^{\dagger}\widetilde{m}_{D})_{11}^{}} \,,
 \end{eqnarray}
where the $\,\vp\,$ factors all cancel out on the right-hand-side, and
for the SM the function $\,F(x)\,$ takes the following form,
%
\beqa
\label{eq:F}
  F(x) ~\equiv~
  x \left[
    1 - (1 + x^2) \ln \frac {1 + x^2}{x^2}
  + \frac 1 {1 - x^2} \right]
\,=\, - \f{3}{\,2x\,} + \O\!\( \f{1}{x^3} \),
\quad (\textrm{for}~\,  x \gg 1\,)\,.
\eeqa
%
In parallel, for the mirror $CP$-asymmetry parameter
$\,\epsilon_{1}'\,$,\, we have,
\begin{eqnarray}
\label{eq:epsilon'-1}
 \epsilon_{1}' ~=~
 \f{\Gamma[N_{1}'\rightarrow \ell_R' H']-\Gamma[N_{1}'\rightarrow\bar{\ell}_R'H^{\prime *}]}
   {\Gamma[N_{1}'\rightarrow \ell_R' H']+\Gamma[N_{1}'\rightarrow\bar{\ell}_R'H^{\prime *}]}
 ~=~ \f{1}{\,4\pi v_{\phi'}^{2}\,}F\!\(\!\frac{M_{2}'}{M_{1}'}\!\)\f{\,\Im\mathfrak{m}
 \left\{ [(\widetilde{m}_{D}^{\prime\,\dagger}\widetilde{m}_{D}')_{12}^{}]^{2}\right\}\,}
 {(\widetilde{m}_{D}^{\prime\,\dagger}\widetilde{m}_{D}')_{11}^{}} \,.
 \end{eqnarray}
Due to the soft breaking relation $\,M_N^{}\varpropto M_N'\,$ in
(\ref{eq:Yl-Ynu-MR-mirror}), we have equal mass-ratios
$\,M_2^{}/M_1^{}=M_2'/M_1'\,$.\,
Using the mirror symmetry requirements (\ref{eq:Yl-Ynu-MR-mirror}) and (\ref{eq:mD-mD'}),
and noting that the VEV factors all drop off in
(\ref{eq:epsilon-1}) and (\ref{eq:epsilon'-1}),  we deduce,
\begin{eqnarray}
\epsilon_{1}' ~=~ \epsilon_{1}^{}\,.
\end{eqnarray}
Hence, the difference between $\,\N_{B}^{}\,$ and $\,\N_{B}'\,$ actually arises
from the parameters $\,\kappa_{f}^{}\,$ and $\,\kappa_{f}'\,$ as in
(\ref{eq:kappaf'/kappaf}).
Now, from (\ref{eq:N-B}), (\ref{eq:N-B'}) and (\ref{eq:kappaf'/kappaf}),
we can deduce the ratio of visible and mirror baryon asymmetries,
\begin{equation}
\label{eq:ratio-etaB'/etaB}
\f{\,\N_{B}'\,}{\N_{B}^{}}
~=~ \frac{\,\xi'\kappa_{f}'\epsilon_1'\,}{\xi\kappa_f^{}\epsilon_1^{}}
~=~ \frac{\,\kappa_{f}'\,}{\kappa_{f}^{}}
~=~ \(\f{\,M_1'\,}{\,M_1^{}\,}\)^{\!1.1\pm 0.1} \, .
\end{equation}

With (\ref{eq:DM-B-ratio}), (\ref{eq:ratio-mN'/mN}) and
(\ref{eq:ratio-etaB'/etaB}), we finally arrive at,
\begin{eqnarray}
\label{eq:predict-DM/M}
\frac{\,\Omega_{\rm DM}\,}{\Omega_{\rm M}} &~ =~  &
\frac{\,\Omega_{B'}\,}{\Omega_B}
~=~ \f{\,\N_{B}'\,}{\N_{B}^{}}\f{\,m_N'\,}{m_N^{}}
~=~ \(\f{\,M_1'\,}{\,M_1^{}\,}\)^{\!(1.1\pm 0.1)}\!
    \(\f{v_{\phi'}^{}}{v_{\phi}^{}}\)^{\!2/9} .
 \end{eqnarray}
Thus, to realize the astrophysical observation of
$\,{\Omega_{\rm DM}}/{\Omega_{\rm M}} = 5.0\pm 0.74\,$
as inferred from (\ref{eq:Omega-B-exp})-(\ref{eq:Omega-DM-exp}) \cite{Omega-B-DM},
we deduce a constraint on the ratio between the visible and mirror
heavy singlet neutrino masses,
\beqa
\label{eq:M1'/M1-v'/v}
\f{M_1'}{\,M_1^{}\,} ~=~
\(\f{\,\Omega_{\DM}^{}\,}{\,\Omega_{\rm M}^{}\,}\)^{\!\f{1}{\varrho}}
\(\f{v_{\phi}^{}}{v_{\phi'}^{}}\)^{\!\f{2}{9\varrho}} ,
\eeqa
where $\,\varrho \equiv 1.1\pm 0.1\,$.\,
As we will show in (\ref{eq:v/v'-range}) of Sec.\,\ref{sec:3-3},
the BBN will put nontrivial constraint on the VEV ratio, $\,\vpb /\vp < 0.70\,$.\,
Combining with the naturalness condition (\ref{eq:v'/v-ratio}), we have,
\beqa
\label{eq:v'/v-ULB}
0.1 ~<~ \f{\vpb}{\vp} ~<~ 0.7\,.\,
\eeqa
Taking the ratio $\,{\vpb}/{\vp}=(0.1,\,0.5,\,0.7)\,$,\, we evaluate the VEV factor
in (\ref{eq:M1'/M1-v'/v}) with $\,\varrho = 1.1\,$  and find,
$\,({\vp}/{\vpb})^{\f{2}{9\varrho}}=(1.6,\,1.4,\,1.1)\,$,\, respectively,
which is rather insensitive to $\,{\vpb}/{\vp}\,$.\,
So, in the numerical analyses of Sec.\,\ref{sec:4}-\ref{sec:6}, we will
set a sample value of $\,\vpb /\vp = \hf\,$.\,
With this model-input and $\,\varrho = 1.1\,$,\,
we derive the constraint for the mass-ratio of heavy singlet neutrinos,
$\,r_N^{}\equiv {M_1^{}}/{M_1'}\,$,\,
\beqa
\label{eq:rN-bound}
0.18 ~<~ r_N^{} ~<~ 0.23 \,,
\eeqa
with a central value $\,r_N^{}=0.2\,$,\, where we have imposed
the $2\sigma$ astrophysical limit on the ratio of
dark matter over matter densities,
$\,4.26 < {\Omega_{\rm DM}}/{\Omega_{\rm M}} < 5.74\,$.\,

\vspace*{3mm}
\subsection{Explicit Seesaw Realizations of Visible/Dark Matter Genesis}
\label{sec:3-2}
\vspace*{1.5mm}

In this subsection,  we will extend two of our seesaw constructions \cite{Ge:2010js,He:2011kn}
to the present visible and mirror neutrino sectors,
where the $\mutau$ and $CP$ symmetry breakings naturally arise from a common origin.
With these, we show how the common origin of visible and dark matter geneses
are explicitly realized in the seesaw formalism.
Also, as given in Eq.\,(\ref{eq:kapa-f-BDP}), we see that $\kappa_{f}^{}$
(or similarly $\,\kappa_{f}'$\,) depends on the mass-parameters $\,\tm_{1}^{}$
(or $\,\tm_1'\,$).
We expect that $\,\tm_1^{}\,$
should be the typical mass scale of light neutrinos, around
$\,\tm_1^{} = O(10^{-1}\!-\!10^{-2})\,$eV.\,
With the seesaw formalisms \cite{Ge:2010js,He:2011kn},
we will derive $\,\tm_1^{}\,$ and $\,\tm_1'\,$ explicitly,
which also justify the conditions $\,\tm_1^{}>m_*^{}\,$ and $\,\tm_1'>m_*'\,$,\,
for the application of $\,\kappa_{f}^{}\,$ formula (\ref{eq:kapa-f-BDP}) and its extension
to $\,\kappa_{f}'$\,.\,

The $\mutau$ symmetry is a $\,\mathbb{Z}_2\,$ invariance of the light neutrino mass-matrix
$\,M_\nu\,$ under the exchange $\,\nu_\mu^{} \leftrightarrow \nu_\tau^{}\,$.\,
The $\mutau$ symmetric limit predicts a unique pattern for atmospheric and reactor
neutrino mixing angles, $\,\theta_{23}=45^\deg\,$ and $\,\theta_{13}=0^\deg\,$.\,
This is strongly supported by the existing neutrino oscillation data
as a good zeroth order symmetry, because
both deviations $\,\theta_{23}-45^\deg\,$ and $\,\theta_{13}-0^\deg\,$ are
constrained to be generically small,
$\,-7.5^\deg < \theta_{23}-45^\deg < 2.9^\deg\,$ and
$\,5.1^\deg < \theta_{13}-0^\deg < 10^\deg\,$ at 90\%\,C.L.\,\cite{nu-data},
which are all within $10^\deg$ range.
A vanishing $\,\theta_{13}\,$ also enforces the absence of Dirac $CP$ violation,
so under the attractive conjecture
that all $CP$ violations arise from a common
origin in the neutrino seesaw, it is deduced
that all $CP$ violations share a common origin
with $\mutau$ breaking\,\cite{Ge:2010js,He:2011kn}.
We also note that under $\mutau$ symmetry, the singlet heavy Majorana neutrinos
can either transform simultaneously or act as $\mutau$ singlets since the low energy
oscillation data only measure the mixings encoded in the light neutrino mass-matrix.
In the first case, we found that the common origin of $\mutau$ and $CP$ violations
can be formulated as the unique soft breaking via dimension-3 mass-term of
heavy singlet Majorana neutrinos \cite{Ge:2010js};
while in the second case (called $\mutau$ blind seesaw),
we found that this common breaking can be uniquely formulated in the Dirac mass term
from Yukawa interactions \cite{He:2011kn}.

In the following, we will extend these two constructions to the present mirror
model where the mirror symmetry enforces the visible and mirror
neutrino seesaws to share the common origin of $\mutau$ and $CP$ breaking,
and thus the common origin for visible and mirror leptogeneses with the same $CP$
phase in the $N_1$ and $N_1'$ decays.
For simplicity, we consider the minimal seesaw with two singlet heavy Majorana neutrinos,
in the visible and mirror sectors, respectively. This ensures one of the light neutrinos
to be massless under seesaw and thus predicts the hierarchical mass-spectrum for
light neutrinos (with normal or inverted mass-ordering).
This is always a good approximation
when the third singlet Majorana neutrino is much heavier than the
other two and thus decoupled from the seesaw Lagrangian. Extensions to general
three-neutrino seesaw were also considered in \cite{Ge:2010js,He:2011kn} where a
massless light neutrino is still predicted even after including
the common $\mutau$ and $CP$ breaking.

\vspace*{2mm}
\subsubsection{Visible/Mirror Seesaws with Common Soft $\mutau$ and $CP$ Breaking}
\label{sec:3-2-1}
\vspace*{1.5mm}

For the common origin of soft $\mutau$ and $CP$ breaking,
it uniquely arises from the dimension-3 mass term of
singlet heavy Majorana neutrinos.
For the visible neutrino seesaw, we have the Dirac and Majorana mass-matrices
including the common soft $\mutau$ and $CP$ breaking \cite{Ge:2010js},
\begin{eqnarray}
\label{eq:mD-MN}
m_{D}=\left(\!
\begin{array}{cc}
a & a\\
b & c\\
c & b
\end{array}\!\right) \!,
\qquad M_{N}^{} =\left(\!\!
\begin{array}{cc}
M_{22} & M_{23}\\[2mm]
M_{23} & M_{22}(1\!-\!\zeta e^{i\omega})
\end{array}\!\!\right) \!,
\end{eqnarray}
where the small breaking is characterized by the
module $\,0<\zeta<1\,$ and $CP$ angle $\,\omega\in [0,\,2\pi)\,$.\,
Accordingly, for the mirror neutrino seesaw we have,
\begin{eqnarray}
\label{eq:mD-mD'-soft}
m_{D}' \,=\, \f{\,v_{\phi'}^{}\,}{v_{\phi}^{}}\,m_{D}^{} \,,
&~~~~~~&
M_{N}' \,=\, \f{1}{\,\rN\,} M_{N}^{}\,.
\end{eqnarray}
Here the minimal soft $P$-breaking is realized by the overall ratio
$\,r_N^{}\approx 0.2 \neq 1\,$,\, and the proportionality of
$\,M_{N}' \,\varpropto\, M_{N}^{}\,$ ensues the same structure of
$\,M_{N}^{}\,$ and $\,M_{N}'\,$,\, i.e.,
the visible and mirrors sectors shares the common origin of
$CP$ violation which will thus serve as the common genesis for
matter and dark matter in the two sectors.
The mass-spectrum of light neutrinos falls into the normal mass-ordering (NMO)
pattern ($\,0=m_{1}^{}<m_{2}^{}\ll m_{3}^{}\,$), where the zero mass-eigenvalue
$\,m_{1}^{}=0\,$ was found to persist even in the generalized three-neutrino seesaw
with common soft $\mutau$ and $CP$ breaking\,\cite{Ge:2010js}.
From systematical derivations\,\cite{Ge:2010js}, we have,
\begin{eqnarray}
\label{eq:m1bar-NMO}
{\tm}_{1}^{} ~\equiv~
\frac{(\,\widetilde{m}_{D}^{\dagger}\widetilde{m}_{D}^{})_{11}^{}}{M_1^{}}
~\simeq~\frac{\,(b-c)^{2}}{M_{1}}~\simeq~\chi_1^{}{m}_{3}^{}
~=~ \chi_{1}^{}\sqrt{\Delta m_{13}^2} \,,
\end{eqnarray}
where the third light neutrino mass $\,m_3^{}=\sqrt{\Delta m_{13}^2}\,$
for the NMO spectrum,
and the mass-squared difference $\,\Delta m_{13}^2 \equiv |m_3^2-(m_2^2+m_1^2)/2|\,$
is measured to be
$\,2.06 \!\times\! 10^{-3} < \Delta m_{31}^2 <2.67 \!\times\! 10^{-3}\,\textrm{eV}^2$\,
at $3\sigma$ level.\ with the central value
$\,\sqrt{\Delta m_{31}^2}=0.048\,\textrm{eV}$ \cite{nu-data}.
The coefficient $\chi_1^{}=\chi(M_1^{},M_Z^{})$ in (\ref{eq:m1bar-NMO})
is a renormalization group running factor which evolves
$\,m_3^{}\,$ from weak scale $M_Z^{}$ up to the
leptogenesis scale $M_1^{}$, and is found to be about $\,1.3-1.4\,$ for
$\,M_1=10^{13}-10^{16}\,$GeV \cite{Ge:2010js}. So we can estimate,
$\,{\tm}_{1}^{}\simeq 0.06-0.07\,\textrm{eV} >\, m_*^{}$\,,\, where
$\,m_*^{} \simeq 1.5 \!\times\! 10^{-3}\,$eV\, is given in (\ref{eq:m*}).

For the mirror neutrino seesaw, we can deduce from (\ref{eq:mD-mD'-soft}) and
(\ref{eq:m1bar-NMO}),
\beqa
\label{eq:m1'bar-NMO}
{\tm}_{1}' ~\equiv~
\frac{(\,\widetilde{m}_{D}^{\prime\,\dagger}\widetilde{m}_D')_{11}^{}}{M_1'}
~=~ \f{\,v_{\phi'}^2\,}{v_{\phi}^2}\,\rN\, {\tm}_{1}^{}\,.
\eeqa
As mentioned earlier, from the BBN constraint (Sec.\,\ref{sec:3-3}) and naturalness consideration,
we have, $\,{v_{\phi'}^{}}/{v_{\phi}^{}}\approx \hf\,$;\, while with the density ratio
of dark matter over matter, we inferred from (\ref{eq:rN-bound}),
$\,r_N^{}\approx 0.2\,$.\,  So we can estimate,
$\,{\tm}_{1}' \approx \tm_1^{}/20 \simeq (3-3.5) \times 10^{-3}\,\textrm{eV}
 >\, m_*'$\,,\,
since $\,m_*'\,$ is given by
\beqa
\label{eq:m*'}
m_*' & = & \f{\,16\pi^{\f{5}{2}}\sqrt{g_*'}\,}{3\sqrt{5}}\f{v_{\phi'}^2}{\MP}
~\simeq~ 3.8\times 10^{-4}\,\textrm{eV}\,,
\eeqa
where we count $\,g_*'=219.75=O(200)\,$ at the temperature $\,T=T'\sim M_1'\,$,\,
which contains $\,106.75 \!\times\! 2\,$
from the SM degrees of freedom plus their mirror partners, and
$\,\f{7}{4}\!\times\! 3\,$ for $\,N_1'\,$ and $(N_1,\,N_2)$,\,
as well as another $1$ by the real scalar $\,\X$\,.

\vspace*{2.5mm}
\subsubsection{Visible/Mirror $\mutau$ Blind Seesaws with Common $\mutau$ and $CP$ Breaking}
\label{sec:3-2-2}
\vspace*{1.5mm}

For the $\mutau$ blind seesaw, the heavy Majorana neutrinos are $\mutau$ singlets,
so we can always start with their mass-eigenbasis under which $\,M_N^{}\,$ and $\,M_N'\,$
are diagonal. Thus the Dirac mass-terms $\,m_D^{}\,$ and $\,m_D'\,$ are the unique place
for common $\mutau$ and $CP$ breaking \cite{He:2011kn},
\beqs
\begin{eqnarray}
\label{eq:mD-MN-blind}
m_{D}^{} ~=\left(\!\begin{array}{cc}
\overline{a} & \overline{a}^{\prime}\\[1.5mm]
\overline{b} & \overline{c}(1-\zeta^{\prime})\\[1.5mm]
\overline{b} & \overline{c}(1-\zeta e^{i\omega})
\end{array}\!\!\right) \!,
&~~~~~~~&
M_N^{}~=\left(\!\!
\begin{array}{cc}
M_{1} & 0
\\[2mm]
0 & M_{2}
\end{array}
\!\!\right) \!,
\end{eqnarray}
for the visible neutrino sector, and
\begin{eqnarray}
\label{eq:mD-mD'-blind}
m_{D}' \,=\, \f{v_{\phi'}^{}}{v_{\phi}^{}}\,m_{D}^{} \,,
&~~~~~~~&
M_{N}' \,=\, \f{1}{\,\rN\,} M_{N}^{}\,,
\end{eqnarray}
\eeqs
for the mirror neutrino sector,
where $\,0<\zeta <1\,$,\, $|\zeta'|<1\,$,\, and $\,\omega\in [0,2\pi)\,$
parameterize the $\mutau$ and $CP$ breaking.
Hence, it is the proportionality $\,m_D^{} \propto m_D'\,$ that
ensues the visible and mirror sectors to share the common origin of
$CP$ violation which will then serve as the common genesis for
matter and dark matter in the two sectors.

From systematical analysis \cite{He:2011kn}, we see that the light neutrino
mass-spectrum falls into the inverted mass-ordering (IMO)
pattern ($\, m_{2}\gtrsim m_{1}\gg m_{3}=0\,$), where we found that
the zero mass-eigenvalue $\,m_{3}^{}=0\,$ persists up to the next-to-leading
order even in the generalized three-neutrino seesaw with common $\mutau$ and $CP$
breaking.  Then we deduce,
\beqs
\begin{eqnarray}
\label{eq:m1bar-IMO}
{\tm}_{1}^{} ~\equiv~
\f{(\widetilde{m}_{D}^{\dagger}\widetilde{m}_{D}^{})_{11}^{}}{M_1^{}}
\,~\simeq~ \chi_1^{}m_1^{}~\simeq\,\chi_{1}\sqrt{\Delta m_{13}^2} ~,
\end{eqnarray}
for the visible seesaw as in \cite{He:2011kn}, and
\beqa
\label{eq:m1'bar-IMO}
{\tm}_{1}' ~\equiv~
\frac{(\,\widetilde{m}_{D}^{\prime\,\dagger}\widetilde{m}_D')_{11}^{}}{M_1'}
~=~ \f{\,v_{\phi'}^2\,}{v_{\phi}^2} \,\rN\, {\tm}_{1}^{}\,,
\eeqa
\eeqs
for the mirror seesaw.

Comparing (\ref{eq:m1bar-IMO})-(\ref{eq:m1'bar-IMO}) in the $\mutau$
blind seesaw with (\ref{eq:m1bar-NMO})-(\ref{eq:m1'bar-NMO})
in the soft breaking seesaw, we can deduce, for both cases,
\beqs
\label{eq:m1bar-m1'bar}
\begin{eqnarray}
\label{eq:m1bar}
{\tm}_{1}^{} & ~\simeq~ & \chi_1^{}\sqrt{\Delta m_{13}^2}
~\simeq~ 0.06-0.07\,\textrm{eV} ~>~ m_*^{} \,,
\\
\label{eq:m1bar/m1'bar}
{\tm}_{1}' & \simeq &
\f{\,v_{\phi'}^2\,}{v_{\phi}^2} \rN {\tm}_{1}^{}
~=~ (3-3.5) \!\times\! 10^{-3}\,\textrm{eV} ~>~ m_*' \,,
\end{eqnarray}
\eeqs
where we take the ratios
$\,{v_{\phi'}^{}}/{v_{\phi}^{}}\approx \hf\,$ and $\,r_N^{}\approx 0.2\,$,\,
as explained above.

\vspace*{3mm}
\subsection{Analysis of the BBN Constraint}
\label{sec:3-3}
\vspace*{1.5mm}

Before concluding this section, we discuss the possible constraint from the
Big-Bang nucleosynthesis (BBN) on the mirror sector.
The observed light elements abundances in the universe agrees well
with the predictions of BBN nucleosynthesis in the SM of particle physics.
This means that at the temperature $\,T\sim 1\,\text{MeV}$,\,
the number of effective degrees of freedom should be $\,g_*^{}=10.75\,$
as contributed by photons, electrons and three species of neutrinos.
Considering the mirror model, we have additional contributions from mirror photons,
mirror electrons and mirror neutrinos to $\,g_*^{}\,$.\,
So the total number of degrees of freedom becomes as $\,\hat{g}_*^{}\,$,
\begin{eqnarray}
\label{eq:gbar*}
\hat{g}_*^{} ~=~ g_*^{}\left[1+\(\f{\,T'}{T}\)^{\!\!4}\right] ,
\end{eqnarray}
where $T$ ($T'$) is the temperature of visible (mirror) sector.
The deviation of $\,\hat{g}_*^{}\,$ from $\,g_*^{}\,$ is normally parametrized
in terms of the effective number of extra neutrino species $\Delta N_\nu$ via
$\,\Delta g_*^{} = \hat{g}_*^{} - g_*^{} = 1.75\Delta N_\nu\,$.\,  So we have,
\begin{eqnarray}
\label{eq:Delta-Nnu}
\Delta N_\nu ~\simeq~ 6.14\(\frac{\,T'}{T}\)^{\!\!4} .
\end{eqnarray}
But, the current BBN analysis gives\,\cite{Izotov:2010ca},
$\,N_\nu=3.80^{+0.80}_{-0.70}\,$ at $2\sigma$ level, for the neutron lifetime
being $\,878.5\pm 0.8\,s$.\,
So this puts a $2\sigma$ upper limit,
$\,\Delta N_\nu < 1.50\,$,\,
and thus imposes the constraint on the mirror temperature $\,T'\,$
in the BBN epoch,
\begin{eqnarray}
\label{eq:mBBN-cond}
T' ~<~ 0.70\,T \,,
\end{eqnarray}
where the coefficient is proportional to $(\Delta N_\nu )^{\f{1}{4}}$, with
only a mild dependence on $\,\Delta N_\nu$.\,
In the literature\,\cite{lowT'} it was assumed that after inflation
the reheating temperatures in the two sectors are different such that the
condition (\ref{eq:mBBN-cond}) will be obeyed during the BBN.
But it is not the case for the current construction
due to the mixed Higgs interactions $\,|\phi|^2|\phi'|^2\,$,\,
$\,\chi|\phi|^2\,$ and $\,\chi|\phi'|^2\,$
in (\ref{eq:V}) which will bring the two sectors
into thermal equilibrium before the BBN starts
(even though the kinetic mixing between photons and mirror photons
in (\ref{eq:Kmix-ga-ga'}) may be negligible).

Our model realizes (\ref{eq:mBBN-cond}) in a different way.
For the above reason, we simply have the equal temperatures $\,T=T'\,$
for the two sectors after inflation and then at the leptogensis scales 
$\,M_{N}$ and $M_{N}'\,$.\,
We observe that the desired temperature difference in (\ref{eq:mBBN-cond})
can be produced through the visible and mirror electroweak phase transitions at
the scales $\,\sim\! (v_\chi^{},\,v_\phi^{},\,v_{\phi'}^{})=O(100\,\textrm{GeV})\,$.\,
For simplicity of illustration, let us first write down the one-loop
effective potential for scalar field $\,\chi\,$ alone
at temperature $T$. Defining the thermal average
$\,\left<\chi\right> =\chi_c^{}\,$,\, we have
\begin{eqnarray}
\label{eq:V-Xc}
V(\Xc) & ~ =~ &
\LB -\hf\mu_{\X}^{2} \X_c^2
    +\f{1}{4}\lambda_{\X}^{}\X_c^{4}\RB
  +\LB \f{\la_\X^{}}{6}\,T^2\,\X_c^{2}-\f{\pi^2}{90}
       \(\!N_B^{}+\f{7}{8}N_F^{}\!\)T^4 \RB
\nn\\[1mm]
 && +\f{3\la_\X^2}{128\pi^2}\,\X_c^{4}\(\ln\f{\X_c^{2}}{M^2}-\f{25}{6}\) ,
\end{eqnarray}
where on the RHS inside the second brackets is the finite temperature correction
and the last line gives the Coleman-Weinberg term.
In the effective potential above, $N_B^{}$ ($N_F^{}$) denotes the number
of bosonic (fermionic) degrees of freedom.
At finite temperature, the Higgs mass-term receives a correction from
the thermal fluctuation and becomes,
\begin{eqnarray}
m^2_\X (T) ~=\, -\mu^2_\X + \f{\,\la_\X^{}}{3}\,T^2 \,.
\end{eqnarray}
Near the critical temperature $\,T_c=\mu_\X^{}\sqrt{\f{3}{\,\la_\X^{}}}\,$,\,
we have the scalar mass $\,m_\X^2\simeq 0\,$,\,
so the effective potential takes the form,
\begin{eqnarray}
\label{eq:V(Xc)}
V(\X_c^{}) ~=~ \f{3\la_\X^2}{128\pi^2}\,\X_c^{4}\(\ln\f{\X^2_c}{M^2}-\frac{25}{6}\)
+ \f{1}{4}\lambda_{\X}^{}\X_c^{4}
- \f{\pi^2}{90}\(\!N_B^{}+\f{7}{8}N_F^{}\!\)T^4 \,.
\end{eqnarray}
Similar kind of scalar potential to (\ref{eq:V(Xc)})
was also used to generate the electroweak scale inflation
in Ref.\,\cite{Knox:1992iy} for a singlet inflaton field at the weak scale.
It found that this
can result in about 30 e-foldings at the weak scale.
We do not need such huge expansion at this stage
since we have the conventional high scale inflation in our scheme.
But from this we see that it is easy for the electroweak phase transition of
a Higgs field to cause a small expansion just about
$\,1 \!\sim\! 2\,$ e-foldings\footnote{Since the expansion is much smaller
here, a precise calculation will require the inclusion of radiation contribution
to the Hubble constant in the phase transition epoch. But such detail is not
needed for our discussion below.}
starting from the temperature\footnote{In our
explicit minimization of the electroweak vacuum in Sec.\,\ref{sec:4-2}, we will realize
three sample patterns of the Higgs VEVs,
$\,v_\chi^{}\!\sim\! v_\phi^{} \!\sim\! 2v_{\phi'}^{}\,$
or $\,v_\chi^{}\!\sim\! 4v_\phi^{} \!\sim\! 8v_{\phi'}^{}\,$.},
$\,T=T' \!\sim\! \max (v_\chi^{},\,v_\phi^{},\,v_{\phi'}^{})\,$.\,
And then $T$ rolls down together with $T'$.
After these three Higgs bosons roll into the potential minimum at lower temperature,
the reheatings\,\cite{KT-book} from the electroweak vacuum energies start.
The vacuum energy density in (\ref{eq:V0}) takes the forms of
$\,v_\X^4\,$,\, $\,v_\X^2v_\phi^2\,$,\, $\,v_\phi^4\,$,\,
$\,v_\X^2v_{\phi'}^2\,$,\, $\,v_\phi^2v_{\phi'}^2\,$
and $\,v_{\phi'}^4\,$.\,
Under the explicit constructions in Sec.\,\ref{sec:4}, we will always have the mass-eigenstate
Higgs $\hat{\X}$ and $\hat{\phi}$ dominantly decay into the visible SM particles,
and the mirror Higgs $\hat{\phi}'$ mainly decay into mirror particles.
So the reheatings
of vacuum energies associated with $\hat{\X}$ and $\hat{\phi}$ will raise the temperature
of visible sector back to  $\,T\!\sim\!\max (v_\chi^{},\,v_\phi^{})\,$,\,
and the reheating with $\hat{\phi}'$ raises the temperature of mirror sector back to
$\,T'\!\sim\! v_{\phi'}^{}\,$.\,
So, the visible and mirror reheatings end up with a temperature relation,
\beqa
\label{eq:T'/T-EWreheating}
\f{\,T'}{T} ~\sim~ \f{\,v_{\phi'}^{}}{\,\max (v_\chi^{},v_\phi^{})\,}
~\sim~ \f{\,\vpb\,}{\vX},\, \f{\,\vpb\,}{\vp} \,.
\eeqa
After reheatings, the temperature difference (\ref{eq:T'/T-EWreheating})
remains along the expansion of the universe till the BBN epoch
at $\,T\!\sim\! 1\,$MeV.\,
Combining the BBN condition (\ref{eq:mBBN-cond}) with (\ref{eq:T'/T-EWreheating}),
we find the VEV ratio subjects to the constraint,
\beqa
\label{eq:v/v'-range}
\f{\,\vpb\,}{\vX},\, \f{\,\vpb\,}{\vp} ~<~ 0.7 \,.
\eeqa
For sufficient reheatings with hundred percent conversion of
the vacuum energies into radiations,
we need Higgs oscillations to decay rapidly, with $\,\Gamma_S^{} \gtrsim H\,$
\cite{KT-book},
where $\,\Gamma_S\,$ denotes the decay width of a given scalar $S$,\, and
$H$ is the Hubble expansion rate at the electroweak phase transition as determined by
\beqa
\label{eq:H-weak}
H ~\simeq~ \sqrt{\f{\,8\pi\! \left<V(S)\right>\,}{3M_{\rm Pl}^2}}
~\sim~ \sqrt{\f{8\pi}{3}}\f{v_S^2}{M_{\rm Pl}^{}}
~\sim~ 10^{-15}\,\textrm{GeV} \,,
\eeqa
where $\,\left<V(S)\right>\sim v_S^4\,$ is the typical
vacuum potential of a scalar $S\,(=\X,\phi,\phi')$ with its VEV
$\,v_S^{}=O(100)\,$GeV at the weak scale.
For our model, the typical decay width of $\,S\,$
is found to be $\,\Gamma_S^{}=O(10^{-5}-1)$GeV (cf.\ Sec.\,\ref{sec:5}), so from (\ref{eq:H-weak})
we have $\,\Gamma_S^{}\gg H\,$,\, showing that the electroweak vacuum energies are
fully converted to radiations.
Since all Higgs bosons decay away by the end of
electroweak vacuum reheating,
the mixed interactions like $\,|\phi|^2|\phi'|^2\,$ could only occur via virtual
processes and are much suppressed.
So the thermal contact between the visible and mirror sectors is negligible, and
thus the temperature difference (\ref{eq:T'/T-EWreheating}) is retained.
There could be potential mixing between visible and mirror photons
as in (\ref{eq:Kmix-ga-ga'}), but we find that in our model their mixing parameter is constrained
down to $\,\ep < 3.4\times 10^{-5}\,$ due to the orthopositronium bound
[cf.\ (\ref{eq:our-AA'-limit}) in Sec.\,\ref{sec:4-3}] and
$\,\ep \leqq 10^{-8}\,$  due to the direct dark matter search limit of TEXONO
[cf.\ Fig.\,\ref{fig:DR-ER}(b) in Sec.\,\ref{sec:6}],
so it will not affect the temperature difference (\ref{eq:T'/T-EWreheating}).

\vspace*{3mm}
\section{Higgs Masses, Couplings and Low Energy Constraints}
\label{sec:4}
\vspace*{1.5mm}

In this section we present realistic numerical samples of our model-predictions
that can be tested at colliders.
For the successful matter and dark matter genesis (Sec.\,\ref{sec:3-1}-\ref{sec:3-2}) and the
realization of BBN (Sec.\,\ref{sec:3-3}), we have derived constraints on the mass ratio of
heavy singlet neutrinos and the ratio of Higgs boson VEVs
between the visible and mirror sectors,
as in (\ref{eq:rN-bound}) and (\ref{eq:v/v'-range}).
With the vacuum minimization conditions for spontaneous mirror parity violation
and electroweak symmetry breaking in Sec.\,\ref{sec:2-3}, we first analyze the viable parameter space.
Then, we will present three numerical samples under (\ref{eq:v/v'-range}), and derive
the corresponding Higgs mass-spectrum and couplings. Finally, we analyze the
constraints from the direct Higgs search and indirect electroweak precision data.

\vspace*{3mm}
\subsection{Analytical Constraints on the Parameter Space}
\label{sec:4-1}
\vspace*{1.5mm}

There are seven free parameters in the Higgs potential (\ref{eq:V}),
including two masses and five couplings. The soft breaking term (\ref{eq:V-soft})
contains an extra coefficient $\,\beta_\X^{}\,$.\,
But, we need to impose following nontrivial physical constraints,
which will largely reduce the number of input parameters of our Higgs potential.
These constraints are:
(i)~the VEV of $\,\phi\,$ must generate the right amount of electroweak symmetry breaking in the
    visible sector, i.e., we have $\,v_{\phi}^{} = (2\sqrt{2}G_F)^{\hf}_{}\simeq 174\,$GeV.
(ii)~for successful mirror dark matter genesis and realization of BBN,
    the ratio of the Higgs VEVs
    should obey the condition (\ref{eq:v/v'-range}).
(iii)~both the mixings between $\,\phi'-\chi\,$ and between $\,\phi-\phi'\,$ should be small,
    so that decays of the mass-eigenstate of $\,\phi\,$ or $\,\chi\,$ into mirror fermions
    and gauge bosons are negligible. This is to ensure that during the reheating of
    electroweak phase transition the decays of the mass-eigenstates of $\,\phi\,$ and $\,\chi\,$
    will mainly heat up the temperature $T$ of the visible sector, but without affecting
    the temperature $T'$ of the mirror sector (cf.\ Sec.\,\ref{sec:3-3}).
This last constraint also implies that the mass-eigenstate of mirror Higgs $\,\phi'\,$ mainly
decouple from the visible sector. These are unique features of our construction and differ
from all previous mirror models in the literature.

We note that among all seven parameters in the Higgs potential (\ref{eq:V}),
we have three with mass-dimension one, $(\mu_{\phi},\, \mu_{\chi},\, \beta_{\chi\phi})$,\,
and the rest four are dimensionless couplings.
The second order derivative of the vacuum potential (\ref{eq:V0})
at the minimum should be positive. This means that the Higgs mass matrix (\ref{eq:Mhiggs-3x3})
should be positive-definite, so we can infer the following conditions,
\begin{subequations}
\begin{eqnarray}
\label{eq:PDcond-1}
&&
\lambda_{\phi}^{+} ~>~ 0\,,~~~~~~ \lambda_{\phi}^{-} ~>~ 0 \,,
\\
\label{eq:PDcond-2}
&&
2\lambda_{\phi}^{+}\lambda_{\phi}^{-}
\left[ -\mu_{\chi}^{2}+ 3\lambda_{\chi}v_{\chi}^{2}+\lambda_{\chi\phi}
(v_{\phi}^{2}+v_{\phi^{\prime}}^{2})\right]
-\lambda_{\phi}^{+}\beta_{\chi\phi}^2
-\lambda_{\phi}^{-}\lambda_{\chi\phi}^{2}v_{\chi}^2 ~>~ 0 \,.
\end{eqnarray}
\end{subequations}
Furthermore, whole potential (\ref{eq:V0}) should be bounded from below.
This is determined by the coefficients of the quartic terms.
Let us rewrite the quartic interactions of (\ref{eq:V}) in terms of quadratical form
$\,\Omega^T{\cal C}\,\Omega\,$ where
$\,\Omega =(|\phi|^2,\,|\phi'|^2,\,\chi^2)\,$ and ${\cal C}$ is a $3\times 3$ matrix including the
relevant quartic Higgs couplings. So, requiring the matrix $\,{\cal C}\,$ to be positive-definite,
we deduce a new condition in addition to (\ref{eq:PDcond-1}),
\beqa
\label{eq:PDcond-3}
\lambda_{\phi}^{+}\lambda_{\chi}^{} ~>~ \f{1}{4}\lambda_{\chi\phi}^2 \,.
\eeqa
Together with (\ref{eq:PDcond-1}), this also leads to $\,\lambda_{\chi}^{} > 0\,$.\,
As explained in Sec.\,\ref{sec:3-3}, to satisfy the BBN constraint 
we require that the mass-eigenstate Higgs
bosons $\,\chih\,$ and $\,\phih\,$ predominantly decay into the visible SM particles.
This means that the mixings between $\,\chi-\phi'\,$ and $\,\phi-\phi'\,$
should be sufficiently small. So, inspecting the scalar mass-matrix in
(\ref{eq:Mhiggs-3x3})-(\ref{eq:Mhiggs-ij}), we require the mixing elements
$\,(m_{\phi\phi'}^2,\,m_{\phi'\chi}^2)\simeq 0\,$,\,
which impose the following constraints,
\begin{subequations}
\label{eq:noMix-cond}
\begin{eqnarray}
\label{eq:noMix-cond-phiphi'}
\lambda_{\phi}^{+} -\lambda_{\phi}^{-} &  \simeq  & 0 \,,
\\
\label{eq:noMix-cond-chiphi'}
\lambda_{\chi\phi}v_{\chi} - \beta_{\chi\phi} &  \simeq  & 0 \,.
\end{eqnarray}
\end{subequations}
We find that for the invisible decays of $\,\chih\,$ and $\,\phih\,$
into mirror particles to have a branching fraction less than 5-10\%,
it is enough to numerically hold the conditions
(\ref{eq:noMix-cond-phiphi'})-(\ref{eq:noMix-cond-chiphi'}) just to a few percent level.

Since the constraint (\ref{eq:v/v'-range}) gives $\,v_{\phi}^{}>v_{\phi'}^{}\,$
and the condition (\ref{eq:PDcond-1}) shows $\,\lambda_\phi^{\pm}>0\,$, the solution
(\ref{eq:vphi-vphi'}) will then require the trilinear coupling
$\,\beta_{\chi\phi}^{}\,$ to be negative.
Combining this with the relation (\ref{eq:noMix-cond-chiphi'}), we thus arrive at,
\begin{subequations}
\begin{eqnarray}
\beta_{\chi\phi}^{} & < & 0 \,,
\\
\lambda_{\chi\phi}^{} & < & 0 \,,
\end{eqnarray}
\end{subequations}
where we have adopted the convention with all Higgs VEVs being positive.

Among eight free parameters in the original Higgs potential
(\ref{eq:V})-(\ref{eq:V-soft}),
three of them $(\mu_\phi^{},\,\mu_\chi^{},\,\beta_{\chi\phi}^{})$ have mass-dimension equal one
and the soft breaking parameter $\,\beta_\X^{}\,$ has mass-dimension equal three,
while the other four are dimensionless couplings
$(\lambda_\phi^+,\,\lambda_{\phi}^-,\,\lambda_\chi^{},\,\lambda_{\chi\phi}^{})$.
The three Higgs VEVs, $(v_\phi^{},\,v_{\phi'}^{},\,v_\chi^{})$, are all constrained
by the vacuum conditions (\ref{eq:v-phi})-(\ref{eq:v-x-0}) and (\ref{eq:vX-soft-sol}).
In the above, we have imposed four physical constraints:
(i) the Higgs VEV $\,v_\phi^{}\simeq 174$\,GeV is
to generate full electroweak symmetry breaking in the visible sector;
(ii) the mirror Higgs VEV  $\,v_{\phi'}^{}$,\,
besides realizing the mirror electroweak symmetry breaking,
should obey the BBN constraint (\ref{eq:T'/T-EWreheating}),
and we will set a natural sample value $\,x\equiv v_{\phi}^{}/v_{\phi'}^{} =2\,$
for convenience of the numerical analysis in Sec.\,\ref{sec:4-2};
(iii) the two additional constraints,
$\,(m_{\phi\phi'}^2,\,m_{\phi'\chi}^2)\simeq 0\,$,\,
will ensure the mass-eigenstates $\,\chih\,$ and $\,\phih\,$ to predominantly decay
into the particles in the visible sector (rather than mirror sector).
These will reduce the eight free parameters of
the Higgs potential (\ref{eq:V-tot}) down to four, which we may choose, for instance,
to be the four dimensionful parameters
\,$(\mu_\phi^{},\,\mu_\chi^{},\,\beta_{\chi\phi}^{},\,\beta_\X^{})$.\,
Using \,$(\mu_\phi^{},\,\mu_\chi^{},\,\beta_{\chi\phi}^{},\,\beta_\X^{})$\, as inputs,
we can then resolve the remaining four parameters in (\ref{eq:V}) as follows,
\beqs
\label{eq:sol4}
\beqa
\label{eq:sol4-lambda+-}
\lambda_\phi^+ &\!\simeq\!& \lambda_{\phi}^-
~\simeq~ \f{\,\mu_\phi^2\,}{\,v_\phi^2\,}
         \f{x^2}{\,x^2\!+\!3\,} \,,
\\[1mm]
\label{eq:sol4-lambdaXP}
\lambda_{\chi\phi}^{} &\!\simeq\!&  \f{\,\beta_{\chi\phi}^{}\,}{v_\chi^{}}
~\simeq\,
-\f{\,\beta_{\chi\phi}^2\,}{2\mu_\phi^2}
 \f{\,x^2\!+\!3\,}{\,x^2\!-\!1\,}
\,,
\\[1mm]
\label{eq:sol4-lambdaX}
\lambda_\chi^{} &\!\simeq\!&
\f{\mu_\chi^2}{\,v_\chi^2\,}\!\left(\!
1 + \f{\,\beta_{\chi\phi}^2v_\phi^2\,}{\,\mu_\chi^2\mu_\phi^2\,}
    \f{\,x^2\!+\!3\,}{\,x^2\!-\!1\,}
\right) -\f{\,\beta_\X^{}\,}{v_\X^3} \,,
\eeqa
\eeqs
where the VEV of $\,\chi\,$ is derived from
(\ref{eq:vphi-vphi'}) and (\ref{eq:sol4-lambda+-}),
\beqa
\label{eq:sol-vX}
v_\chi^{} ~\simeq~
\frac{2\mu_{\phi}^{2}}{\,-\beta_{\chi\phi}^{}\,}\frac{x^{2}\!-\!1}{\,x^2\!+\!3\,}
 \,,
\eeqa
which is positive due to $\,x>1\,$ and $\,\beta_{\chi\phi}^{}<0\,$.\,
As expected, we see that under the physical constraints, all four dimensionless couplings
\,$(\lambda_\phi^+,\,\lambda_{\phi}^-,\,\lambda_\chi^{},\,\lambda_{\chi\phi}^{})$\,
are now expressed as functions of the four dimensionful parameters
$\,(\mu_\phi^{},\,\mu_\chi^{},\,\beta_{\chi\phi}^{},\,\beta_\X^{})\,$
of (\ref{eq:V-tot}),  in addition to the physically constrained
Higgs vacuum expectation value $v_\phi^{}$
and the ratio $\,x\equiv v_\phi^{}/v_{\phi'}^{}\,$.\,
The above analytical solutions will hold to a numerical precision of a few percent
for our viable parameter space. Although we will use the exact numerical solutions for
the phenomenological analyses below, the above allows us to analytically understand the
viable parameter space and provides us with nontrivial consistency checks.

We also note that under the approximation $\,(m_{\phi\phi'}^2,\,m_{\phi'\chi}^2)\simeq 0\,$,\,
the scalar mass-matrix (\ref{eq:Mhiggs-3x3}) reduces to
a $2\times 2$ matrix form with $\,\phi-\X$\, mixing,
\begin{eqnarray}
\label{eq:Mhiggs-2x2}
{\cal M}_{\phi\X}^2 ~=
\left(\!
\begin{array}{cc}
m_{\phi\phi}^{2}  & m_{\phi\chi}^{2}
\\[1.5mm]
m_{\phi\chi}^{2} & m_{\chi\chi}^{2}
\end{array}\!\right) \!.
\end{eqnarray}
Using the approximate condition (\ref{eq:noMix-cond}),
we simplify the elements of mass-matrix (\ref{eq:Mhiggs-2x2})
as,
%
\begin{eqnarray}
\label{eq:Mhiggs-phichi}
m_{\phi\phi}^{2} & \!\!\simeq\!\! & \mu_{\phi}^{2}\frac{8 x^2}{x^2 +3}\,,
~~~~~~
\nn\\[1mm]
m_{\chi\chi}^{2} & \!\!\simeq\!\! & \mu_{\chi}^{2}
+ v_{\phi}^{2}\f{\,\beta_{\chi\phi}^2\,}{2\mu_\phi^2}
 \f{\,x^2\!+\!3\,}{\,x^2\!-\!1\,}
 \(\f{\,5x^2\!-\!1\,}{\,2x^2}
   +\frac{3}{2}\frac{\beta_{\chi}^{}}{\beta_{\chi\phi}^{}v_{\phi}^2}\) ,
\\[2mm]
m_{\phi\chi}^{2} & \!\!\simeq\!\! & 2\be_{\chi\phi}v_{\phi}^{} \,.
\nn
\end{eqnarray}
%
The mass matrix ${\cal M}_{\phi\X}^2$ can be diagonalized by an orthogonal rotation
from the gauge-eigenbasis \,$(\phi,\,\chi)$\, to the mass-eigenbasis
\,$(\hat{h},\,\hat{\chi})$,\,
where the rotation angle $\,\theta_{\X}^{}\,$ is given by
\begin{equation}
\tan(2\theta_{\X}^{}) ~=~
\f{2m_{\phi\chi}^2}{\,m_{\chi\chi}^2\!-m_{\phi\phi}^2\,} \,.
\end{equation}
Then we can readily derive the approximate mass-eigenvalues for
all three Higgs bosons $(\hh,\,\hh',\,\Xh)$,
\beqs
\label{eq:hh'X-mass}
\beqa
\label{eq:hX-mass}
m_{h,\X}^{2} & \!\simeq\!& \frac{1}{2}\left[ (m_{\phi\phi}^{2}\!+m_{\chi\chi}^{2})
\pm \sqrt{(m_{\phi\phi}^{2}\!-m_{\chi\chi}^{2})^2 +4m_{\phi\chi}^4\,}\,\right]\!,
\\[1mm]
\label{eq:h'-mass}
m_{h'}^{2} & \!\simeq\!&
\mu_{\phi}^{2}\frac{8}{\,x^2\!+\!3\,} \,,
\eeqa
\eeqs
where in (\ref{eq:hX-mass}) the larger (smaller) mass-eigenvalue
corresponds to \,$+\,(-)$\, sign in the bracket.
These analytical formulas will be used for consistency checks
of our exact numerical samples in Sec.\,\ref{sec:4-2}.

\vspace*{3mm}
\subsection{Higgs Mass-Spectrum and Couplings: Three Numerical Samples}
\label{sec:4-2}
\vspace*{1.5mm}

With the guidelines from Sec.\,\ref{sec:4-1}, we can construct realistic numerical samples of
our model-predictions. Inspecting (\ref{eq:sol4-lambda+-}) and (\ref{eq:sol-vX}) and
setting the ratio $\,x\equiv v_{\phi}^{}/v_{\phi'}^{} = 2\,$,\, we have
\beqa
\vX ~\simeq~ \f{\,2\lambda_\phi^- v_\phi^2\,}{\,-\beta_{\chi\phi}^{}\,}
             \f{\,x^2\!-\!1\,}{x^2}
~\simeq~ \f{v_\phi^2}{\,\f{2}{3}|\beta_{\chi\phi}^{}|/\lambda_\phi^-\,} \,.
\eeqa
So, for the natural choice of $\,|\beta_{\chi\phi}^{}|/\lambda_\phi^- =O(v_\phi^{})\,$,\,
we have $\,\vX = O(v_\phi^{})=O(v_\phi^{} - v_{\phi'}^{})\,$.
This means that the visible and mirror electroweak symmetry breakings together
with the spontaneous mirror parity violation all naturally happen at the weak scale,
around $O(10^2\,\textrm{GeV})$. Furthermore, for all quartic Higgs couplings
$\,\lambda_i \lesssim O(0.1-1)\,$ in perturbative region
of the Higgs potential (\ref{eq:V-tot}),
we expect that the three Higgs bosons $(\phi,\,\phi',\,\X)$ should have
masses around $O(10^2\,\textrm{GeV})$, which are significantly below 1\,TeV.

To avoid the BBN constraint in Sec.\,\ref{sec:3-3}, we have required the two Higgs mass-eigenstates
$\,\hat{\X}$\, and $\,\hat{\phi}$\, to predominantly decay into the visible sector. (Hereafter,
for convenience we will use the notations $\,\hh$\,,\, $\hph$\, and $\,\Xh\,$ to denote the
mass-eigenstates of $\,\phi\,$,\, $\phi'$\, and $\,\chi$\,,\,
respectively, unless specified otherwise.)
Numerically, we find it sufficient to have the
branching ratios of $\,\Xh$\, and $\,\hh$\, decays into the visible sector larger than
about 90\%. So more than 90\% of the vacuum energies associated
with $\,\Xh\,$ and $\,\hh\,$ will be converted to the visible sector and less
than 10\% to the mirror sector. Thus the temperature $\,T\,$ of visible sector will be
reheated up to $\,T \propto (90\%)^{\f{1}{4}}\max (v_\chi^{},\,v_\phi^{})
\simeq 0.97\max (v_\chi^{},\,v_\phi^{})\,$, and the $\,T'$\, of mirror sector will be reheated
up to  $\,T' \propto (10\%)^{\f{1}{4}}\max (v_\chi^{},\,v_\phi^{})
\simeq 0.56\max (v_\chi^{},\,v_\phi^{})\,$ due to the $\,\Xh$\, and $\,\hh$\, decays.
This leads to a ratio $\,T'/T \sim 0.58\,$,\,
which still obeys the BBN constraint (\ref{eq:mBBN-cond}).

Taking all these into consideration, we systematically explore the viable parameter space
via numerical analysis. To cover the main parameter space, we have constructed
three sample inputs, called Sample-A, -B and -C, respectively, which are summarized
in the Table\,\ref{tab:input-ABC}.
We see that the mass-parameters $\mu_{\phi}^{}$ and $\mu_{\X}^{}$,
as well as the dimensionful cubic coupling (over the dimensionless
quartic couplings $\lambda_\phi^\pm$\,),
$\,\beta_{\chi\phi}^{}/\lambda_\phi^\pm\,$,\, are all of $\,O(10^2\,\textrm{GeV})$.\,
The four quartic Higgs couplings are in the natural range of $\,O(1-0.01)\,$.\,

\begin{table}[h]
\vspace*{3mm}
\begin{center}
\begin{tabular}{c||c|c|c||c|c|c|c||c}
\hline\hline
&&&&&&&\\[-2.7mm]
Sample & $\mu_{\phi}^{}$\,(GeV)  & $\mu_{\chi}^{}$\,(GeV) & $\beta_{\chi\phi}^{}$\,(GeV)
& $\lambda_{\phi}^{-}$  & $\lambda_{\phi}^{+}$  & $\lambda_{\chi\phi}^{}$
& $\lambda_{\chi}^{}$ & $\beta_{\chi}^{\frac{1}{3}}$\,(GeV)
\\
&&&&&&&&\\[-3.4mm]
\hline
&&&&&&&& \\[-3.6mm]
 A  &$70$ & $113$ & $-35$ & $0.094$ & $0.0923$ & $-0.28$ & $2.03$ & $-30 $
\\
\hline
&&&&&&&& \\[-3.6mm]
 B  & $ 60$ & $255$ & $-21$ & $0.068$ & $0.0696$ & $-0.154$ & $3.42$ & $-30 $
\\
\hline
&&&&&&&& \\[-3.6mm]
C  &$ 62$ & $56.6$ & $-5$ & $0.077$ & $0.0747$ & $-0.0074$ & $0.0075$ & $-20 $
\\
\hline\hline
\end{tabular}
\caption{Three samples of input parameters for the Higgs potential
(\ref{eq:V-tot}) in our model.}
\label{tab:input-ABC}
\par
\end{center}
\vspace*{-4mm}
\end{table}

\begin{figure}[t]
\begin{centering}
\includegraphics[width=8.6cm,height=7.6cm,clip=true]{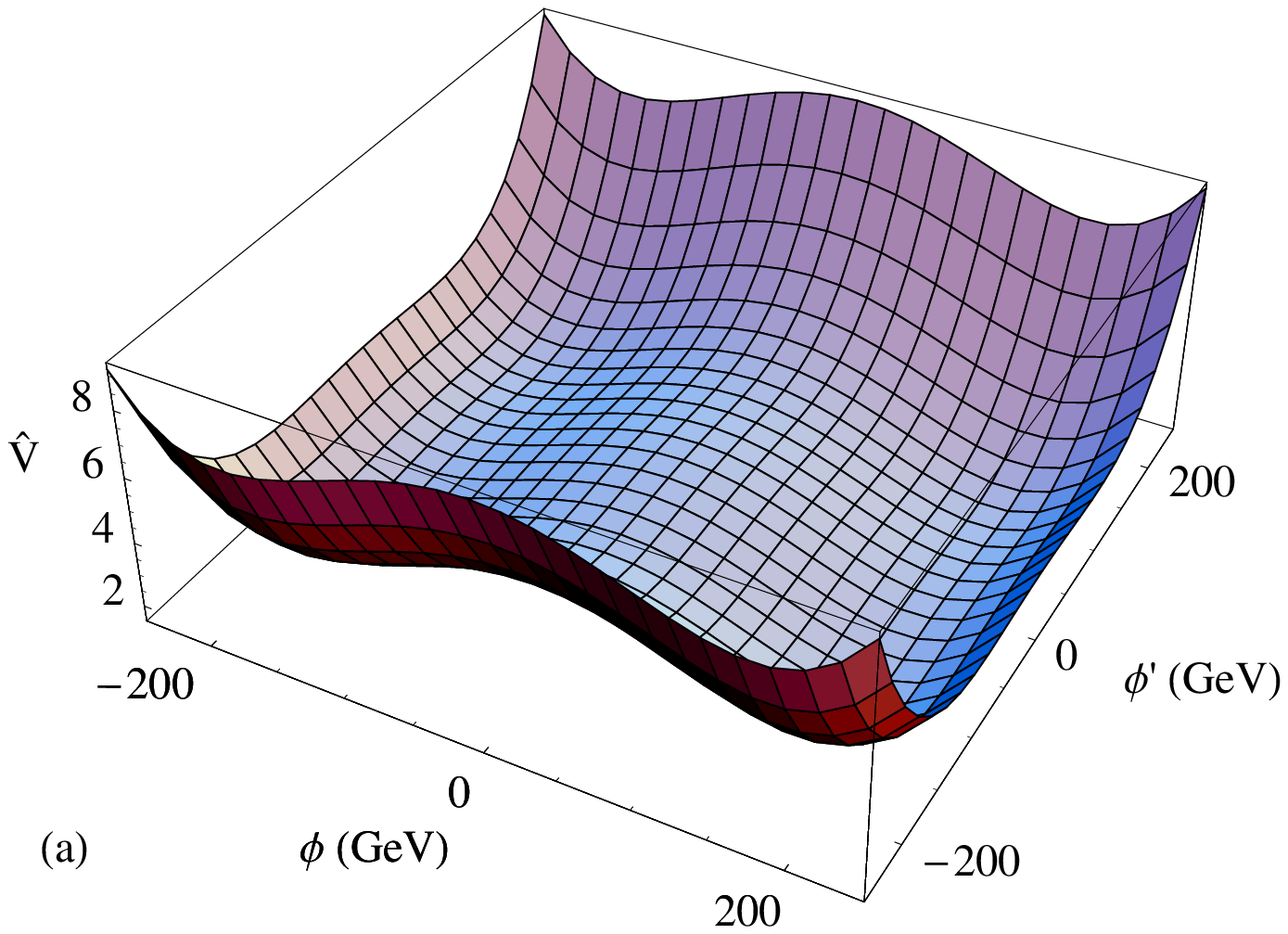}
\\
\hspace*{-7mm}
\includegraphics[width=8.3cm,height=6.5cm,clip=true]{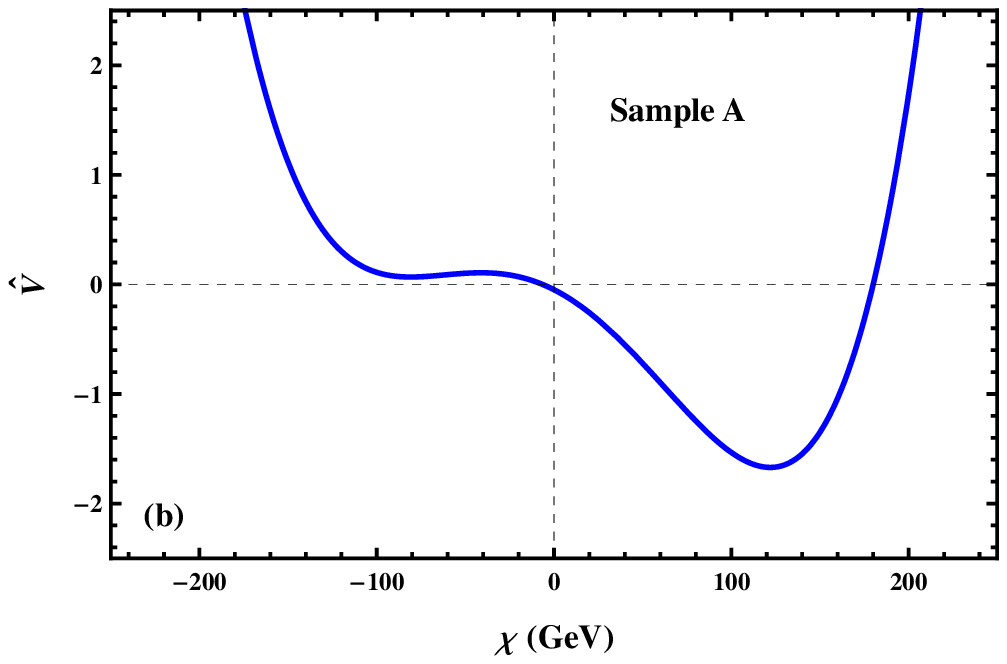}
\includegraphics[width=8.3cm,height=6.66cm,clip=true]{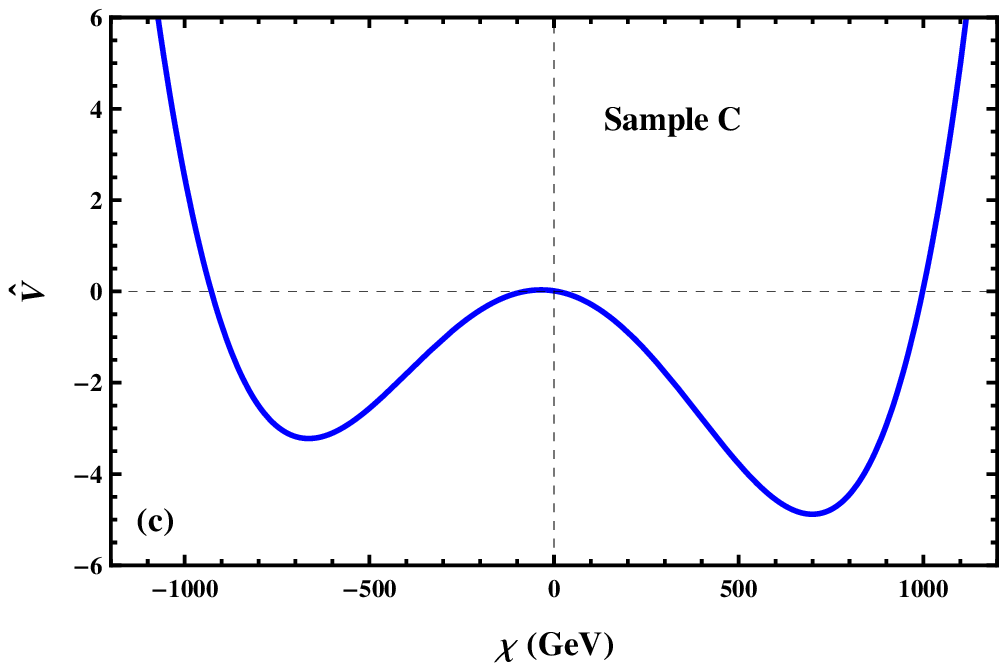}
\caption{Vacuum structure of the Higgs potential $\,\Vh\,$.\,
Plot-(a) depicts $\,\Vh\,$ as a function of
$\,\phi\,$ and $\,\phi'\,$,\, for Sample-A; and the features for Sample-B
and -C appear very similar.  Plot-(b) displays the potential $V$
in the same unit as a function of $\,\chi\,$ in Sample-A; while plot-(c) shows
$\,V\,$ versus $\,\chi\,$ in Sample-C.
The potential $\,\Vh\,$ is in unit of $10^8$\,GeV$^4$.
}
\label{fig:V}
\end{centering}
\end{figure}

In Fig.\,\ref{fig:V}, we analyze the vacuum structure of the Higgs potential $V$
in (\ref{eq:V-tot}) or (\ref{eq:V0-tot}).
For the plot-(a), we display the Higgs potential $\,\Vh\,$
as a function of visible and mirror Higgs fields, $\phi$ and $\phi'$,
where $\,\Vh\,$ is plotted in unit of $10^8$\,GeV$^4$ and
the singlet scalar $\X$ is set to its extremal value $\,\vX\,$.\,
Here we choose Sample-A as an example for illustration, but we find
that the features of Sample-B and -C appear very similar to the plot-(a).
We see that the potential minimum occurs at
$(|\phi|,\,|\phi'|)=(v_\phi^{},\,\vpb)=(174,\,87)$\,GeV, with a
ratio $\,\vp /\vpb = 2\,$.\,
In Fig.\,\ref{fig:V}(b) and (c), we depict the Higgs potential $V$
(in unit of $10^8$\,GeV$^4$) as a function of the $P$-odd singlet Higgs field $\X\,$,\,
for Sample-A and -C, respectively. In these two plots, we have set the
other two Higgs fields $\,\phi\,$ and $\,\phi'\,$ to their extremal values
$(v_\phi^{},\,\vpb)$.  Note that the two minima in the potential are asymmetric for
$\,\X\,$,\, and the true minimum is given by the right one with
$\,\vX =122\,$GeV in Sample-A and $\,\vX =699\,$GeV in Sample-C.
This is expected because the asymmetry is
generated by the unique cubic term, $\,\beta_{\X\phi}^{}(|\phi|-|\phi'|^2)\X\,$,\,
in the Higgs potential (\ref{eq:V}),
which is {\it linear} in $\,\X\,$ and realizes the
spontaneous mirror parity violation.
Since $\,\beta_{\X\phi}^{}<0\,$ and $\,\vp >\vpb\,$, we see that this cubic term
becomes positive for $\,\X < 0\,$  and negative for $\,\X > 0\,$.\,
This explains why in Fig.\,\ref{fig:V}(b) and (c) the right minimum is lower than
the left one, and thus serves as the true minimum of the potential.
The asymmetry between the two minima in Fig.\,\ref{fig:V}(b)
is much larger than that in Fig.\,\ref{fig:V}(c),
because the size of the cubic coupling $\,|\beta_{\X\phi}^{}|\,$ in Sample-A
is a factor-7 bigger than that in Sample-C (cf.\ Table\,\ref{tab:input-ABC}).
In addition, we have made the same plot of $\,\Vh\,$ versus $\,\X\,$ for Sample-B and
find its shape is between the plot-(b) and plot-(c), so we do not display Sample-B
here.

\begin{figure}[t]
\begin{centering}
\hspace*{-7mm}
\includegraphics[width=8.3cm,height=6.5cm,clip=true]{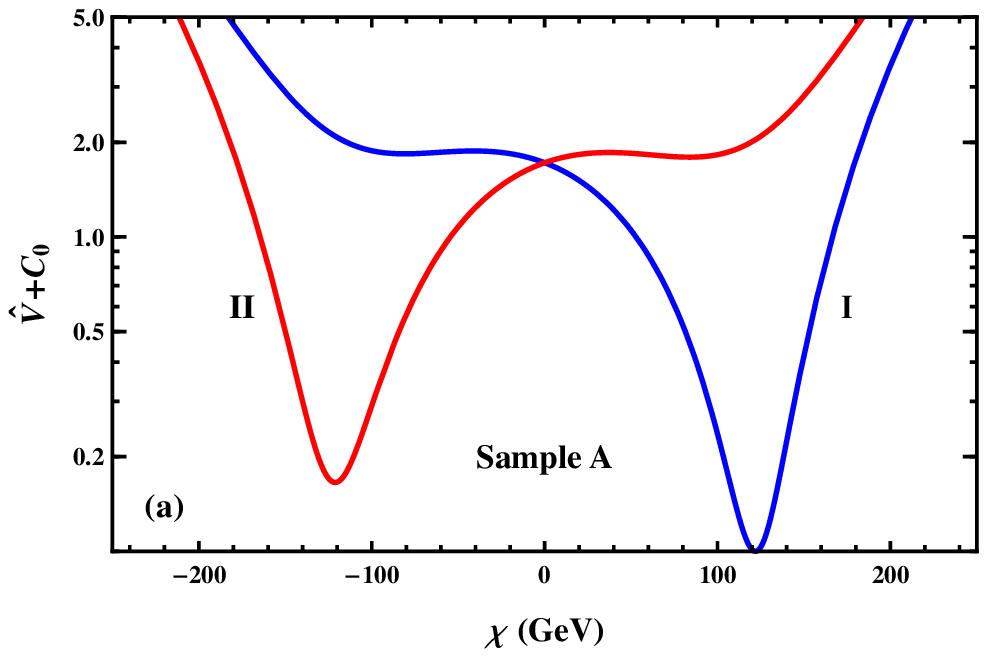}
\includegraphics[width=8.3cm,height=6.5cm,clip=true]{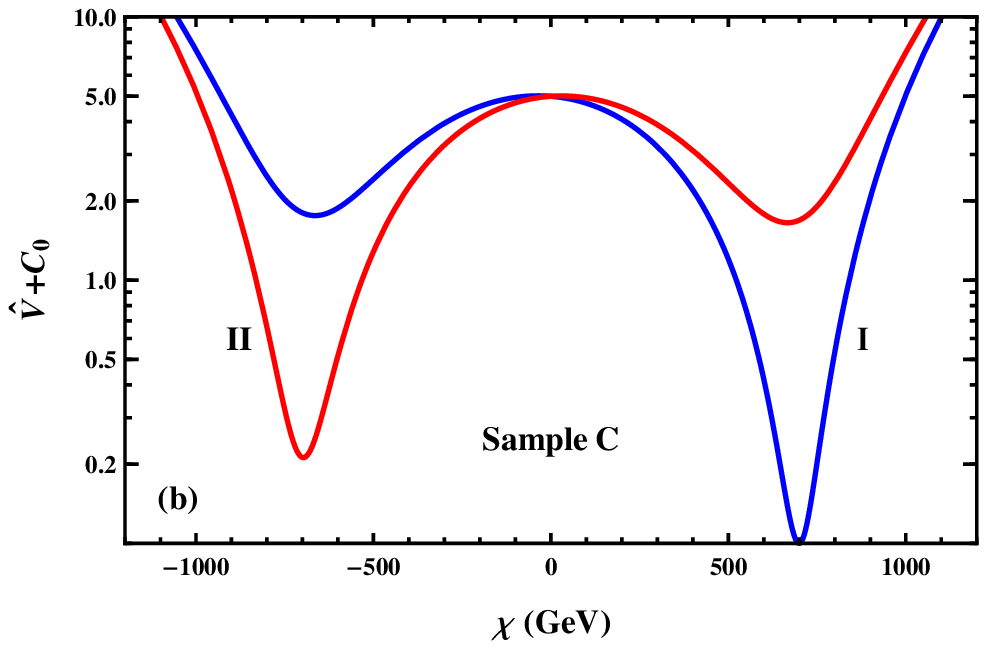}
\caption{Non-degenerate vacua of the Higgs potential $\,\Vh\,$ due to minimal soft breaking.
Plot-(a) depicts $\,\Vh\,$ as a function of $\,\X\,$ for Sample-A,
and plot-(b) is for Sample-C. In each plot,
the curve-I (blue) represents $\,\Vh(\phi,\phi',\X)\,$ and
the curve-II (red) denotes $\,\Vh(\phi',\phi,\X)\,$.\,
For convenience of plotting, a pure constant shift $C_{0}$ is added to $\,\Vh\,$ (cf.\ text).
The potential $\,\Vh\,$ is in unit of $10^8$\,GeV$^4$. }
\label{fig:V-soft}
\end{centering}
\vspace*{2.5mm}
\end{figure}

We note that there are two degenerate field-configurations in the vacuum potential
(\ref{eq:V0}),
\beqa
\label{eq:V-degen}
\left<V(\phi,\phi',\X)\right> ~=~ \left<V(\phi',\phi,-\X)\right> \,,
\eeqa
and they transform into each other under mirror parity.
But due to the minimal soft breaking term $\,\langle\De\Vh_{\textrm{soft}}^{}\rangle\,$
in (\ref{eq:Vsoft-vac}), these two vacuum states become non-degenerate, and
the full potential $\,\langle\Vh\rangle\,$ splits between these two vacuum
configurations,
\beqa
\label{eq:Vtot-split}
\langle\Vh (\phi,\phi',\X)\rangle - \langle\Vh(\phi',\phi,-\X)\rangle
~=~ 2\langle\De\Vh_{\textrm{soft}}^{} (\X)\rangle \,.
\eeqa
So this provides the simplest way to evade the domain wall problem.
To explicitly check the non-degeneracy between
the two vacuum configurations under soft breaking (\ref{eq:Vsoft-vac}), we plot
$\,\Vh (\phi,\phi',\X)\,$ and $\,\Vh (\phi',\phi,\X)\,$ in Fig.\,\ref{fig:V-soft}
as curve-I (blue) and curve-II (red), respectively.
The plot-(a) is for Sample-A and plot-(b) is for Sample-C.
The potential $\,\Vh\,$ is in unit of $10^8$\,GeV$^4$, and for convenience of plotting
we have shifted the potential $\,\Vh\,$ by a pure constant
$\,C_0=1.77\times 10^8$\,GeV$^4\,$
($\,C_0=4.98\times 10^8$\,GeV$^4\,$) in plot-(a) [plot-(b)].
It is clear that in each plot the minima of curve-I and curve-II are no longer
degenerate, and the true vacuum minimum is given by the one in curve-I (blue).

\vspace*{2.5mm}

Next, we systematically derive the outputs for all three samples,
as summarized in Table\,\ref{tab:output-ABC}.
For each sample, we solve the global minimum of
the Higgs potential $\,\Vh\,$ numerically,
and thus determine the three vacuum expectation values \,$(\vp,\,\vpb,\,\vX)$.\,
From Table\,\ref{tab:output-ABC}, the Higgs vacuum expectation value
$\,v_\phi^{}=174$\,GeV\, just generates the observed electroweak symmetry breaking
in the visible sector, while $\,v_{\phi'}^{}=v_\phi^{}/2$\, holds for all three samples.
So they all give the same prediction for the mirror dark matter density according to
(\ref{eq:predict-DM/M}).
The VEV of the $P$-odd Higgs singlet $\,\X\,$ significantly varies among
the three samples, it is around $O(v_\phi^{})$ in Sample-A and -B,
but is about a factor-4 larger than $v_\phi^{}$ in Sample-C.

We further diagonalize the $3\times 3$ Higgs mass-matrix
(\ref{eq:Mhiggs-3x3})-(\ref{eq:Mhiggs-ij}) for each sample, and
derive their mass-eigenvalues as shown in Table\,\ref{tab:output-ABC}.
All three samples have a SM-like Higgs boson $\hh$,
with masses falling into the range of $120-140$\,GeV;\,
while the Higgs boson $\hh'$ is mainly
from the mirror Higgs doublet $\phi'$ and has a mass around $67-75$\,GeV,
which is about half of the $\,\hh$\, mass.
This is quite expected since we have the ratio of two-Higgs-doublet
VEVs, $\,v_\phi^{}/v_{\phi'}^{}=2$,\, based upon the condition (\ref{eq:v/v'-range}).
Finally, for the Higgs boson $\,\Xh\,$,\,  we have
$\,m_\X^{} < 2 m_h^{}$\, in Sample-A, and  $\,m_\X^{} > 2 m_h^{}$\, in Sample-B.
But in contrast with both Samples A and B, the Sample-C has
$\,m_\X^{} < \hf m_h^{}\,$.\,
As will be shown in Sec.\,\ref{sec:5}, these three samples will lead to distinctive
new Higgs signatures for the LHC discovery.

For diagonalizing the $3\times 3$ Higgs mass-matrix ${\cal M}^2$ in
(\ref{eq:Mhiggs-3x3})-(\ref{eq:Mhiggs-ij}),
we introduce the orthogonal rotation matrix $U$,
which connects the gauge-eigenbasis $(\phi,\,\phi',\,\X)$
to the mass-eigenbasis \,$(\hh,\,\hh',\,\Xh)$.\,
So we have  $\,U^T{\cal M}^2 U = {\cal D}^2\,$,\,
where the diagonal mass-matrix
$\,{\cal D}^2= \textrm{diag}(m_\phi^2,\,m_{\phi'}^2,\,m_\X^2)\,$.\,
The predicted mass-eigenvalues are summarized in Table\,\ref{tab:output-ABC},
and we derive the rotation matrix $U$ for all three samples as follows,
\beqs
\label{eq:U}
\begin{eqnarray}
\label{eq:UA}
\textrm{Sample-A:} &~~&
U \,=\left(\!\begin{array}{ccc}
0.8408   & 0.00630 & -0.5413	\\[1mm]
-0.00534 & 1       &  0.00335 	\\[1mm]
 0.5413  & 0.00007 &  0.8408
 \end{array}\right) \!,
\\[2mm]
\label{eq:UB}
\textrm{Sample-B:} &&
U \,=\left(\!\begin{array}{ccc}
0.9921 & -0.00679 & -0.1254 \\[1mm]
0.00658 & 1 & -0.00212 \\[1mm]
0.1254 & 0.00128 & 0.9921
 \end{array}\!\right) \!,
\\[2mm]
\label{eq:UC}
\textrm{Sample-C:} &&
U \,=\left(\!\begin{array}{ccc}
0.9929   &  0.00617  &  0.1187      \\[1mm]
-0.00977 &  0.9995   &  0.0298      \\[1mm]
-0.1184  & -0.0307   &  0.9925
\end{array}\right)  \!.
\end{eqnarray}
\eeqs
We see that the $(1,3)$-element $\,U_{\phi\X}\,$,\,
which characterizes the mixing between $\phi$ and $\chi$,
is \,54.1\%,\, 12.5\%\, and \,11.9\%\, in Sample-(A,\,B,\,C), respectively.
On the other hand, the $(1,2)$-element $\,U_{\phi h'}\,$
and $(2,3)$-element $\,U_{\phi'\X}\,$
represent mixings between $\phi-\phi'$ and $\phi'-\X$, respectively;
they are always around \,2-3\%\, or smaller for the three samples;
so they are negligible for our phenomenology studies below.
Furthermore, the (1,1)-element $\,U_{\phi h}\,$ describes the transition of
$\,\phi\,$ into its mass-eigenstate $\hh\,$;\,
this equals 84\% in Sample-A, and is more than 99\% in both Sample-B and -C.
It is clear that the mass-eigenstate $\,\hh\,$ mainly arises from
visible Higgs doublet $\,\phi\,$,\,  while the mass-eigenstate $\,\Xh\,$
largely comes from $\,\X\,$ and has sizable mixings with $\,\phi\,$.
For comparison we summarize these three elements of \,$U$\,
into the last columns of Table\,\ref{tab:output-ABC}.

\begin{table}[t]
\vspace*{3mm}
\begin{center}
\begin{tabular}{c||c|c|c||c|c|c||c|r|l}
\hline\hline
&&&&&&&&\\[-2.6mm]
~Sample~  & $v_{\phi}^{}$ & ~$v_{\phi'}^{}$~ & $\vX$ & $m_h^{}$ & $m_{h'}^{}$ & $m_{\chi}^{}$
        & $U_{\phi h}$ & $U_{\phi h'}$\hspace*{1.5mm}
        & \hspace*{2.5mm} $U_{\phi\chi}$
\\
&&&&&&&&\\[-3.6mm]
\hline
&&&&&&&& \\[-3.0mm]
A &$174$ & $87$ & $122$ & $122$ & $75.1$ & $203$ & $0.841$ & $0.0063$ &
$-0.541$~
\\
&&&&&&&& \\[-3.8mm]
\hline
&&&&&&&& \\[-3.0mm]
B & $ 174$ & $87$ & $147$ & $125$ & $64.5$ & $277$ & $0.992$ & $-0.0068$ &
$-0.125$~
\\
&&&&&&&& \\[-3.8mm]
\hline
&&&&&&&& \\[-3.0mm]
C & ~$174$~ & ~$87$~ & ~$699$~ & ~$136$~ & ~$67.8$~ & ~$59.4$~ & ~$0.993$~ & ~$0.0062$~ &
$+0.119$~
\\
\hline\hline
\end{tabular}
\caption{Outputs of the three samples, including all Higgs VEVs and Higgs masses,
in unit of GeV.  The three mixing elements
$\,U_{\phi h}$\,,\, $U_{\phi h'}$\, and $\,U_{\phi\chi}$\, in the rotation matrix
$\,U$ are also listed, which characterize the transformations of $\,\phi\,$
into the mass-eigenstate $\,\hh\,$,\, $\hh'$\, and $\,\Xh\,$,\, respectively.}
\label{tab:output-ABC}
\end{center}
\vspace*{-3mm}
\end{table}

Then, using the mixing matrix (\ref{eq:U}) we further derive all the
mass-eigenbasis couplings of Higgs bosons with themselves,
with the gauge bosons, and with the fermions, respectively.
These are summarized in Table\,\ref{tab:higgs-self-coupling} and
Table\,\ref{tab:h-VV-ff-coupling}.
For all cubic scalar couplings in Table\,\ref{tab:higgs-self-coupling},
we have factorized out a common dimension-one VEV parameter,
$\,v = \sqrt{2}v_{\phi}^{} \simeq 246\,$GeV, and the numbers shown are
all dimensionless.

\begin{table}
\vspace*{3mm}
\begin{center}
\begin{tabular}{c||c|c|c|c||c|c|c|c|c}
\hline\hline
&&&&&&&&& \\[-2.5mm]
~Sample~ & $\Xh\Xh\Xh$ & $\Xh\Xh\hh$ & $\Xh\hh\hh$ & $\hh\hh\hh$
& ~$\Xh\Xh\Xh\Xh$~ & ~$\Xh\Xh\Xh\hh$~ &  ~$\Xh\Xh\hh\hh$~
& ~$\Xh\hh\hh\hh$~ & ~$\hh\hh\hh\hh$~
\\
&&&&&&&&& \\[-3.8mm]
\hline
&&&&&&&&& \\[-3.0mm]
 A & $0.586$ & $1.360$ & $0.429$ & $0.182$ & $0.243$ & $0.653$ & $0.704$ & $0.184$ & $0.052 $
\tabularnewline
&&&&&&&&& \\[-3.8mm]
\hline
&&&&&&&&& \\[-3.0mm]
 B & $ 2.010$ & $0.713$ & $-0.058$ & $0.126$ & $0.829$ & $0.428$ & $0.048$ & $-0.019$ & $0.033 $
\tabularnewline
&&&&&&&&& \\[-3.8mm]
\hline
&&&&&&&&& \\[-3.0mm]
 C & ~$0.020$~ & ~$-0.009$~ & ~$0.035$~ & ~$0.151$~ & $0.002$ & $-0.001$ & $0.002$ & $0.018$ & $0.037 $
\tabularnewline
\hline\hline
\end{tabular}
\end{center}
\vspace*{-3mm}
\caption{Predicted Higgs boson self-couplings in the mass-eigenbasis,
where for the trilinear couplings we have factorized out a common VEV parameter,
$\,v = \sqrt{2}v_{\phi}^{} \simeq 246\,$GeV, so the listed numbers are all dimensionless.}
\label{tab:higgs-self-coupling}
\end{table}
\begin{table}
\vspace*{3mm}
\begin{center}
\begin{tabular}{c||c|c|c|c||c|c}
\hline\hline
&&&&&& \\[-3.4mm]
\rule{0em}{1.3em} Sample~ & $\hh VV$ & $\Xh VV$ &
$\hh f\bar{f}$ & $\Xh f\bar{f}$ &
~$\hh' V'V'$~ & ~$\hh'f'\bar{f}'$~
\\
&&&&&& \\[-3.8mm]
\hline
&&&&&& \\[-3.0mm]
 A & ~$0.841$~ & ~$-0.541$~ & ~$0.841$~ & ~$-0.541$~  & $ 0.5$ & $1$
\tabularnewline
&&&&&& \\[-3.8mm]
\hline
&&&&&& \\[-3.0mm]
 B & $ 0.992$ & $-0.125$ & $0.992$ & $-0.125$ & $ 0.5$ & $1$
\tabularnewline
&&&&&& \\[-3.8mm]
\hline
&&&&&& \\[-3.0mm]
 C & ~$0.993$~ & $~~0.119$ & ~$0.993$~ & $~~0.119$ & $0.5$ & $1$
\tabularnewline
\hline \hline
\end{tabular}
\caption{Predicted Higgs couplings with visible/mirror gauge bosons and fermions.
We use $\,V(=W^\pm ,\,Z^0)$\, and $\,f\,$ to denote the visible weak gauge bosons and
fermions (either quark or lepton), respectively; while $\,V'$ and $\,f'$ are their
corresponding mirror partners. For $\,\hh VV$,\, $\Xh VV$,\, and $\,\hh' V'V'$ couplings,
we have divided them by a common coupling which equals the SM value of $\,\hh VV$ coupling;
similarly, for $\,\hh f\bar{f}$,\, $\Xh f\bar{f}$,\, and $\,\hh' f'\bar{f}'$ couplings,
we divide them by a common coupling which equals the SM value of $\,\hh f\bar{f}$\, coupling.
}
\label{tab:h-VV-ff-coupling}
\end{center}
\vspace*{-2mm}
\end{table}

In Table\,\ref{tab:h-VV-ff-coupling}, we use $V$ to represent the visible weak gauge
bosons $(W^\pm ,Z^0)$, while $V'$ denotes their mirror partners
$(W^{\pm\prime},Z^{0\prime})$.
Similarly, we use $f$ and $f'$ to denote ordinary fermions (either quark or lepton)
and mirror fermions, respectively.
For $\hh VV$,\, $\Xh VV$ and $\hh' V'V'$ couplings, we have divided them by
a common coupling (taken as the SM value of $\hh VV$ coupling);
while for $\hh f\bar{f}$,\, $\Xh f\bar{f}$ and $\hh' f'\bar{f}'$ couplings,
we divide them by a common coupling (chosen as the SM value of $\hh f\bar{f}$ coupling).
We see that the $\,\hh\,$ couplings to gauge bosons and to fermions have significant
deviation ($16\%$) from the SM values in Sample-A, while those in Samples B and C
are fairly close to the corresponding SM values.  But the $\,\Xh\,$ couplings
to gauge bosons and to fermions vary a lot among the three samples.
Relative to the SM value of $\,\hh VV\,$ or $\,\hh f\bar{f}\,$ coupling,
the largest $\,\Xh VV\,$ or $\,\Xh f\bar{f}\,$ coupling is about 54\% in Sample-A,
and reduces to about 13\% and 12\% in Sample-B and -C.
The ratio of the mirror coupling $\,\hh'V'V'$\,
over the SM value of $\,\hh VV$\, coupling equals
$\,\f{\,\vpb}{\vp} \simeq \f{1}{2}\,$ to high precision, as shown in the sixth column
of Table\,\ref{tab:h-VV-ff-coupling}, where the invoked gauge couplings cancel
in this ratio because the mirror parity requires identical gauge couplings between the
visible and mirror gauge groups. Furthermore, all mirror Yukawa couplings equal
the corresponding SM Yukawa couplings,
so the last column of Table\,\ref{tab:h-VV-ff-coupling}
has little deviation from one since the element $\,U_{\phi'h'}\simeq 1\,$ holds to
high accuracy and the mixings of $\,\hh'\,$ with the other two Higgs bosons
are negligible in our model, as shown in (\ref{eq:U}). This also means that the ratio
of every mirror fermion mass over the corresponding SM fermion mass is given by,
$\,\f{\,m_{f'}^{}}{m_f^{}} \simeq \f{\,\vpb}{\vp} \simeq \f{1}{2}\,$,\,
to good precision in the present model.
Finally, we note that the masses of visible and mirror weak gauge bosons obey the
relation, $\,\f{\,M_{V'}}{M_V} = \f{\,\vpb}{\vp}\,$,\,
where $\,V=(W,\,Z)\,$ and $\,V'=(W',\,Z')\,$.

\vspace*{3mm}
\subsection{Low Energy Precision Constraints}
\label{sec:4-3}
\vspace*{1.5mm}

Inspecting Table\,\ref{tab:output-ABC}, we see that the mirror Higgs boson
$\,\hh'\,$ is rather light in all three samples, around $67-75$\,GeV, while
the singlet Higgs boson $\,\Xh\,$ becomes the lightest scalar of mass about $59$\,GeV
in Sample-C. It is thus important to analyze the lower energy direct and indirect
precision constraints on our model.

For the mirror Higgs boson $\,\hh'$,\, its coupling to the visible gauge bosons $WW/ZZ$
and fermions $f\bar{f}$ could be generated via the $\phi-\phi'$ mixing,
i.e., the mixing element $\,U_{\phi h'}=O(10^{-2})\,$ in (\ref{eq:U}).
The LEP collaboration\,\cite{Barate:2003sz} has searched for Higgs boson
in the reaction $\,e^-e^+\to Z h\,$ with Higgs decay via $\,h\to b\bar{b}\,$.\,
So we can analyze a similar channel
for searching the mirror Higgs $\hh'$ at LEP via
$\,e^-e^+\to Z \hh'\,$ with $\,\hh'\to b\bar{b}\,$.\,
Then, we immediately realize that the production cross section
is suppressed by a factor $\,U_{\phi h'}^2\,$ and this same factor
$\,U_{\phi h'}^2\,$ enters again the decay branching fraction of
$\,\hh'\to b\bar{b}\,$.\,  Hence, the expected final signals of mirror $\,\hh'\,$
must be suppressed by a factor of $\,U_{\phi h'}^4\,$ relative to that of the
SM Higgs boson with the same mass, which is $\,U_{\phi h'}^4\sim 10^{-9}\,$
for all three samples. It is clear that the LEP data\,\cite{Barate:2003sz}
actually place no bound on such a nearly invisible light mirror Higgs boson.

\begin{figure}[t]
\begin{centering}
\includegraphics[width=11.3cm,height=7.7cm,clip=true]{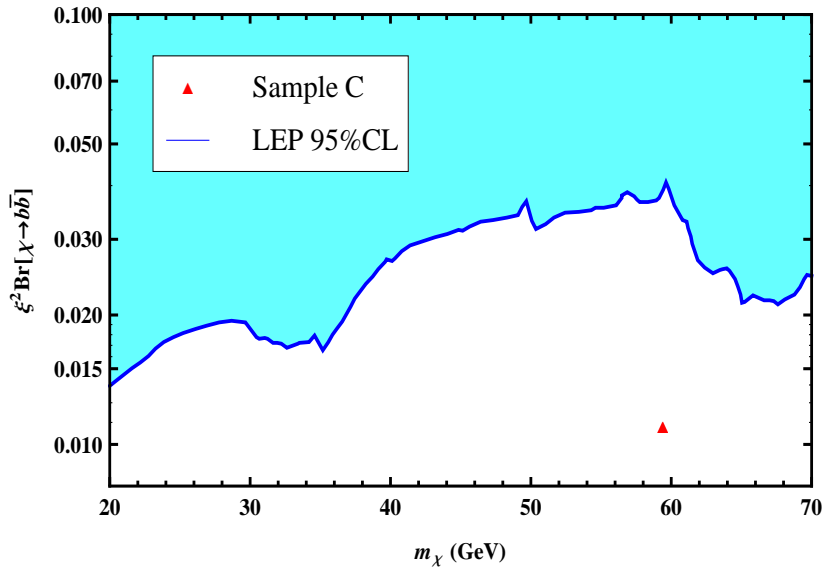}
\vspace*{-2mm}
\caption{LEP upper bound on the product
$\,\xi^{2}\textrm{Br}[\Xh\to b\bar{b}]\,$ is depicted by the blue curve,
where the shaded regions above the curve is excluded at 95\%\,C.L.
The prediction of Sample-C is marked by the red triangle.}
\label{fig:LEP}
\end{centering}
\end{figure}

Then, we analyze the possible LEP direct search limit on the $P$-odd singlet
Higgs boson $\,\Xh$\,.\,
Inspecting Table\,\ref{tab:output-ABC}, we see that the $\Xh$ mass in
Sample-A and -B lies in the range of $\,200-290$\,GeV, and thus beyond the
kinematical capability of LEP.  Only Sample-C predicts a rather light $\X$ with
mass at $59.4$\,GeV which is potentially accessible by LEP.
The relevant reaction for $\,\Xh\,$ detection is via
$\,e^-e^+\to Z \Xh\,$ via the decay $\,\Xh\to b\bar{b}\,$.\,
This channel invokes the $\,\Xh ZZ\,$ and $\,\Xh b\bar{b}\,$ couplings, which are
suppressed by the mixing element $\,U_{\phi\X}\,$ in (\ref{eq:U}) relative to
the SM couplings of $\,hZZ\,$ and $\,hb\bar{b}\,$, respectively.
So we deduce the following relation for the ratios of $\,\Xh\,$ couplings over the
corresponding SM Higgs couplings,
\beqa
\xi ~\equiv~ \f{C_{\X ZZ}^{}}{C_{h ZZ}^{\textrm{SM}}}
~=~ \f{C_{\X f\bar{f}}^{}}{C_{h f\bar{f}}^{\textrm{SM}}}
~=~ U_{\phi\X} \,.
\eeqa
For Sample-C, we have computed the mixing element
$\,U_{\phi\X} \simeq 0.12\,$ as in (\ref{eq:U}), and the decay branching ratio, 
$\,\textrm{Br}[\Xh\to b\bar{b}]=80.5\%\,$,\,  as will be summarized
in Table\,\ref{tab:Higgs-BR} of the next section.
So we can derive a product for Sample-C, relevant to the LEP constraint,
\begin{equation}
\xi^{2}\,\textrm{Br}[\Xh\rightarrow b\bar{b}] ~=~ 0.011 \,.
\label{eq:product-C}
\end{equation}
For any non-standard Higgs boson,
the LEP experimental analysis\,\cite{Barate:2003sz} already put nontrivial limit
on the product of $\,\xi^2\,$ with the Higgs decay branching fraction into
$\,b\bar{b}\,$.\,  We display the LEP upper bound\,\cite{Barate:2003sz}
in Fig.\,\ref{fig:LEP}, where the shaded regions above the curve is excluded at 95\%\,C.L.\
and the prediction (\ref{eq:product-C}) of Sample-C is marked as the red triangle.
We find that Sample-C is fully consistent with the LEP limit.

\vspace*{2.5mm}

Next, we analyze the indirect electroweak precision constraints on the
Higgs sector of our model. The effects of new physics can be generally formulated
into the oblique corrections, as characterized by the parameters $(S,T,U)$
\cite{Peskin:1991sw}.
Since our model contains light Higgs bosons with masses comparable to the
$Z$ mass $m_Z^{}$, we will adopt a more precise set of formulas to compute
the $(S,T,U)$ as in \cite{Forshaw:2001xq}.
Thus, for the SM Higgs boson, we have the oblique corrections,
\beqs
\label{eq:ST-SM}
\begin{eqnarray}
S_{\SM}^{}[m_h^{}] &\!=\!&
\f{1}{\pi}\left[\f{3m_h^2}{8m_Z^2}-\frac{m_h^4}{12m_Z^4}
+\f{m_h^2}{m_Z^2}\ln\f{m_h^2}{m_Z^2}
\( \f{\,3m_Z^2\!-\!m_h^2\,}{4m_Z^2}+ \frac{m_h^4}{24m_Z^4}
   +\f{3m_Z^2}{4(m_Z^2\!-\! m_h^2)} \)
\right.
\nonumber\\[2.5mm]
&& \left. \hspace*{6mm}
+\( 1-\f{m_h^2}{3m_Z^2}+\f{m_h^4}{12m_Z^4}\)
 \f{m_h^{}q(m_h^{})}{m_Z^2} \right] \!,
\\[3mm]
T_{\SM}^{}[m_h] &\!=\!&
\frac{3}{\,16\pi s_Z^2 c_Z^2\,}
\left[\f{m_h^2}{\,m_Z^2\!-\! m_h^2\,} \ln\f{m_h^2}{m_Z^2}
-\f{c_Z^2 m_h^2}{\,c_Z^2 m_Z^2\!-\! m_h^2\,}
\ln\f{m_h^2}{\,c_Z^2 m_Z^2\,}\right] \!,
\end{eqnarray}
\eeqs
with the function $\,q(m_h)$\, defined as,
\begin{eqnarray*}
q(m_h^{}) ~=~ \left\{
\ba{ll}
\dis\sqrt{4m_Z^2-m_h^2\,}\arctan
{\frac{\sqrt{4m_Z^2-m_h^2}}{m_h^{}}} \,,
~~&~~  m_h^{}\leqq 2m_Z^{} \,,
\\[6mm]
\dis\sqrt{m_h^2-4m_Z^2\,}\,\ln\f{2m_Z}{\,m_h^{} +\sqrt{m_h^2-4m_Z^2}\,} \,,
~~&~~ m_h^{}\geqq 2m_Z^{} \,,
\ea
\right.
\end{eqnarray*}
where the weak mixing angle $\theta_W^{}$ is defined at the $Z$-pole, and we
use the notations $\,s_Z^2=\sin^2\theta_W^{}|_Z^{}$\, and $\,c_Z^2=1-s_Z^2\,$.\,
The Higgs correction to the oblique parameter $U$ is much smaller and thus
negligible in the analysis below.
For the case of large Higgs mass $\,m_h^2\gg m_Z^2\,$,\, it is justified to expand
the above formulas. So we can reproduce the conventional approximate results
\cite{Peskin:1991sw} under the large Higgs mass expansion, as a consistency check,
\beqs
\beqa
S_{\SM}^{}[m_h^{}] &\!\!\approx\!\!& \f{1}{\,12\pi\,}\ln\f{m_h^2}{m_Z^2} \,,
\\[1mm]
T_{\SM}^{}[m_h^{}] &\!\!\approx\!\!&\! -\f{3}{\,16\pi c_Z^2\,}\ln\f{m_h^2}{m_Z^2} \,.
\eeqa
\eeqs

In the present mirror model, the three Higgs bosons have
mixings in their gauge-eigenbasis, and we transform them into the
mass-eigenbasis via the mixing matrix $\,U$\, in Eq.\,(\ref{eq:U}).
So we can derive the new contributions of all Higgs bosons to $\,S\,$ and $\,T\,$,
\beqs
\label{eq:ST-new}
\begin{eqnarray}
\Delta S &\!=\!& U_{\phi h}^2 S_{\SM}^{}[m_h^{}] + U_{\phi h'}^2 S_{\SM}^{}[m_{h'}^{}]
      +U_{\phi\chi}^2 S_{\SM}^{}[m_{\X}^{}] -S_{\SM}^{}[m_{h}^{\textrm{ref}}]
\,,
\\[1.5mm]
\Delta T &\!=\!& U_{\phi h}^2 T_{\SM}^{}[m_h^{}] + U_{\phi h'}^2 T_{\SM}^{}[m_{h'}^{}]
      +U_{\phi\chi}^2 T_{\SM}^{}[m_{\X}^{}] -T_{\SM}^{}[m_{h}^{\textrm{ref}}]
\,,
\end{eqnarray}
\eeqs
where we have subtracted the SM Higgs contributions at the reference point
$\,m_{h}^{\textrm{ref}}\,$,\, and $\,U_{ij}$\, denotes the relevant element of the
mixing matrix $\,U$.

\begin{figure}[t]
\begin{center}
\includegraphics[width=12cm,height=8.1cm,clip=true]{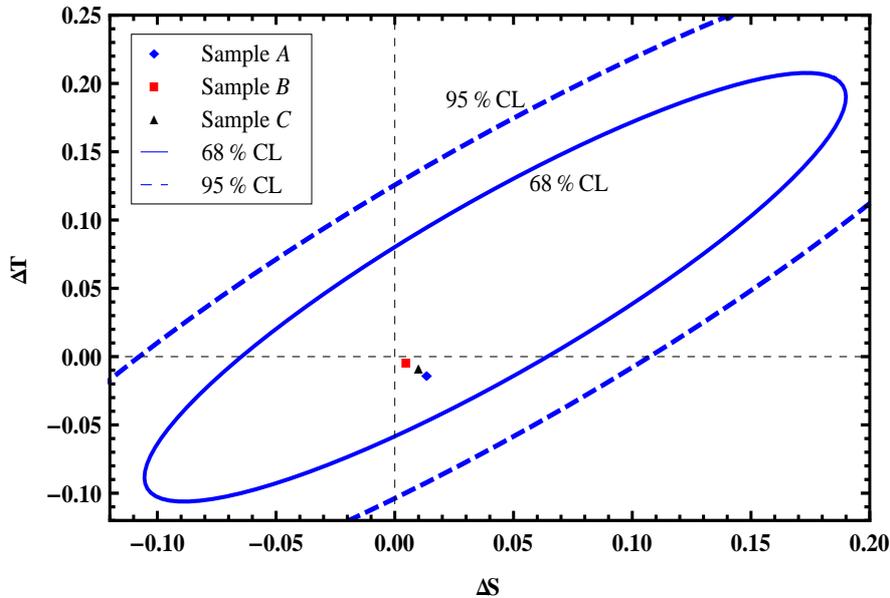}
\vspace*{-3mm}
\caption{Electroweak precision constraints in the $\Delta S-\Delta T$ plane,
with a SM reference point $\,m_h^{\textrm{ref}} = 120$\,GeV.
The predictions of our Samples A, B, and C are marked by the
blue diamond, red square and black triangle, respectively.}
\label{fig:ST}
\end{center}
\vspace*{-3mm}
\end{figure}

From (\ref{eq:ST-new}) and (\ref{eq:ST-SM}),
we explicitly compute the oblique corrections in our mirror model,
for the three Samples in Table\,\ref{tab:output-ABC}, and arrive at
\beqs
\label{eq:ST-ABC}
\beqa
\textrm{Sample-A:} &~~&
(\Delta S,\,\Delta T) = (0.0134,\, -0.0138)\,,
\\
\textrm{Sample-B:} &~~&
(\Delta S,\,\Delta T) = (0.0048,\, -0.0043)\,,
\\
\textrm{Sample-C:} &~~&
(\Delta S,\,\Delta T) = (0.0100,\, -0.0088)\,,
\eeqa
\eeqs
where we have set the SM reference point $\,m_{h}^{\textrm{ref}}=120\,$GeV.
Then, we analyze the electroweak precision data\,\cite{PDG} and make a
precision fit by using the method of Peskin and Wells\,\cite{Peskin-Wells}.
We choose the three most accurately measured observables \cite{PDG},
$\,\Gamma_{\ell}[Z]$,\,  $M_W$,\, and $\sin^2\theta_W^{\textrm{eff}}$,\,
for the precision $\,\Delta S-\Delta T\,$ fit with $\,\Delta U=0\,$
and the SM reference point $\,m_{h}^{\textrm{ref}}=120\,$GeV.\,
For computing the SM contributions, we have followed the approach of Marciano
\cite{Mac} and take into account the allowed experimental ranges of
the top mass $m_t^{}$ and fine structure constant $\alpha(m_Z^{})$ \cite{PDG}.
The resultant constraints on $\,\Delta S-\Delta T\,$  are shown
at 68\%\,C.L.\ and 95\%\,C.L.\ in Fig.\,\ref{fig:ST}, respectively.
Our precision fit is in good agreement with the recent more elaborated
systematical analysis by the Giffter Group\,\cite{Gfit}.
In Fig.\,\ref{fig:ST}, we have marked the predictions (\ref{eq:ST-ABC})
of our Sample-(A,\,B,\,C) by the blue diamond, red square
and black triangle,  respectively.
We see that they are fully consistent with the precision constraints.

Finally, we analyze the low energy precision constraint on the kinetic mixing parameter
$\,\epsilon\,$ in \ref{eq:Kmix-ga-ga'}.
For the case with unbroken mirror parity, the electroweak scales are exactly the same
in the two sectors ($\,\vp =\vpb\,$) as in \cite{kmix-more}.
It was suggested by Carlson and Glashow that this mixing can be probed through the oscillations of
orthopositronium (o-Ps) and mirror orthopositronium (o-Ps$'$). The oscillation effect becomes maximal
when o-Ps and o-Ps$'$ have the same masses. But this does not apply to the case with spontaneously
broken mirror parity ($\,\vp \neq\vpb\,$).
If $\,\vp\ll\vpb\,$ as realized in \cite{An:2009vq}, there is basically no constraint available.
But, for $\,\vp \gtrsim \vpb\,$ as realized in our construction, we can derive nontrivial limit
from the invisible decays of the o-Ps. The visible decays of o-Ps mainly go to $\,3\gamma$\,
channel, and the decay width is given by \cite{Czarnecki:1999mt},
\begin{equation}
\label{eq:oPs-3photon}
\Gamma[\textrm{o-Ps}\rightarrow 3\gamma] ~\simeq~
4.63 \times\! 10^{-18} \,\mathrm{GeV} \,,
\end{equation}
which represents the total width to good accuracy.
The SM can provide invisible decays of orthopositronium via
$\,\textrm{o-Ps}\to\nu_e^{}\bar{\nu}_e^{}\,$,\,
which has a partial decay width\,\cite{Czarnecki:1999mt},
\begin{equation}
\label{eq:o-Ps-to-neutrino}
\Gamma [\textrm{o-Ps}\to \nu_e^{}\bar{\nu}_e^{}]
~=~ \frac{\,G_{F}^{2}\alpha^{3}m_{e}^{5}\,}{24\pi^2}\left(1+4\sin^2\theta_W^{}\right)^{2}
~\simeq~ 2.88\times\! 10^{-35}\,\text{GeV} \,,
\end{equation}
and thus the corresponding branching fraction,
$\,\textrm{Br}\left[\textrm{o-Ps}\to \nu_e^{}\bar{\nu}_e^{}\right]
   \simeq 6.22\times\! 10^{-18}\,$.\,
This is negligibly small as it is far below the present experimental upper bound
on the branching ratio of all possible invisible decays \cite{AA'mix-Exp2007},
\beqa
\label{eq:Br-oP-inv-Exp}
\text{Br}[\text{o-Ps}\rightarrow \text{invisible}]
~<~ 4.2\times\! 10^{-7} \,,
\eeqa
which holds at 90\%\,C.L.

In our model, the mirror-particle-induced invisible decays of o-Ps predominantly go to
mirror electron-positron pairs, $\,\textrm{o-Ps}\rightarrow e'^{+}e'^{-}\,$.\,
So, we can compute its partial decay width,
\beqa
\label{eq:o-Ps-to-e'e'}
\Gamma\!\left[\textrm{o-Ps}\rightarrow e'^{-}e'^{+}\right]
~=~ \frac{16\pi\alpha^{2}\epsilon^{2}}{3} \frac{\left|\psi(0)\right|^{2}}{M^{2}}\left(1+\frac{m_{e'}^{2}}{2m_{e}^{2}}\right)
\sqrt{1-\frac{m_{e'}^{2}}{m_{e}^{2}}\,} \,,
\eeqa
where $\,M\simeq 2m_{e}^{}\,$  is the mass of o-Ps,  and
$\,|\psi(0)|^{2}=\frac{\alpha^{3}m_{e}^{3}}{8\pi}\,$
is the square of the wave function at the origin.
For the spontaneous mirror parity violation, we have
$\,\f{m_{e}^{}}{\,m_{e'}^{}\,}=\f{\vp}{\,\vpb\,}\equiv x\,$.\,
So, from (\ref{eq:oPs-3photon}) and (\ref{eq:o-Ps-to-e'e'}),
we can derive the branching fraction of the invisible decay channel
$\,\textrm{o-Ps}\rightarrow e'^{+}e'^{-}\,$,\,
\beqa
\label{eq:Br-oPs-e'e'}
\text{Br}\left[\text{o-Ps}\rightarrow e'^{-}e'^{+}\right]
~\simeq~  379.3\, \epsilon^{2} \!\(\!1 + \frac{1}{2x^2}\)\sqrt{1\!-\!\f{1}{x^2}\,} ~,
\eeqa
which is proportional to $\,\epsilon^2\,$.\,
Following our numerical samples in Sec.\,\ref{sec:4-2}, we have $\,x=\f{\vp}{\,\vpb\,}=2\,$,\,
and thus $\,m_{e}^{}=2m_{e'}^{}\,$.\, So, from the experimental limit (\ref{eq:Br-oP-inv-Exp}),
we can infer the upper bound of $\,\epsilon$\, at 90\%\,C.L.,
\begin{equation}
\label{eq:our-AA'-limit}
\epsilon ~<~ 3.4\times\! 10^{-5} \,.
\end{equation}

\vspace*{3mm}
\section{New Higgs Signatures at the LHC}
\label{sec:5}
\vspace*{1.5mm}

In this section, we further derive decay widths and branching fractions of the
non-standard Higgs bosons in the present model. We identify their major LHC production and
decay channels.  Then, we analyze the predictions for new Higgs signatures at the LHC.
As shown in the previous section, our mirror model construction generically
predicts light Higgs bosons with distinct mass-spectrum and non-standard couplings.
Especially, the $P$-odd scalar $\,\Xh\,$, characterizing spontaneous
mirror parity violation, has a mass equal $277$\,GeV in Sample-B, which
is more than twice of the mass $m_h^{}=125$\,GeV of the SM-like Higgs boson
$\,\hh\,$;\,  while $\,\Xh\,$ is as light as about 59\,GeV for Sample-C
and is less than half of the $\,\hh\,$ mass of $136$\,GeV.
Note that $\,\phi\,$ can also have sizable mixing with $\,\X\,$,\, which
is about 54.1\%, 12.5\%, and 11.9\% in Sample-(A,\,B,\,C), respectively.
These will lead to new Higgs production and decay channels,
and can be experimentally searched at the LHC.
In addition, since the mirror Higgs boson $\,\hh'\,$ in our construction is always light,
with a mass around \,$67-75$\,GeV, which is about half of the $\,\hh\,$ mass
due to the VEV ratio of $\,\vpb /\vp = \hf\,$.\,
But, as the mixing of $\,\hh'\,$ with $\,\hh\,$ and $\,\Xh\,$ is always below \,$2-3\%$\,,\,
it largely decouples from the visible sector and dominantly decays into invisible
mirror partners. The above distinct features also make our Higgs phenomenology fully
different from all previous mirror models in the
literature\,\cite{Barbieri:2005ri}\footnote{An interesting recent
study\,\cite{Englert:2011yb} analyzed the LHC discovery of the Higgs portal from the
SM to a hidden sector where the hidden Higgs boson mixes with the SM Higgs boson
via the quartic interaction term.}.
For instance, our model always forbids mirror Higgs $\,\hh'\,$ decays into
visible Higgs $\,\hh\,$ via $\,\hh'\to\hh\hh\,$,\, while the inverse channel,
$\,\hh\to\hh'\hh'\,$,\, is either disallowed or practically negligible.

\begin{table}[h]
\vspace*{3mm}
\begin{center}
\begin{tabular}{c||cc|cc|cc}
\hline\hline
&&&&&\\[-3.8mm]
\rule{0mm}{3.5mm} Sample & \multicolumn{2}{c|}{A} & \multicolumn{2}{c|}{B} & \multicolumn{2}{c}{C}
\tabularnewline
\hline
&&&&&\\[-3.8mm]
\rule{0mm}{3.5mm} Higgs 			 &	 	$\hh$		 & $\Xh$ 		& $\hh$ 		
 &	 $\Xh$		 &      $\hh$ 	     & $\Xh$
\tabularnewline
\hline
\hline
&&&&&\\[-3.5mm]
$\Gamma$\,(MeV) 	& $2.63$ & $454$ & $4.10$ & $110$ & $7.49$ & $0.0226$
\tabularnewline
\hline	
&&&&&\\[-3.7mm]
$WW$ 			& $0.157^{*}$ & $0.728$ & $0.209^{*}$ & $0.615$ & $0.358^{*}$
& $0$
\tabularnewline
\hline
&&&&&\\[-3.7mm]
$ZZ$ & $0.0185^{*}$ & $0.268$ & $0.0263^{*}$ & $0.269$ & $0.0499^{*}$
& $0$
\tabularnewline
\hline
&&&&&\\[-3.7mm]
\rule{0mm}{3.5mm}$\hh\hh$	
&$0$ & $0$ & $0$ & $0.113$ & $0$ & $0$
\tabularnewline
\hline
&&&&&\\[-3.8mm]
\rule{0mm}{3.5mm}$\Xh\Xh$			    & $0$ & $0$ & $0$ & $0$ & $0.102$ & $0$
\tabularnewline
\hline
&&&&&\\[-3.7mm]
\rule{0mm}{3.5mm}$b\,\bar{b}$ 				& $0.617$ & $0.0022$ & $0.565$
& $6.4\!\times\!10^{-4}$ & $0.332$ & $0.805$
\tabularnewline
\hline
&&&&&\\[-3.8mm]
\rule{0mm}{3.5mm}$\tau\,\bar{\tau}$	    &$0.0672$ & $2.7\!\times\!10^{-4}$
& $0.0619$ & $8.2\!\times\!10^{-5}$ & $0.0369$ & $0.0759$
\tabularnewline
\hline
&&&&&\\[-3.8mm]
\rule{0mm}{3.5mm}$c\,\bar{c}$ 	&$0.0311$ & $1.1\!\times\!10^{-4}$ & $0.0285$
& $3.2\!\times\!10^{-5}$ & $0.0167$ & $0.0411$
\tabularnewline
\hline
&&&&&\\[-3.8mm]
\rule{0mm}{3.5mm}$g\,g$ 					& $0.0866$ & $9.0\times10^{-4}$ & $0.0843$
& $5.7\!\times\!10^{-4}$ & $0.0593$ & $0.0284$
\tabularnewline
\hline
&&&&&\\[-3.8mm]
\rule{0mm}{3.5mm}$\gamma\,\gamma$ & $0.0022$ & $5.2\!\times\!10^{-5}$ & $0.0023$
& $1.5\!\times\!10^{-5}$ & $0.0018$ & $4.4\!\times\!10^{-4}$
\tabularnewline
\hline
&&&&&\\[-3.8mm]
\rule{0mm}{3.5mm}$Z\,\gamma$ 			& $0.0012$ & $1.7\!\times\!10^{-4}$
& $0.0015$ & $6.3\!\times\!10^{-5}$ & $0.0020$ & $0$
\tabularnewline
\hline\hline
\end{tabular}
\caption{Total decay widths and major decay branching fractions of Higgs bosons
$\,\hh\,$ and $\,\Xh\,$ in Sample-(A,\,B,\,C). For $WW$ and $ZZ$ decay channels,
the numbers marked by a superscript $^*$ denote that one of the weak gauge boson
in the final state is off-shell.}
\label{tab:Higgs-BR}
\end{center}
\vspace{-2mm}
\end{table}

Based on the Higgs mass-spectrum in Table\,\ref{tab:output-ABC} and Higgs
couplings in Table\,\ref{tab:higgs-self-coupling}-\ref{tab:h-VV-ff-coupling},
we systematically compute the Higgs decay widths and branching fractions.
These results are summarized in Table\,\ref{tab:Higgs-BR}.
As shown in (\ref{eq:U}), the mixing of the mirror Higgs boson $\,\hh'\,$
with $\,\hh\,$ or $\,\Xh\,$ is always below about $2-3\%$ and thus negligible for
the current analysis, so we do not to list the $\hh'$ decays in
Table\,\ref{tab:Higgs-BR}.  Also, we find that invisible decays
of $\,\hh\,$ and $\,\Xh$\, into the mirror gauge bosons or fermions
are much suppressed and always below 4\%, which are not useful for the
current Higgs searches at the LHC. So for clarity of Table\,\ref{tab:Higgs-BR},
we omit them as well.
We further note that the Higgs $\,\hh$\, in Sample-C has a new on-shell decay channel
with  $\,\textrm{Br}[\hh\to\Xh\Xh]\simeq10\%\,$,\, and the Higgs $\Xh$ in Sample-B has
the new channel with $\,\textrm{Br}[\Xh\to\hh\hh]\simeq 11\%\,$. This means that
their branching fractions have sizable deviations from that of the SM Higgs
boson with the same mass. For the other four cases in Table\,\ref{tab:Higgs-BR},
the branching fractions of $\,\hh\,$ or $\,\Xh\,$ appear quite similar to that
of the SM Higgs boson, up to a few percent of corrections due to their
invisible decays into mirror partners. But in all cases of Table\,\ref{tab:Higgs-BR},
the decay widths can significantly differ from the SM due to the relevant
suppression factor $U_{ij}^2$ from the mixing matrix (\ref{eq:U}).

To derive Higgs decay width, we have included
QCD radiative corrections. For $\,q\bar{q}$\, final state,
the leading order Higgs width is
$\,\Gamma_{qq}^{(0)}=\f{\,3G_Fm_q^2\,}{\,4\sqrt{2}\pi\,}m_h^{}\,$,\,
and including the $\,O(\alpha_s^2)$\, and $\,O(\alpha_s^3)$\, QCD corrections
gives \cite{Hdecay},
\beqa
\Gamma_{qq}^{} &\!=\!& \Gamma_{qq}^{(0)} \times K_{qq} \,,
~~~~~~
K_{qq} ~=~ \frac{\overline{m}_{q}^{2}(m_{h}^{2})}{m_{q}^{2}}
\left[ 1+\Delta_{qq}+\Delta_{H}^{2}\right] \!,
\nn
\\[0.5mm]
\Delta_{qq} &\!=\!& 5.67 \frac{\overline{\alpha}_{s}}{\pi}
 +(35.94-1.36 n_f^{})\frac{\overline{\alpha}_{s}^{2}}{\pi^{2}}
 +(164.14-25.77 n_f^{} +0.26 n_f^2)\frac{\overline{\alpha}_{s}^{3}}{\pi^{3}} \,,
\\[1.5mm]
\Delta_{H}^{2} &\!=\!& \frac{\overline{\alpha}_{s}^{2}}{\pi^{2}}\left( 1.57-\frac{2}{3}\log{\frac{m_{h}^{2}}{m_{t}^{2}}} +\frac{1}{9}\log^{2}{\frac{\overline{m}_{q}^{2}}{m_{h}^{2}}}\right) \!,
\nn
\eeqa
where the running quark mass $\,\overline{m}_{q}^{}(m_{h}^{2})\,$,\,
the strong coupling constant $\,\overline{\alpha}_{s}\equiv \alpha(m_{h}^{2})\,$,\,
and the light fermion flavor-number $\,n_f^{}\,$ are defined at $\,\mu = m_h^{}\,$
under $\overline{\textrm{MS}}$ scheme.
For gluon $gg$ final state, QCD corrections enhance the leading order
Higgs width $\,\Gamma_{gg}^{(0)}\,$ by a factor $K_{gg}^{}$ \cite{Hdecay},
\beqa
\Gamma_{gg} &\!=\!& \Gamma_{gg}^{(0)} \times K_{gg}^{} \,,
\nn\\[2mm]
K_{gg}^{} &\!=\!& 1 +\frac{215}{12}\frac{\over{\alpha}_{s}(m_h^2)}{\pi} +\frac{\over{\alpha}^2_{s}(m_h^2)}{\pi^{2}}
\(156.8-5.7\log{\frac{m_t^2}{m_h^2}}\) \!.
\eeqa
We have verified these formulas numerically and reached full agreement with
\cite{Hdecay}.  For instance, in the mass range of \,$m_h^{}=100-300$\,GeV,\,
the QCD corrections amount to $\,K_{qq}^{}\simeq 0.63-0.39\,$
for the $qq$ final state, and $\,K_{gg}^{}\simeq 1.87-1.74\,$
for the $gg$ final state.
With the above, we systematically summarize our calculations in Table\,\ref{tab:Higgs-BR}. From this table, we note that the Higgs boson $\,\hh\,$ mainly decays
to $\,WW^*\,$ and $\,b\bar{b}\,$, with branching fractions equal to
$(15.7\%,\,20.9\%,\,35.8\%)$ for $\,WW^*\,$ channel and
$(61.7\%,\,56.5\%,\,33.2\%)$ for $\,b\bar{b}\,$ channel,
in Sample-(A,\,B,\,C), respectively.
On the other hand, we find that the Higgs boson $\,\Xh\,$ mainly decays to
$WW$ and $ZZ$ channels for Sample-(A,\,B), with decay branching fractions
$(72.8\%,\,61.5\%)$ in $WW$ channel and
$(26.8\%,\,26.9\%)$ in $ZZ$ channel.
For Sample-C, $\Xh$ dominantly decays to $b\bar{b}$ with a branching fraction
80.5\%, while its decay branching ratios for the final states
$\,\tau\bar{\tau}\,$,\, $\,c\bar{c}\,$ and $\,gg\,$
equal 7.6\%, 4.1\% and 2.8\%, respectively.

Next, we study the production and decays of the visible Higgs bosons
$\,\hh\,$ and $\,\Xh\,$.\,
The Higgs boson $\,\hh\,$ is SM-like in the sense that
its gauge and Yukawa couplings to $\,WW/ZZ\,$ and $\,f\bar{f}\,$
are close to the SM values, but still can have sizable deviation in Sample-A
(cf.\ Table\,\ref{tab:h-VV-ff-coupling}). It has a mass
$\,m_h^{}=(122,\,125,\,136)$\,GeV in Sample-(A,\,B,\,C), respectively.
Its main production channel should be the gluon-gluon fusion with decays into
two photons, $\,gg\to\hh\to \gamma\gamma\,$.\,
We also consider other two channel with the off-shell decays,
$\,gg\to\hh\to WW^*,ZZ^*\,$.\,
For the on-shell production of $\,\hh$\,,\, we compute the cross section times
branching fraction of $\,\hh\to \ga\ga\,$ or
$\,\hh\to VV^*\,$ ($V=W,Z$),\,
relative to that of the SM Higgs boson with the same mass.
This gives the ratios,
\beqs
\beqa
\label{eq:h-2photon-sup}
U_{\phi h}^2
\f{\textrm{Br}[\hh\to \ga\ga]}{~\textrm{Br}[h\to \ga\ga]_{\textrm{SM}}^{}\,}
&\!\simeq\!& (0.693,\, 0.964\, ,0.844)\,,
\\[2mm]
\label{eq:h-VV-sup}
U_{\phi h}^2
\f{\textrm{Br}[\hh\to VV^*]}{~\textrm{Br}[h\to VV^*]_{\textrm{SM}}^{}\,}
&\!\simeq\!& (0.693,\, 0.964\, ,0.844)\,,
\eeqa
\eeqs
for Sample-(A,\,B,\,C), respectively.
The two ratios in (\ref{eq:h-2photon-sup}) and (\ref{eq:h-VV-sup}) exactly
coincide for each sample since their left-hand-sides are actually equal
due to
$\,\Gamma[\hh\to\ga\ga]/\Gamma[\hh\to\ga\ga]_{\text{SM}}^{}
 = \Gamma[\hh\to VV^*]/\Gamma[\hh\to VV^*]_{\text{SM}}^{}
 = U_{\phi h}^2\,$.\,
We see that for Sample-A and -C, the $\hh$ signals in $\gamma\gamma$ channel
(and $VV^*$ channels)
are suppressed by 31\% and 16\% relative to that of the SM predictions,
respectively, while the $\hh$ signal rate is lower by 3.6\% in Sample-B.
So, detecting our $\,\hh\to\gamma\gamma\,$ signals in Sample-A and -C
is significantly {\it harder than that of the SM Higgs boson,}
and it requires {\it higher integrated luminosity at the LHC.}
The same is also true for our signals in $\,\hh\to WW^*,ZZ^*\,$ channels.
It is expected that with a total integrated luminosity of 10\,fb$^{-1}$ for two
experiments at 7\,TeV and combining all available channels, the SM Higgs boson exclusion
will be extended to $\,m_h^{}=114-600$\,GeV at $3\sigma$ level \cite{LHC-now}.
For Sample-C, since $\,\hh\,$ has a mass larger than twice of $\,\Xh\,$,\,
we also have the decay channel $\,\hh\to\Xh\Xh\to b\bar{b}b\bar{b}\,$,\,
as will be discussed below.

\begin{figure}[t]
\begin{center}
\includegraphics[width=16.5cm,height=3.4cm,clip=true]{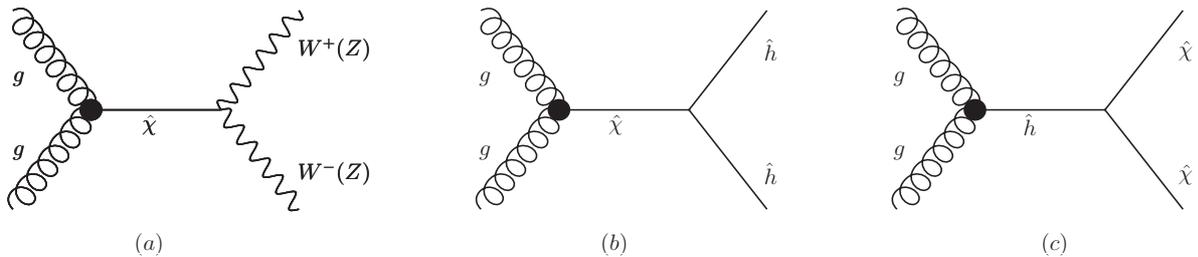}
\caption{Predicted new production processes of Higgs boson via gluon-gluon fusions
at the LHC. The diagram (a) is for Sample-A; the diagrams (a) and (b) are for Sample-B;
and the diagram (c) is for Sample-C. The big black-dot denotes the gluon-gluon-Higgs
vertex as contributed by the fermion and gauge triangle diagrams in each case.}
\label{fig:LHC}
\end{center}
\end{figure}

Then, we consider the Higgs boson $\,\Xh\,$ which has a large $P$-odd component.
The largest channels are still the gluon-gluon fusion processes:
(i).\ $\,gg\to \Xh\to WW (ZZ)\,$ with $WW\to \ell\nu\ell\nu,\,\ell\nu jj$,
or $ZZ\to\ell\ell jj,\,\ell\ell\nu\nu,\,\ell\ell\ell\ell$,  for Sample-(A,\,B),
as shown in Fig.\,\ref{fig:LHC}(a);
(ii).\ and another reaction,
$\,gg\to \Xh\to \hh\hh\,$ with $\,\hh\hh\to b\bar{b}b\bar{b}\,$, for Sample-B,
as illustrated in Fig.\,\ref{fig:LHC}(b);
(iii).\  for Sample-C, we consider the gluon-gluon fusion via
$\,gg\to \hh\to \Xh\Xh\,$ with $\,\Xh\Xh\to b\bar{b}b\bar{b}\,$,\,
which is depicted in Fig.\,\ref{fig:LHC}(c).

For each fusion process of Fig.\,\ref{fig:LHC},
we have computed the production cross section of $\,\hh\,$ and $\,\Xh\,$
for the relevant samples,
by including the full NLO QCD corrections as in \cite{Hdecay}.
Then, we multiply the production cross section by the decay branching fraction
of each final state.  For the final state decays into $b$-jets,
we have taken a $b$-tagging efficiency equal 60\% \cite{b-tag}
in our analysis; while for the final decay products being leptons,
we will just select electrons and muons, $\,\ell =e,\mu\,$.\,
We consider the LHC's center of mass energy at 7\,TeV for the current run, and
at 14\,TeV for its next phase \cite{LHC-2}.
These are summarized in Table\,\ref{tab:CS-BR}.
For the $\,\hh\to\ga\ga\,$ channel, we also list the results for the SM Higgs
boson (with the same mass) in parentheses as a comparison.  
For the $\Xh$ production in the $WW$ and $ZZ$ channels,
we find that the cross section times branching ratio has the following suppression
relative to that of the SM Higgs boson with the same mass,
for Sample-(A,\,B),
\beqs
\label{eq:X-WW/ZZ-sup}
\beqa
\label{eq:X-WW-sup}
U_{\phi \X}^2
\f{\textrm{Br}[\Xh\to WW]}{~\textrm{Br}[h\to WW]_{\textrm{SM}}^{}\,}
&\!\simeq\!& (0.290,\, 0.0139) \,,
\\[1.5mm]
\label{eq:X-ZZ-sup}
U_{\phi \X}^2
\f{\textrm{Br}[\Xh\to ZZ]}{~\textrm{Br}[h\to ZZ]_{\textrm{SM}}^{}\,}
&\!\simeq\!& (0.301,\, 0.0141) \,.
\eeqa
\eeqs
This shows that the signal rate of $\,\Xh\,$ over that of the SM Higgs boson
is $29-30\%$\, for Sample-A and decreases to about \,$1.4\%$\, for Sample-B,
in both $WW$ and $ZZ$ channels.
So, detecting new Higgs boson $\,\Xh\,$ in these channels
will require higher integrated luminosities at the 7\,TeV LHC.
From Table\,\ref{tab:CS-BR}, we see that
the process $\,gg\to\Xh\to\hh\hh\to b\bar{b}b\bar{b}\,$ for Sample-B
has lower rate and is hard to detect at the 7\,TeV LHC;
but the 14\,TeV LHC will have larger signal rate by a
factor of $\,4.5$.\,
For Sample-C, the $\Xh$ boson only weighs about 59.4\,GeV,
and thus the best channel should be
$\,gg\to\hh\to\Xh\Xh\to b\bar{b}b\bar{b}\,$,
which has large signal rates even at the 7\,TeV LHC, about  $111$\,fb,
as shown in Table\,\ref{tab:CS-BR}.
The major concern would be the SM $4b$-backgrounds since
the signal contains relatively soft $b$-jets from the light $\,\Xh\,$ decays.

\begin{table}[t]
\begin{center}
\begin{tabular}{c|c||c|c||c|c||c|c|c||c}
\hline
\hline
\multicolumn{2}{c||}{} & \multicolumn{2}{c||}{} &
\multicolumn{2}{c||}{} & \multicolumn{3}{c||}{} &
\\[-2.4mm]
\multicolumn{2}{c||}{$gg\to\hh$ or $\Xh$}
& \multicolumn{2}{c||}{$\hh\to\gamma\gamma$}
& \multicolumn{2}{c||}{~$\Xh\to WW$~}
& \multicolumn{3}{c||}{$\Xh\to ZZ$}
& \multicolumn{1}{c}{~$\to\hh\hh$ or $\Xh\Xh$~}
\tabularnewline
\hline
\multicolumn{2}{c||}{} & \multicolumn{2}{c||}{} &&&&&&
\\[-2.9mm]
\multicolumn{2}{c||}{Final State} &
\multicolumn{2}{c||}{$\ga\ga$~ (SM)}
& $\ell\nu\ell\nu$ & $\ell\nu jj$ & $\ell\ell jj$ & $\ell\ell\nu\nu$
& $\ell\ell\ell\ell$ & $b\bar{b}b\bar{b}$
\tabularnewline
\hline
\hline
\multirow{2}{*}{Sample-A} & 7\,{TeV} &
\multicolumn{2}{c||}{$ 26.0$  ($37.5$)} & $50.2$ & $319$ & $38.2$ & $10.9$ & $1.84$ & $\slash$ \tabularnewline
\cline{2-10}
 & ~14\,{TeV}~ &
\multicolumn{2}{c||}{$ 84.3$  ($122$)} & $195$ & $1230$ & $148$ & $42.4$ & $7.13$ & $\slash$ \tabularnewline
\hline
\multirow{2}{*}{Sample-B}
 & 7\,{TeV} &
\multicolumn{2}{c||}{~$ 34.7$  ($36.0$)~~} & ~$1.22$~ & ~$7.75$~ & ~$1.11$~ & ~$0.316$~ & ~$0.0532$~ & ~$0.203~ $
 \tabularnewline
\cline{2-10}
 & 14\,{TeV} &
\multicolumn{2}{c||}{$ 113$ ($118$)} & $5.41$ & $34.3$ & $4.89$ & $1.40$ & $0.236$ & $0.901 $
 \tabularnewline
\hline
\multirow{2}{*}{Sample-C}
 & 7\,{TeV} &
\multicolumn{2}{c||}{$ 23.6$  ($28.0$)} & $\slash$ & $\slash$ & $\slash$ & $\slash$
& $\slash$ & $111 $
 \tabularnewline
\cline{2-10}
 & ~14\,{TeV}~ &
\multicolumn{2}{c||}{~$79.2$  ($93.9$)~~} & $\slash$ & $\slash$ & $\slash$ & $\slash$
& $\slash$ & $373$
 \tabularnewline
\hline\hline
\end{tabular}
\caption{Higgs signatures for the LHC discovery via fusion processes
$\,gg\to \hh\to \ga\ga,\Xh\Xh\,$ and $\,gg\to \Xh\to WW,ZZ,\hh\hh\,$.\,
For each sample in every channel, the cross section times decay branching ratios
are shown in unit of \,fb\,. For $\,\hh\to\ga\ga\,$ channel, we also list
the signal rates of the SM Higgs boson in parentheses for comparison.}
\label{tab:CS-BR}
\end{center}
\vspace*{-3.5mm}
\end{table}

For this channel $\,gg\to\hh\to\Xh\Xh\to b\bar{b}b\bar{b}\,$,\,
we note that it may also be probed at Fermilab Tevatron,
which has recorded about $10$\,fb$^{-1}$ data in both CDF and
D0 detectors by the end of this summer\,\cite{Tevatron}.
For Sample-C, we find the production cross section of $\,gg\to\hh\,$
with $\,m_h^{}=136\,$GeV to be about $736$\,fb at Tevatron. Including the
decay branching fractions $\,\textrm{Br}[\hh\to\Xh\Xh] = 10.2\%\,$ and
$\,\textrm{Br}[\Xh\to b\bar{b}] = 80.5\%\,$,\, and a $60\%$ $b$-tagging efficiency,
we estimate the effective signal cross section to be $\,6.3$\,fb\,.\,
For a $10$\,fb$^{-1}$ data, we would expect about $63$ events for the $4b$ final states
from this process in each detector, so we encourage the Tevatron colleagues to
analyze such $4b$ events from their complete data set.
But one should keep in mind that since Sample-C predicts
a rather light singlet Higgs boson $\,\Xh\,$ weighing about $59.4$\,GeV,
the $b$-jets in its decay products will be relatively soft, with energy less than $30$\,GeV
and transverse momentum not much larger than $15\sim 20$\,GeV.
This differs a lot from the $b$-jets out of direct $\hh$ decays in the process of
Fig.\,\ref{fig:LHC}(b) for Sample-B,
where $\,\hh\,$ weighs about $125$\,GeV and the resultant
$b$-jets are hard\footnote{For models with large $\,\hh b\bar{b}\,$ Yukawa couplings,
the $\,\hh b\bar{b}\,$ associate production process is useful at the
Tevatron and LHC\,\cite{hbb}. There the $\hh\to b\bar{b}$ decays generate much harder
$b$-jets since $\hh$ usually has its mass obey the LEP limit,
$\,m_h^{}>114.4\,$GeV.}.   This makes it harder to reconstruct
such a light $\,\Xh\,$ resonance of Sample-C above the background $b$-jets.
At the LHC, the backgrounds with relatively soft $b$-jets are expected to be larger
and thus more challenging. We encourage systematical Monte Carlo analyses for both
Tevatron and LHC detectors to optimize the signals of
$\,gg\to\hh\to\Xh\Xh\to b\bar{b}b\bar{b}\,$
and pin down their $4b$ backgrounds.

\vspace*{2mm}
Before concluding this section, we notice that after this work initially appeared in
arXiv:1110.6893 on October\,31, 2011, the ATLAS and CMS collaborations announced
new results for the Higgs searches at the LHC\,(7\,TeV)
on December\,13, 2011\,\cite{LHC7new}, which showed some interesting excesses
of events for a Higgs boson mass around $125$\,GeV in the diphoton channel,
although statistically inconclusive\,\cite{LHC7new}.\,
We have shown that a SM-like Higgs boson with mass about 125\,GeV
is just in the favored parameter space of the present model, as given by our Sample-B
(cf.\ Table\,\ref{tab:input-ABC}-\ref{tab:output-ABC} in Sec.\,\ref{sec:4-2}).
The Higgs boson $\,\hh\,$ weighs 122\,GeV and 136\,GeV in Sample-A and -C,
respectively. These two samples predict significantly lower diphoton signals than
the SM Higgs boson [cf.\ (\ref{eq:h-2photon-sup})]. If the current excesses
of events at the LHC\,(7\,TeV) are disconfirmed by the upcoming data, our
Sample-A and -C can provide additional viable Higgs candidates.
We expect that the new LHC runs at the collision energy of 8\,TeV \cite{LHC2012}
will further test the predicted Higgs signals of both $\,\hh\,$ and $\,\Xh\,$
in the Samples (A,\,B,\,C) during this year.

\vspace*{3mm}
\section{Direct Detection of GeV-Scale Mirror Dark Matter}
\label{sec:6}
\vspace*{1.5mm}

In this section, we first estimate the abundance of
mirror helium dark matter [Sec.\,\ref{sec:6-1}].
Then, we analyze direct detections of the GeV-scale mirror helium dark matter
in Sec.\,\ref{sec:6-2}, especially, the new constraints from TEXONO \cite{TEX}
and the upcoming tests by CDEX \cite{JP}.
We will study processes via the Higgs-exchange-induced scattering
and the $\ga\!-\!\ga'$ mixing-induced scattering.
We reveal that the cross section of $\ga\!-\!\ga'$ mixing induced scattering
is enhanced in the low recoil-energy region relative to that of the Higgs-exchange,
and is thus sensitive to the direct detections.

\vspace*{3mm}
\subsection{Abundance of Mirror Dark Matter Particles}
\label{sec:6-1}
\vspace*{1.5mm}

In the visible world, the lightest baryon is proton, and after the ordinary BBN
the matter will be mainly composed of ordinary hydrogen atoms.
As discussed in Sec.\,\ref{sec:3-3} [cf.\ (\ref{eq:T'/T-EWreheating})],
the temperature $\,T'\,$ of the mirror world is lower than
the corresponding temperature $\,T\,$ in the visible world by about a factor-2
after the electroweak phase transition. This will cause significant difference
in the mirror BBN. To be concrete, we know that before the mirror BBN,
the mirror protons and neutrons will convert into each other via reactions,
$\,n' \leftrightarrow p' + e^{\prime -} + \bar{\nu}_e'\,$,\,
$\,n' + \nu'_e \leftrightarrow  p' + e^{\prime\, -} \,$,\, and
$\,n' + e^{\prime\, +} \leftrightarrow p' + \bar{\nu}_e' \,$.\,
As the universe expands, the temperature decreases and the cross sections of
these processes become smaller. When the reaction rate becomes comparable to
the Hubble expansion rate $H$, these reactions will be frozen and the mirror neutrons
will decay freely until the mirror BBN starts, during which the mirror protons and neutrons
form the mirror nucleus. Let us denote the freeze-out temperature of mirror sector as $\,T_f'$,\,
then from the distribution of kinetic equilibrium we can infer the ratio
between the number densities of mirror protons and neutrons at freeze-out,
\beqa
\label{eq:n'/p'-Tf'}
\frac{\,n_{n'}^{}}{\,n_{p'}^{}}
~\simeq~ \exp \left(\! -\frac{\,\Delta m'\,}{\,T'_f\,} \!\right) ,
\eeqa
where the mass-difference $\,\Delta m' =m_{n'}^{} - m_{p'}^{}\,$.
The \,$n-p$\, mass-difference $\,\Delta m = m_{n}^{} - m_{p}^{}\,$
is mainly caused by the mass-difference between
the current quarks $d$ and $u$, namely, $\,m_{d}^{}-m_{u}^{}\varpropto \vp\,$.\,
Similarly, the \,$n'-p'$\, mass-difference $\,\Delta m'\,$ mainly arises from
$\,m_{d'}^{}-m_{u'}^{}\varpropto \vpb\,$.\,
Thus, we expect the ratio, $\,\Delta m'/\Delta m\,\sim \vpb/\vp \sim 1/2$\,,\,
in our construction.

Then, we need to estimate the freeze-out temperature $\,T_{f}'$\,
for mirror protons and neutrons. We note that in the visible sector of the
universe, the neutrons and protons freeze out at the temperature
$\,T_{f}^{}\sim 0.8\,$MeV \cite{KT-book}. Then, from the freeze-out to the start of
BBN (at $\,T_{\rm NUC}^{}\sim 0.1$\,MeV), the neutrons decay freely in this period,
and decrease the neutron-to-proton ratio from about $\,1/6$\, to $\,1/7\,$.\,
For the nucleosynthesis, essentially all neutrons combine with protons into $^4$He\,,\,
the resulting mass-abundance of $^4$He is \cite{KT-book},
\beqa
\label{eq:Y-4He}
Y_{\textrm{He}^4}^{} ~\simeq~  \frac{\,4(n_{n}^{}/2)\,}{\,n_{n}^{}\!+\!n_{p}^{}\,}
~=~
\frac{2({\,n_{n}^{}}/{n_{p}^{}})_{\rm NUC}^{}}
         {\,1 \!+\! ({\,n_{n}^{}}/{n_{p}^{}})_{\rm NUC}^{}\,}
~\simeq~ 25\% \,.
\eeqa
This means that the visible universe is dominated by the hydrogens which have
a mass-fraction about 75\%\,.\,
As mentioned above, in the mirror sector the equilibrium of
mirror neutrons and protons is maintained
by the $\beta$-decay, inverse decay and the collision process,
among which the collision process is most relevant.
The collision rate $\,\Gamma_{p'e'\rightarrow n'\nu'}$\, (per nucleon per time)
can be expressed as,
\beqa
\label{eq:rate-pe-nnu}
\Gamma_{p'e'\rightarrow n'\nu'}^{} \,=\, (\tau_{n'}^{}\lambda_{0}^{})^{-1}\!\!
\int_{q}^{\infty}\!\!\! dy\,
\frac{y(y\!-\!q)^{2}(y^{2}\!-\!1)^{1/2}}
     {\,\left[1\!+\!\exp(yz)\right]\left[1\!+\!\exp((q\!-\!y)z_{\nu'}^{})\right]\,} \,,
\eeqa
where the ratios $\,q={\Delta m'}/{m_{e'}^{}}$,\,  $y={E_{e'}^{}}/{m_{e'}^{}}$,\,
$z={m_{e'}^{}}/{T'}$\,,\, and $\,z_{\nu'}^{}={m_{e'}^{}}/{T_{\nu'}^{}}$\,.\,
In (\ref{eq:rate-pe-nnu}), $\tau_{n'}^{}$ is the mean lifetime of mirror neutrons,
\beqa
\label{eq:tau-n'}
\tau_{n'}^{-1} ~=~ \Gamma_{n'\rightarrow p'e'\bar{\nu}'}  ~=~
\frac{G_{F}^{\prime\,2}}{\,2\pi^{3}\,}
\left(1+3g_{A}^{\prime\,2}\right) m_{e'}^{5} \lambda_{0}^{} \,,
\eeqa
where
$\,\lambda_{0}^{} \equiv \int_{1}^{q} dy\, y(y-q)^{2} (y^{2}-1)^{\hf}\simeq 1.636$\,,\,
and $\,g_A'\,$ is the axial-vector coupling of mirror nucleon. Since the mirror
and visible strong forces have the same coupling strength as required by the
mirror parity, we have $\,g_A'=g_A^{}\simeq 1.26\,$.\,
The collision rate (\ref{eq:rate-pe-nnu}) can be evaluated numerically, and in
the high/low temperature limits, it is approximated as,
\beqa
\Gamma_{p'e'\rightarrow n'\nu'} \,=\,
\begin{cases}
~\tau_{n'}^{-1}(T'/m_{e'}^{})^{3} \exp\left(-\Delta m'/T'\right)\,,
&~~~~ T'\ll\Delta m' ,\,m_{e'}^{}\,,
\\[3mm]
\dis~\frac{\,7\pi\,}{60}\(1+3g_{A}^{\prime\,2}\)G_{F}^{\prime\,2}\,T^{\prime\,5} \,,
&~~~~ T' \gg\Delta m' ,\,m_{e'}^{}\,,
\end{cases}
\eeqa
similar to that for the visible sector \cite{KT-book}.
Then, the freeze-out temperature $\,T_{f}'$\, can be estimated
by matching the collision rate and the Hubble expansion rate,
\begin{equation}
\label{eq:freeze-out}
\Gamma_{p'e'\rightarrow n'\nu'}^{} (T_{f}') ~\sim~ H(T_{f}') \,,
\end{equation}
where
\beqa
\label{eq:H-BBN}
H(T') ~=~ \sqrt{\frac{4\pi^3}{45}}
\frac{\,\,\sqrt{\hat{g}_{*}'}\,T^{\prime\,2}\,}{M_{\rm Pl}} \,,
\eeqa
and $\,\hat{g}_{*}'\,$ is the effective relativistic massless degrees of freedom
in the mirror sector,
$\,\hat{g}_{*}'=10.75[1 + (T/T')^4]\,$.
For our construction $\,T/T'\simeq \vp/\vpb\,$ and
$\,x\equiv \vp/\vpb =2\,$,\, so we have $\,\hat{g}_{*}'\simeq 182.75\,$.\,
In the visible sector, we have
$\,\hat{g}_{*}^{}=10.75\left[1+(T'/T)^4\right]\simeq 11.4\,$ for $\,T'/T=1/2\,$,\,
where, as expected, the effective contribution from the mirror sector
is mainly negligible.
So, the condition $\,\Gamma_{pe\rightarrow n\nu}^{} (T_{f}^{}) \,\sim\, H(T_{f}^{})\,$
determines the freeze-out temperature of visible protons and neutrons,
$\,T_f^{}\sim 0.8$\,MeV,\, as in the standard cosmology\,\cite{KT-book}.
We further note that the Fermi constants in the visible and mirror sectors are connected
by $\,G_F'/G_F^{}=(\vp/\vpb)^2$\,.\,
Taking all these into account, we estimate the freeze-out temperature
of the mirror neutrons and protons, $\,T'_{f} \sim 0.5\,$MeV\,.\,
With these, we can infer the mirror neutron-to-proton ratio at
the freeze-out from equation (\ref{eq:n'/p'-Tf'}),
\begin{equation}
\left(\frac{\,n_{n'}^{}}{\,n_{p'}^{}}\right)_{\!\text{freeze-out}}
\simeq~ \exp \left(\! -\frac{\,\Delta m'\,}{\,T'_f\,} \!\right)
~\simeq~ 28\% \,.
\end{equation}
Since the visible and mirror strong forces have the same strength, it expected that
the mirror nucleosynthesis starts at the same temperature as the visible sector (though
at an earlier time), i.e., $\,T'_{\rm NUC}=T_{\rm NUC}^{}\sim 0.1$\,MeV.
For the radiation-dominated epoch, we have
$\,H(T')=(4\pi^{3}/45)^{\hf}\sqrt{\hat{g}_*'}T^{\prime\,2}/M_{\rm Pl}$\, and
$\,t=[2H(T')]^{-1}_{}$.\,
Thus, we can estimate the time from $\,T^{'}=T_{f}^{'}\sim 0.5$\,MeV
to $\,T^{'}=T_{\rm NUC}^{'}\sim 0.1$\,MeV as, $\,\Delta t\sim 17.2\,$sec,
which is less than half minute. Using (\ref{eq:tau-n'}) for the lifetime of
mirror neutrons and the corresponding formula for visible neutrons, we estimate,
$\,\tau_{n'}^{}/\tau_n^{}= (G_F^{}/G_F')^2(m_e^{}/m_{e'}^{})^5
= \vp/\vpb =2\,$,\, and thus $\,\tau_{n'}^{}=2\tau_n^{}\simeq 1757\,$sec.
Thus, the fraction of decayed mirror neutrons from the freeze-out epoch
to nucleosysthesis epoch is about
$\,1-\exp(-\frac{17.2}{1757})\simeq 0.97\%\,$,\, which is negligible.
Hence, we have,
\beqa
\left(\frac{\,n_{n'}^{}}{\,n_{p'}^{}}\right)_{\text{NUC}} \simeq \, \left(\frac{\,n_{n'}^{}}{\,n_{p'}^{}}\right)_{\text{freeze-out}}
\simeq ~ 28\% \,.
\eeqa
Finally, we can estimate the mass-abundance of mirror helium $^4$He$'$\,,
\beqa
\label{eq:Y-4He'}
Y_{\textrm{He}^{4\prime}}^{}
~\simeq~  \frac{\,4(n_{n'}^{}/2)\,}{\,n_{n'}^{}\!+\!n_{p'}^{}\,}
~=~ \frac{2({\,n_{n'}^{}}/{n_{p'}^{}})_{\rm NUC}^{}}
         {\,1 \!+\! ({\,n_{n'}^{}}/{n_{p'}^{}})_{\rm NUC}^{}\,}
~\simeq~ 44\% \,.
\eeqa
This shows that the mirror sector has much larger amount of mirror helium than
the ordinary helium in the visible sector [cf.\ (\ref{eq:Y-4He})].
As we will analyze shortly, the ultra-low-energy germanium detectors of
TEXONO\,\cite{TEX} and CDEX\,\cite{JP} experiments will be most sensitive to the
the mirror heliums as the dark matter particles, since they are significantly
heavier than the mirror hydrogens.

Then, we estimate the mass of mirror helium dark matter.
From Eq.\,(\ref{eq:ratio-mN'/mN}), we can infer the ratio
between the mirror and visible nucleon masses,
\beqa
\label{eq:mN-mN'}
\f{\,m_{N'}^{}\,}{m_N^{}} ~=\, \(\f{\vpb}{\vp}\)^{\!\!2/9}
\simeq~ 0.60 - 0.92 \,,
\eeqa
where we have used VEV limit $\,0.1< {\vpb}/{\vp} <0.7\,$ in (\ref{eq:v'/v-ULB}),
which is based on the BBN constraint (\ref{eq:v/v'-range})
and the naturalness condition (\ref{eq:v'/v-lowerB}).
This means that the mirror helium \,$^4$He$'$\, should weigh about $60-92\%$
of the ordinary \,$^4$He\,,\, and thus has a mass around \,$3\,$GeV,
\beqa
\label{eq:M-He4-He4'}
M_{\textrm{He}^{4\prime}}^{} ~\simeq~ (0.60 - 0.92)M_{\textrm{He}^4}^{}
~\simeq~ 2.3-3.5\,\textrm{GeV} \,,
\eeqa
where our sample value $\,{\vpb}/{\vp}=\hf\,$ corresponds to
$\,M_{\textrm{He}^{4\prime}}^{} \,\simeq\, 0.86M_{\textrm{He}^4}^{}
\,\simeq\, 3.2\,\textrm{GeV}\,$.\,

\vspace*{3mm}
\subsection{Direct Detection of Mirror Helium Dark Matter}
\label{sec:6-2}
\vspace*{1.5mm}

In this subsection,
we study direct detections of the GeV-scale mirror helium dark matter,
especially, the new constraint from TEXONO \cite{TEX} and the upcoming probe
by CDEX \cite{JP}. We will study the Higgs-exchange-induced scattering
process and the $\ga\!-\!\ga'$ mixing-induced scattering process, respectively.

We first analyze the direct detection of Higgs-exchange-induced scattering.
As shown in (\ref{eq:mN-Lambda3}),
the mass of ordinary nuclei depends on the Higgs vacuum expectation value via
$\,m_{N}^{} \propto v_{\phi}^{2/9}(\Lambda^{(6)})^{21/27}$,\,
so the coupling of the Higgs boson with proton or neutron can be estimated
\cite{Shifman:1978zn} by using trace anomaly.
One may shift the vacuum expectation value as
$\,v_{\phi}\rightarrow v_{\phi}+\phi\,$,\, and consequently
the Yukawa coupling of Higgs boson with nuclei can be derived by variation,
\begin{eqnarray}
\label{eq:Coup-phiNN}
\lambda_{\phi NN}^{} ~=~
\frac{\,\partial m_{N}^{}\,}{\,\partial v_{\phi}^{}\,}
~=~ \frac{\,2m_N^{}\,}{9v_{\phi}^{}}   \,.
\end{eqnarray}
Also, from the trace anomaly we have \cite{TA-f},
\begin{eqnarray}
\label{eq:Coup-phiNN-TA}
\lambda_{\phi NN}^{} ~=~
\dis \f{1}{\vp}\langle N|\sum_{q}m_q^{}\bar{q}q|N\rangle
~\equiv~ \f{\,f m_N^{}\,}{\vp} \,,
\end{eqnarray}
where the coefficient $f$ characterizes the contribution of
trace anomaly and may be varied in the range,  $\,0.14<f<0.66\,$,\,
with a central value $\,f=0.30\,$ \cite{TA-f}.
This is consistent with (\ref{eq:Coup-phiNN}) where we have $\,f=\f{2}{9}\simeq 0.22\,$.
Similar to (\ref{eq:Coup-phiNN}), for the mirror Higgs coupling to the mirror nuclei,
we can deduce from (\ref{eq:mN'-Lambda3}),
\beqa
\lambda_{\phi' N'N'}^{} ~=~ \f{\,f' m_{N'}^{}\,}{\vpb} \,,
\eeqa
with $\,f'=f=\f{2}{9}\,$.\, In the following analysis we will set $\,f'=f=0.3\,$
for simplicity. With the Yukawa couplings $\,\lambda_{\phi NN}^{}\,$ and
$\,\lambda_{\phi' N'N'}^{}\,$ given above, we can estimate scattering cross
section of the mirror nucleus with the ordinary nucleus via Higgs exchange.

As shown earlier, we have estimated that after the mirror BBN, the mirror dark matter
mainly consists of the mirror helium \,$^4$He$'$ (with a mass fraction about $44\%$)
and the mirror hydrogen \,H$'$ (with a mass fraction about $56\%$).
The mirror hydrogen is significantly lighter than the mirror helium
according to (\ref{eq:mN-mN'})-(\ref{eq:M-He4-He4'}),
and thus harder to directly detect.
Hence, we will consider the mirror \,$^4$He$'$\, dark matter for the present analysis,
and estimate its scattering cross section in the detector.

We derive the Higgs-exchange-induced differential cross section as follows,
\begin{eqnarray}
\label{eq:dsigma}
d\sigma &\!=\!&
\frac{1}{\,4\pi v_0^2\,}
\left[\lambda_{\phi'p'p'}^{}Z' +\lambda_{\phi'n'n'}^{}(A'-Z')\right]^{2}
\left(\sum_{i}\frac{\,U_{\phi'i}U_{\phi i}\,}{m_{\phi_i}^{2}}\right)^{\!\!2}
\times
\nn\\[2mm]
&& \hspace*{11mm} \left[\lambda_{\phi pp}Z+
\lambda_{\phi nn}(A-Z)\right]^2F_{A^{\prime}}^{2}(Q)F_{A}^{2}(Q)\,dQ^{2} \,,
\end{eqnarray}
where $\,v_0^{}\,$ denotes the velocity of incident dark matter relative to the earth,
$\,(Z',\,A')=(2,\,4)\,$ for mirror helium nucleus, and
the subscript $\,i\,$ runs over the scalar mass eigenstates.
The function $\,F_A(Q)\,$ [$\,F_{A'}(Q)\,$]\, is the form factor of ordinary [mirror] nucleus,
defined as $\,F_A(Q)=\f{\,3j_1^{}(Q\,r_A^{})\,}{Q\,r_A^{}}\,e^{-\hf (Q s)^2}_{}\,$\,,\,
where $\,s=0.9\,$fm, and $\,r_A^{}\simeq 1.14A^{1/3}$\, \cite{FQ}.
Thus, it is found to monotonously increase as $\,Q^2\,$ decreases, and
$\,F_A(Q),\, F_{A'}(Q)\to 1\,$ for $\,Q^2\to 0$\,.

To compare with experiments of direct dark-matter detection, we should normalize
the above cross section to the cross section of mirror dark matter scattering on a proton.
So we apply (\ref{eq:dsigma}) and derive,
\begin{eqnarray}
d\sigma_p^{}\! &=\!&
\frac{1}{\,4\pi v_0^2\,}
  \lambda_{\phi pp}^2\left[\lambda_{\phi'p'p'}^{}Z'
 +\lambda_{\phi'n'n'}^{}(A'-Z')\right]^{2}
\left(\sum_{i}\frac{\,U_{\phi'i}U_{\phi i}\,}{m_{\phi_i}^{2}}\right)^{\!\!2}
 F_{A'}^{2}(Q)\,dQ^{2}
\nn\\[2mm]
&\simeq\! &
\frac{\,(\lambda_{\phi pp}^{}\lambda_{\phi' p'p'}^{}A')^2}{\,4\pi v_0^2\,}
\left(\sum_{i}\frac{\,U_{\phi'i}U_{\phi i}\,}{m_{\phi_i}^{2}}\right)^{\!\!2}
dQ^{2}
\,,
\end{eqnarray}
where in the second step we have used the relation
$\,\lambda_{\phi'p'p'}^{}\simeq \lambda_{\phi'n'n'}^{}\,$
due to $\,m_{p'}^{}\simeq m_{n'}^{}$, as well as
$\,F_{A'}^2(Q)\simeq 1\,$ due to $\,Q^2\simeq 0\,$.\,
Note that $\,Q < Q^{}_{\max}=2\mu_p^{}v_0^{}\,$,\,
where $\,\mu_p^{}\simeq 0.7\,$GeV is the reduced mass of ordinary proton with the
mirror helium $^4$He$'$ dark matter particle, and $\,v_0^{}\,$ is the dark matter
velocity relative to the earth. So, $\,v_0^{}\,$ should be smaller than the sum
of the dark matter's escape velocity ($\simeq 650$\,km/s) and
the relative velocity of sun ($\simeq 230$\,km/s) in the Milky Way.
Thus, we can derive,
$\,Q < Q^{}_{\max} < 4.1\,$MeV, for our case.
We have numerically checked that for $\,A=4\,$ and $\,Q \leqq 5\,$MeV,\,
the form factor $\,F_A^2(Q) \geqq 0.9991\,$ and thus
$\,F_A(Q)\simeq 1\,$ holds to high accuracy.
The form factor $\,F_{A'}(Q)\,$ for mirror nuclei should be similar, so we
expect that $\,F_{A'}^2(Q)\simeq 1\,$ holds well for $\,A'=4\,$ and $\,Q < 4.1\,$MeV,\,
in the case of mirror helium $^4$He$'$.
Integrating over $\,Q^2$\,,\, we arrive at,
\begin{eqnarray}
\sigma_p^{} ~\simeq\, \int_0^{Q^2_{\max}} \!\!dQ^{2}\,
\f{\,d\sigma_p^{}\,}{dQ^2}
~=~
 \frac{\,\mu_{p}^2\,(\lambda_{\phi pp}^{}\lambda_{\phi' p'p'}^{}A')^2\,}{\pi}
 \left(\sum_{i}\frac{\,U_{\phi'i}U_{\phi i}\,}{m_{\phi_i}^{2}}\right)^{\!\!2} ,
\end{eqnarray}
where $\,Q^{}_{\max}=2\mu_p^{}v_0^{}\,$.\,
Using the model-parameters of Sample-(A,\,B,\,C), we finally derive,
\begin{eqnarray}
M_{\textrm{He}^{4\prime}}^{} ~\simeq~
3.2\,\textrm{GeV},
&~~~~~~~&
\sigma_p^{} ~\simeq~
(1.4,\,3.4,\,7.6) \times\! 10^{-50}\,\textrm{cm}^{2}\,,
\end{eqnarray}
for Sample-(A,\,B,\,C), respectively.
The $\,\sigma_p^{}\,$ appears quite below the sensitivities of
current dark matter direct search experiments.

Alternatively, we note that the mirror dark matter may also be detected
via $\gamma\!-\!\gamma'$ mixing term (\ref{eq:Kmix-ga-ga'}).
The cross section of a mirror nucleus $(A',Z')$  scattering on an
ordinary nucleus $(A,Z)$ is,
\begin{eqnarray}
\label{eq:dsigma-photon-mix}
d\sigma ~=~
\frac{~4\pi\epsilon^2\alpha^{2}Z^{\prime2}Z^2\,}{Q^4\,v^2_0}
F_{A'}^{2}(Q)F_{A}^{2}(Q)\,dQ^{2} \,.
\end{eqnarray}
Due to the $\,Q^{4}$\, factor in the denominator, this differential cross section
receives a large enhancement in the low recoil-energy region relative
to the above cross section via Higgs exchanges. This will overcome
the large $\epsilon^2$ suppression in (\ref{eq:dsigma-photon-mix}) since
the $\gamma\!-\!\gamma'$ mixing parameter subjects to the experimental limit
$\,\epsilon < 3.4\times\! 10^{-5}\,$ in Eq.\,(\ref{eq:our-AA'-limit}).
This may be used to explain \cite{Foot:direct,An:2010kc,Panci:2011lrf}
the recent results from DAMA/LIBRA\,\cite{DAMA}, CoGeNT\,\cite{COG},
and CRESST\,\cite{CRE} experiments for the dark matter detection.
We further note that the ultra-low-energy germanium detectors of
TEXONO\,\cite{TEX} at Kuo-Sheng (KS) lab and of CDEX\,\cite{JP}
at Jinping deep underground lab (CJPL)
have a low recoil-energy threshold and are sensitive to the
light dark matter in \,$1\!-\!10$\,GeV\, mass range \cite{JP}.
This should be an ideal place to look for the GeV-scale mirror dark matter
as in (\ref{eq:M-He4-He4'}), via the $\gamma\!-\!\gamma'$ mixing induced scattering.

\begin{figure}[t]
\begin{centering}
\includegraphics[width=8.3cm,height=7.05cm,clip=true]{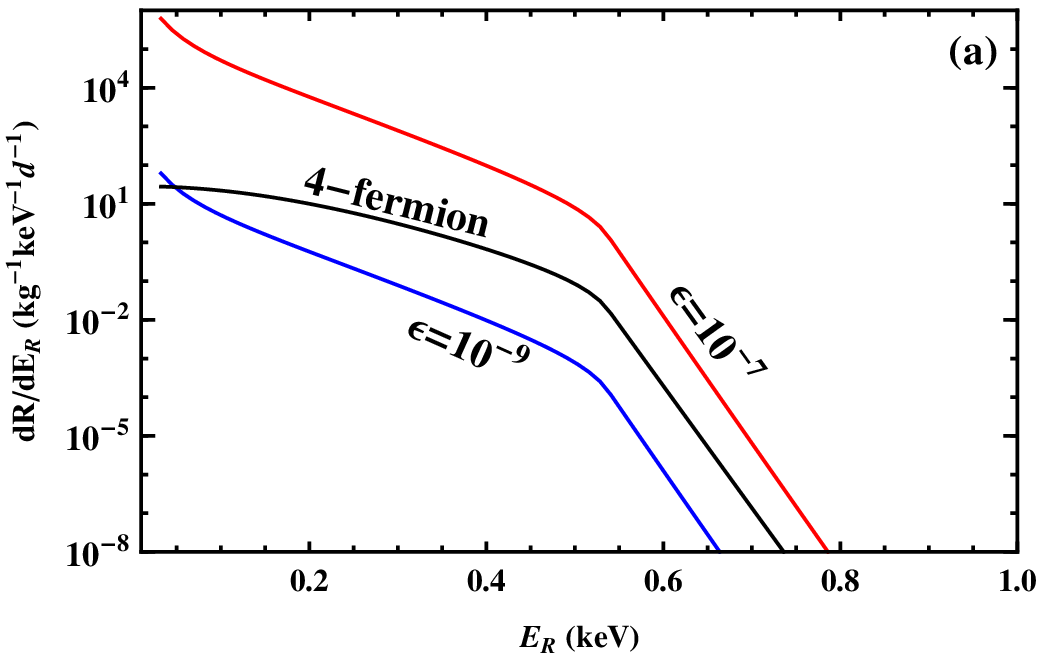}
\includegraphics[width=8.3cm,height=7.25cm,clip=true]{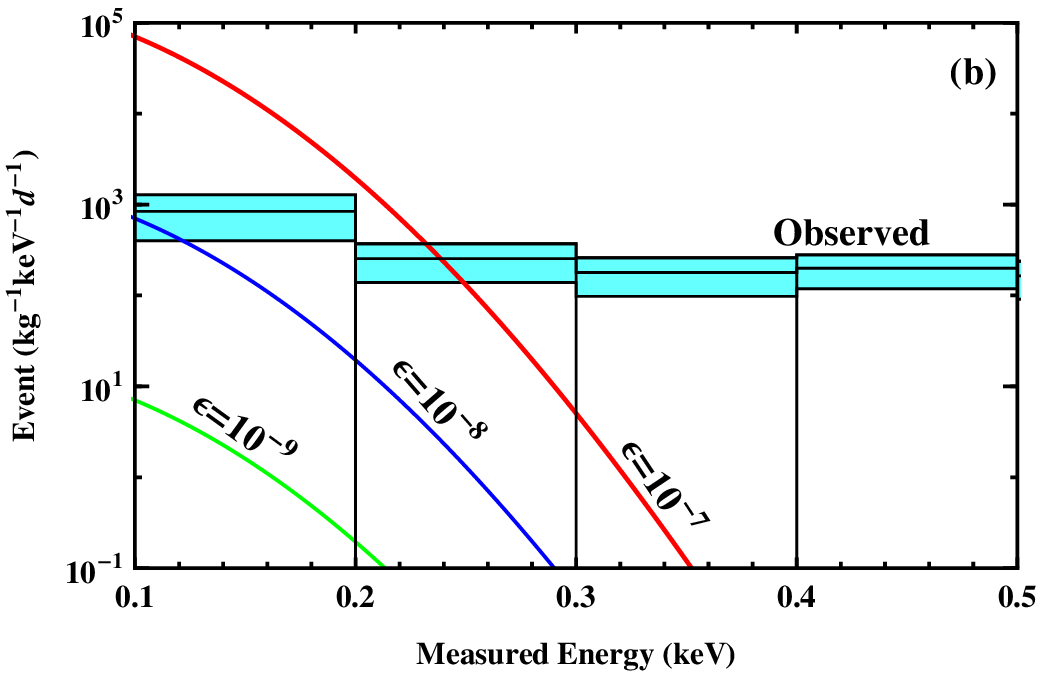}
\caption{Event rate distributions versus recoil energy.
Plot-(a) shows the event rate distributions (in unit of kg$^{-1}$keV$^{-1}$day$^{-1}$)
as a function of recoil-energy $E_R$ (in keV), for two different values of the
$\gamma\!-\!\gamma'$ mixing parameter $\,\ep\,$ (red and blue curves).\,
As a comparison, the distribution from a 4-Fermion interaction with an assumed
$\,\sigma_p^{} \simeq 10^{-38}$\,cm$^2$\, is shown by the black curve.
Plot-(b) depicts the event rate distributions as a function of
quenched recoil-energy, for three sample values of the mixing parameter $\,\ep$\,.\,
The observed event rate of TEXONO\,\cite{TEX} is shown by the black histogram,
and the shaded areas (light blue) are the experimentally allowed region
within $\,\pm 1\sigma\,$ errors.}
\label{fig:DR-ER}
\end{centering}
\vspace*{-3mm}
\end{figure}

For our analysis, we simulate the event rate distributions over the recoil energy $\,E_R\,$
for both $\gamma\!-\!\gamma'$ mixing induced interaction and the usual 4-Fermion interaction.
We show the results in Fig.\,\ref{fig:DR-ER} for the event rate
(in unit of kg$^{-1}$keV$^{-1}$day$^{-1}$) versus the recoil-energy (in keV),
for germanium detectors.
Since the mirror helium has the typical thermal energy (temperature) of
\,$O(\text{keV})$\, which is much larger than the ionization energy
(about $20-50$\,eV) of the $^4$He$'$ atoms,
we expect that all the mirror helium atoms get ionized.
So the mean mass of the particles composing the thermal mirror gas component
of the halo should be about one third of the $^4$He$'$ nuclei mass \cite{Foot:2004ej}.
According to (\ref{eq:M-He4-He4'}),
we have chosen here a sample mirror dark-matter mass as 3.2\,GeV.
The reduced mass for mirror helium with germanium is, $\,\mu'\simeq 3.1\,$GeV.\,
So we have $\,Q<Q_{\max}=2\mu'v_0\simeq 18.2\,$MeV, and for $\,A'=4\,$,
the mirror form factor $\,F_{A'}^2(Q)> 0.9879\,$.\,
The form factor $\,F_{A}^2(Q)\,$ for germanium in (\ref{eq:dsigma-photon-mix})
will be evaluated precisely.
In Fig.\,\ref{fig:DR-ER}(a), we have shown the rate distributions for the
$\gamma\!-\!\gamma'$ mixing parameter $\,\ep =10^{-7}\,$ (red curve) and
$\,\ep =10^{-9}\,$ (blue curve), respectively.
As a reference of comparison, we have also plotted the distribution
from a 4-Fermion interaction with an assumed
$\,\sigma_p^{} \simeq 10^{-38}$\,cm$^2$ (black curve).
We see that for the low recoil-energy region, the event rate of $\gamma-\gamma'$ interaction
is much larger than that of the 4-Fermion interaction. The TEXONO experiment\,\cite{TEX}
already put stringent limits on both spin-independent and spin-dependent cross-sections
for dark matter mass around \,$3-6$\,GeV,\,
where an energy threshold of \,($220\pm 10$)\,eV\, was achieved at an efficiency of \,50\%\,
with a four-channel ultralow-energy germanium detector, each with an active mass of 5\,g\,.\,

To compare with TEXONO detection\,\cite{TEX}, we show our simulated signals
and the observed experimental data of TEXONO in Fig.\,\ref{fig:DR-ER}(b).
Here the energy quenching factor is $\,0.2\,$ for the germanium detector,
and the energy resolution is given by \cite{TEX},
$\,\Delta E = (18.64 \sqrt{E} + 60) \times 10^{-3}$\,.\,
In Fig.\,\ref{fig:DR-ER}(b), we plot the predicted event rate distributions as a function of
quenched recoil-energy, for three sample values of the mixing parameter
$\,\ep = (10^{-7},\,10^{-8},\,10^{-9})$,\, in red, blue and green curves, respectively.
The observed event rate of TEXONO \cite{TEX} is depicted by the black histogram,
and the shaded areas (light blue) represent the experimentally allowed region
within $\,\pm 1\sigma\,$ errors.
From Fig.\ref{fig:DR-ER}(b), we see that the red curve with
$\,\epsilon = 10^{-7}\,$
is significantly above the experimental observation (black histogram with errors)
around the threshold, and is thus already excluded by TEXONO data.
But,  the blue and green curves in Fig.\ref{fig:DR-ER}(b) are
fully consistent with data. Using the TEXONO data\,\cite{TEX}, we can further derive
a $2\sigma$ upper bound on the range of $\gamma\!-\!\gamma'$ mixing parameter,
$\,\epsilon < 2.7\times 10^{-8}$\,.\,
Our predictions can be further explored
by the exciting on-going CDEX experiment in Jinping \cite{JP}.

\vspace*{4mm}

Before concluding this section, we clarify two issues related to the mirror
baryonic dark matter in general, although they are not particular to
our present model. The first one concerns the stability of the dark halo. As
the dark halo is assumed to be spherical and isothermal, if the dark matter is composed
of mirror elements (mainly H$'$ and He$'$), they would be ionized when the temperature $T'$
is much higher than their ionization energy. Thus the bremsstrahlung and other processes
can radiate off energies of the dark halo, so that it could not maintain its temperature
$T'$. This is known as the radiative cooling problem. This issue can be resolved by a
proper heating mechanism which prevents the collapse of dark halo \cite{MRev,Foot:2004wz}.
It was shown\,\cite{Foot:2004wz} that the energy released from both ordinary
and mirror types of supernovas are candidates for such heating sources.
Mirror supernovas can supply the energy if they occur at a rate of around one per year.
Alternatively, ordinary supernovas can do the job to heat the mirror dark matter
if the photon-mirror-photon kinetic mixing (\ref{eq:Kmix-ga-ga'}) is about
$\epsilon\sim 10^{-9}$.
This mixing can release a significant fraction of the total energy given
by supernova explosions into $e'^{\pm}$ and $\gamma'$, and these energies
can be absorbed by the halo.
In comparison with the visible sector,
since the mirror sector has lower temperature ($T'<T$)
and thus earlier mirror BBN, different light-element abundances,
lighter particle masses ($m'<m$), and much larger dark-matter density
($\Omega_{\rm DM}^{}\simeq 5\Omega_{\rm B}^{}$),
there is {\it no macroscopic mirror symmetry.}
So, it is quite expected that a significant asymmetry between the heating rates
in both sectors exists, which can explain
why the ordinary matter has collapsed into the disk and the mirror matter has not.

The second issue concerns structure formation of baryonic dark matter.
The standard model of cosmology suggests that the early universe is
extremely homogeneous, while the large structures we see today
(such as galaxies and clusters) arise from small
primordial inhomogeneities that grow via gravitational instability. The primordial acoustic perturbation cannot grow until the recombination of the protons and the electrons,
which occurs at a temperature around $\,T_{\rm dec}^{} \simeq 0.25$\,eV.
(Prior to the photon decoupling, the radiation pressure prevents the growth of perturbations).
But, the CMB data show that these perturbations do not have enough time to grow into galaxies.
So the standard model cosmology requires the primordial perturbations of cold dark matter
(instead of baryonic matter) to provide the seed of the large structure formation.
For mirror models with unbroken mirror electromagnetism,
mirror baryonic density can only begin to grow after mirror photon decoupling occurs
(roughly at \,$T'_{\rm dec} \approx 0.25$\,eV).
But, as the BBN constraint requires $\,T'<T\,$,\,
this means that mirror photons decouple \emph{earlier} than the visible photons.
It is shown\,\cite{MRev,mirror-LSS} that for $\,T'$ sizably below $T$\,,\,
the large scale structure formation with mirror dark matter closely resembles the
conventional cold dark matter scenario.
On the other hand, since mirror baryons can couple to visible photons through
the $\ga\!-\!\ga'$ kinetic mixing, they become millicharged particles,
having electric charges equal to $\epsilon$ times that of their visible partners.
Thus, there is a possibility that electric force may suppress the primordial perturbation
of the mirror baryons. As shown in \cite{McDermott:2010pa} by using the CMB data,
this imposes a constraint on the $\ga\!-\!\ga'$ mixing,
$\,\epsilon < (4-6)\times 10^{-6}\,$,\,
for the mirror baryonic dark matter in the mass-range of $\,2-4$\,GeV.
This is consistent with our model and is weaker than the limit derived
from the direct detections in the above Fig.\,\ref{fig:DR-ER}(b).

\vspace*{3mm}
\section{Conclusions}
\label{sec:7}
\vspace*{1.5mm}

The possible existence of a hidden mirror world in the universe is
a fundamental way to restore parity symmetry in weak interactions.
It naturally provides the lightest mirror nucleon as a
unique GeV-scale dark matter candidate.

We conjecture that {\it the mirror parity is respected by the fundamental interaction
Lagrangian,} so its violation only arises from spontaneous breaking of the Higgs vacuum,
and the possible soft breaking can only be linear or bilinear terms;
we further conjecture that {\it all possible soft breakings simply arise from
the gauge-singlet sector.}
With this conceptually simple and attractive conjecture,
we have studied spontaneous mirror parity violation in Sec.\,\ref{sec:2},
which quantitatively connects the visible and mirror neutrino seesaws
with the common origin of CP violation.
We presented systematical analysis of the minimal Higgs potential (\ref{eq:V-tot}),
which includes the visible/mirror Higgs doublets $\,\phi\,$ and $\,\phi'\,$
as well as a $P$-odd singlet scalar $\,\X\,$.\,
The singlet $\,\X\,$ develops a nonzero VEV $\,\vX\,$ at weak scale and generates
$\,\vp \neq \vpb\,$ as in (\ref{eq:vphi-vphi'}) and (\ref{eq:v-x-0})
[or (\ref {eq:vX-soft-sol})],
leading to spontaneous breaking of the mirror parity (cf.\ Fig.\,\ref{fig:V}).
The domain wall problem is resolved by a unique non-interacting soft breaking term
in (\ref{eq:V-soft}), and the usual vacuum degeneracy is removed
(cf.\ Fig.\,\ref{fig:V-soft}).
We have realized both the visible and dark matter
geneses from a common origin of CP violation in the neutrino seesaw
via leptogenesis (Sec.\,\ref{sec:3-1}).  We presented two explicit seesaw schemes which
generate successful visible and mirror leptogeneses with the common
CP violation (as well as $\mutau$ breaking) in Sec.\,\ref{sec:3-2}.
We found that the right amounts of visible and dark matter densities
\,($\,\Omega_{\textrm{DM}}:\Omega_{\rm M}\simeq 5:1\,$)\,
are generated in the parameter space with a natural ratio of Higgs VEVs
[cf.\ (\ref{eq:v'/v-ULB})] and a proper mass-ratio of singlet Majorana neutrinos
[cf.\ (\ref{eq:rN-bound})] between the visible and mirror sectors.
The constraints from BBN on the visible and mirror sectors are
further analyzed in Sec.\,\ref{sec:3-3}.

In Sec.\,\ref{sec:4-1}-\ref{sec:4-2}, we analyzed the analytical parameter space of the model
and explicitly constructed three numerical samples from the vacuum minimization,
which predict distinctive Higgs mass-spectrum and couplings,
as shown in Table\,\ref{tab:input-ABC}-\ref{tab:h-VV-ff-coupling} and Fig.\,\ref{fig:V}.
We also studied in Sec.\,\ref{sec:4-3} the low energy direct and indirect constraints on
the present model. We note that although the light mirror Higgs boson $\,\hh'\,$ safely
hides itself due to its small mixing of $\,O(10^{-2})\,$ with the visible Higgs $\hh$,\,
the $P$-odd singlet Higgs $\,\Xh\,$ (which generates unequal VEVs of $\,\hh\,$ and $\,\hh'\,$
and thus the spontaneous parity violation) has significant mixings with $\,\hh\,$,\,
as shown in (\ref{eq:U}). The Higgs boson $\,\X\,$ is particularly light in Sample-C
and thus the LEP production channel $\,e^-e^+\to Z\Xh\,$ (with $\,\Xh\to b\bar{b}\,$)
is open. But we found that the Sample-C prediction is well within
the LEP Higgs search limit (Fig.\,\ref{fig:LEP}).
We further analyzed the indirect electroweak precision constraints via oblique
corrections and found that the new contributions from our Higgs sector satisfy the
precision $\,\Delta S-\Delta T\,$ limit in Fig.\,\ref{fig:ST}.

In Sec.\,\ref{sec:5} we further studied the distinctive new Higgs signatures of the predicted
non-standard Higgs bosons at the LHC.
We systematically computed the Higgs decay widths and
branching fractions for all three samples as summarized in Table\,\ref{tab:Higgs-BR}.
Our construction always predicts a light mirror Higgs boson $\,\hh'\,$,\,
weighing about half of the $\hh$ mass due to the VEV condition (\ref{eq:v'/v-ULB})
with our sample value  $\,\vpb /\vp =\hf\,$;\,
but its mixing with $\,\hh\,$ is only of $\,O(10^{-2})$\, to satisfy the BBN constraint.
So, different from all previous mirror models, the decay channel
$\,\hh'\to\hh\hh\,$ and its inverse process $\,\hh\to\hh'\hh'\,$
are either forbidden or negligible. The mass of $\,\hh\,$ lies in the range around
$\,120-140\,$GeV, and its main LHC-production channel is
$\,gg\to \hh\,$  (with $\,\hh\to\ga\ga\,$ and $\,\hh\to VV^*\,$). 
As Eq.\,(\ref{eq:h-2photon-sup})
shows, relative to that of the SM Higgs boson,
the $\,\hh\,$ signal rate is suppressed by about
(31\%,\,4\%,\,14\%) in Sample-(A,\,B,\,C), respectively.
Besides, Sample-C has a new production
channel $\,gg\to \hh\to\Xh\Xh\to 4b\,$ [Fig.\,\ref{fig:LHC}(c)] with large signal rate
at the LHC (Table\,\ref{tab:CS-BR}), and is also potentially detectable at Tevatron.
We encourage systematical Monte Carlo analyses to pin down the $4b$  backgrounds
in this channel at both the LHC and Tevatron.
For the $P$-odd Higgs boson $\,\Xh\,$,\, its main production channels are
gluon-gluon fusions as shown in Fig.\,\ref{fig:LHC}(a)-(b) besides Fig.\,\ref{fig:LHC}(c),
and its signal rates for each final state are summarized in Table\,\ref{tab:CS-BR}.
 The $\,\Xh\,$ signal rates for both $\,WW\,$ and $\,ZZ\,$ final states are sizable
at the 7\,TeV and 14\,TeV LHC, but as Eq.\,(\ref{eq:X-WW/ZZ-sup}) shows,
they are suppressed relative to that of the SM Higgs boson (with the same mass)
by a factor about \,$29-30\%$\, and $\,1.4\%$\, in Sample-(A,\,B), respectively.
So, a higher integrated luminosity is required for their detection.
The other fusion channel $\,gg\to \Xh\to\hh\hh\to 4b\,$ in Fig.\,\ref{fig:LHC}(b)
is open for Sample-B, but with a relatively low signal rate
as shown in Table\,\ref{tab:CS-BR}.
The approved LHC runs with 8\,TeV collision energy \cite{LHC2012} will further
probe the predicted Higgs signals of our Samples A, B and C in this year.

Finally, in Sec.\,\ref{sec:6}, we have studied direct detections of the mirror dark matter,
which mainly consists of the mirror helium \,$^4$He$'$\, [with a mass-abundance about 44\%
as estimated in (\ref{eq:Y-4He'})] and the mirror hydrogen \,H$'$ (about 56\%).\,
The mass of mirror helium is around $\,3$\,GeV\, [cf.\ (\ref{eq:M-He4-He4'})].
Analyzing the scattering cross section of mirror helium with the nuclei in the
(germanium or xenon) detector via Higgs-exchanges shows that the signal is
quite below the sensitivities of the current dark matter direct search experiments.
But, it is important to note that the $\gamma\!-\!\gamma'$ mixing induced scattering
is enhanced in the low recoil-energy region relative to that of the Higgs-exchange
[Fig.\,\ref{fig:DR-ER}(a)].
We found that TEXONO experiment \cite{TEX} already puts nontrivial constraint on
the parameter space of $\gamma\!-\!\gamma'$ mixing, as shown in Fig.\,\ref{fig:DR-ER}(b).
It reveals that the parameter region with $\gamma\!-\!\gamma'$ mixing
$\,\epsilon \gtrsim 10^{-7}$\, is significantly excluded by TEXONO;
but the parameter space with
$\,\epsilon < 2.7\times 10^{-8}$\, is fully consistent with TEXONO data
at the $2\sigma$ level.
The on-going CDEX direct search experiment at CJPL deep underground lab
also has a low recoil-energy threshold and is sensitive to the
light dark matter in the mass range of \,$1\!-\!10$\,GeV\, \cite{JP}.
It thus provides the ideal place to further explore the GeV-scale mirror dark matter.
A summary of this work was presented in Ref.\,\cite{Cui:2012mq}.

\vspace*{6mm}
\noindent
\addcontentsline{toc}{section}{Acknowledgments\,}
{\bf\large Acknowledgments}
 \\[2mm]
 We thank Robert Foot, Peihong Gu, Rabindra N.\ Mohapatra,
 Qaisar Shafi, and Raymond R.\ Volkas for valuable discussions,
 and Liang Han for discussing the LHC and Tevatron Higgs searches.
 We are also grateful to Henry Wong, Qian Yue and Shin-Ted Lin for discussing the
 dark matter direct detections by TEXONO\,\cite{TEX} and CDEX\,\cite{JP} experiments, and
 especially, to Henry Wong for providing us the original data of TEXONO\,\cite{TEX}
 [used in our Fig.\,\ref{fig:DR-ER}(b)], and to him and Shin-Ted Lin for explaining
 their analysis in \cite{TEX}.
 HJH thanks the Center for High Energy Physics at Peking University for kind support.
 This research was supported by the NSF of China (under grants 10625522, 10635030, 11135003)
 and the National Basic Research Program of China (under grant 2010CB833000),
 and by Tsinghua University. JWC was supported in part by the Chinese Postdoctoral Science
 Foundation (under grant 20090460364).

\newpage

\end{document}